INVESTIGATING THE OPTIMAL NEURAL NETWORKS PARAMETERS FOR DECODING

By

JOSHUA TSHIFHIWA MAUMELA

A mini-dissertation submitted for the partial fulfilment of the requirements for the degree

BACCALAUREUS INGENERIAE

In

ELECTRICAL AND ELECTRONIC ENGINEERING SCIENCE

At the

UNIVERSITY OF JOHANNESBURG

STUDY LEADER: Mrs R HEYMANN

JUNE / NOVEMBER 2010

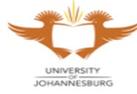

# ABSTRACT

Neural Networks have been proved to work as decoders in telecommunications, so the ways of making it efficient will be investigated in this thesis. The different parameters to maximize the Neural Network Decoder's efficiency will be investigated. The parameters will be tested for inversion errors only.

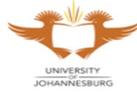

# ACKNOWLEDGEMENTS

I would like to give thanks to these people of whom without, this thesis would not have had been possible:

My Study Leader: Mrs Reolyn Heymann
Dr Fulufhelo Nelwamondo
My Parents: Mr Jonas R And Mrs Ndifelani E Maumela
My Brother: Dr Chris Munaka Maumela
Mr H Tshivhandekano
Fellow TRG 2010 Students
And more family and friends
With more regards and gratitude towards
Mr SO Ngwenya

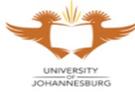

# TABLE OF CONTENTS

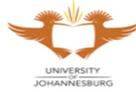

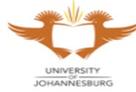

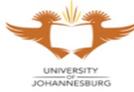

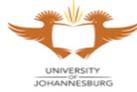

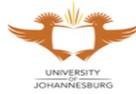

# List of Figures

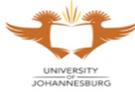

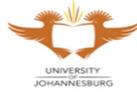
# List of Tables

# List of Symbols and Abbreviations

| | |
|---|---|
| ECSA | : Engineering Council of South Africa |
| BCE | : Before Common Era |
| BER | : Bit Error Rate |
| R | : Code Rate |
| $m$ | : Convolutional Encoder Memory Order |
| $k$ | : Number of Information Symbols in a Codeword |
| $n$ | : Number of Code Symbols |
| BN | : Biological Neuron |
| AN | : Artificial Neuron |
| ANN | : Artificial Neural Network |
| BNN | : Binary Neural Network |
| ADAM | : Advanced Distributed Associated Memory |
| RAM | : Random Access Memory |
| ECSA | : Engineering Council of South Africa |
| "group" | : This refers to the bits after they are separated from the main n-bit stream after errors were introduced in the channel. |
| NND | : Neural Network Decoder |

# Chapter 1: Introduction

## 1.1 Introduction

Telecommunications involve the process of sending and receiving information over a long distance. The olden forms of telecommunications involved the use of smoke, mirrors, pigeons and drums [**1**]. There had to be a certain code, secret or standard, to be followed when this communications were followed. The sender and the receiver had to all know what each code meant.

The 21$^{st}$ century telecommunications has really changed from the ones used in the olden days, which possibly trail back the BCE. According to the Concise Oxford dictionary, telecommunications can now be defined as "*the communication over a distance by cable, telegraph, telephone or broadcasting.*" [**2**].

Telecommunications can also be defined as the "*branch of electrical engineering concerned with the technology of electronic communications at a distance.*" [**3**]. The modern telecommunication system transmits information in either analogue or digital form. Transmitted information is converted into electrical energy for transmission. The main focus of information transmission is now in the digital field of information transmission.

## 1.2 Problem Statement

*"To err is human"* [**4**], due to statements and acknowledgements of such statements it was realised that there is nothing which can be perfectly engineered to be perfect. Every message being sent has the probabilities of getting errors.

The big and significant errors include deletions, insertions and inversions. A deletion is when one bit in the digital information being transmitted is accidentally deleted. This leads to the wrong information being received at the receiver side. Example

*Transmitted digital information:*

$$[1\ 0\ 1\ 0\ 1\ 0\ 1\ 0]$$

*Received digital information after deletion:*

$$[1\ 0\ 1\ 0\ 1\ 1\ 0]$$

An insertion is the case where an extra bit is accidentally added to the transmitted information. Example

*Transmitted digital information:*

$$[1\ 0\ 1\ 0\ 1\ 0\ 1\ 0]$$

*Received digital information after insertion*

$$[1\ 0\ 1\ 0\ 1\ 0\ 1\ 0\ 1]$$

With insertions and deletions errors all bits shift even though only one error might have been experienced.

Inversion is the case where the bits change from 1 to 0 or from 0 to 1
Example

*Transmitted digital information:*

$$[1\ 0\ 1\ 0\ 1\ 0\ 1\ 0]$$

*Received digital information after inversion:*

$$[1\ 0\ 1\ 0\ 1\ 0\ 1\ 1]$$

This errors need to be corrected as most of the information being transmitted is very important such that the wrong information can be very dangerous.

## 1.3 Project Objective

This thesis will focus in finding the optical parameters to be used when using neural networks as error decoders. These optimal values will be investigated for weightless or binary neural networks. The efficiency of these parameters should be visible at the end of this project. The other parameters had been tested before, but they are not usually slow and use too much memory, so this thesis should be able to point out why certain values should be chose or why and how to get these optimal values to improve the speed and decrease the memory usage by the decoder.

## 1.4 Scope of project

This project will not include the investigation of other decoding methods, but they will be discussed to give an idea of how decoders work and why use the neural network as the decoder. Neural network is the one which will be concentrated more on in this dissertation. The project scope indicates all the steps to be taken when solving the problem related to this thesis. Different kinds of neural networks will be investigated to see how they work in decoding.

This project however will only be limited to computer simulations as there will not be a physical circuit implemented. The simulations will not include the source codes, but only the channel decoding. The source coding will be assumed to have been effectively done.

## 1.5 Methodology Overview

The methodology will indicate all the steps to be taken to meet the constraints of the project and the steps to be taken if the results don't meet the constraints defined.

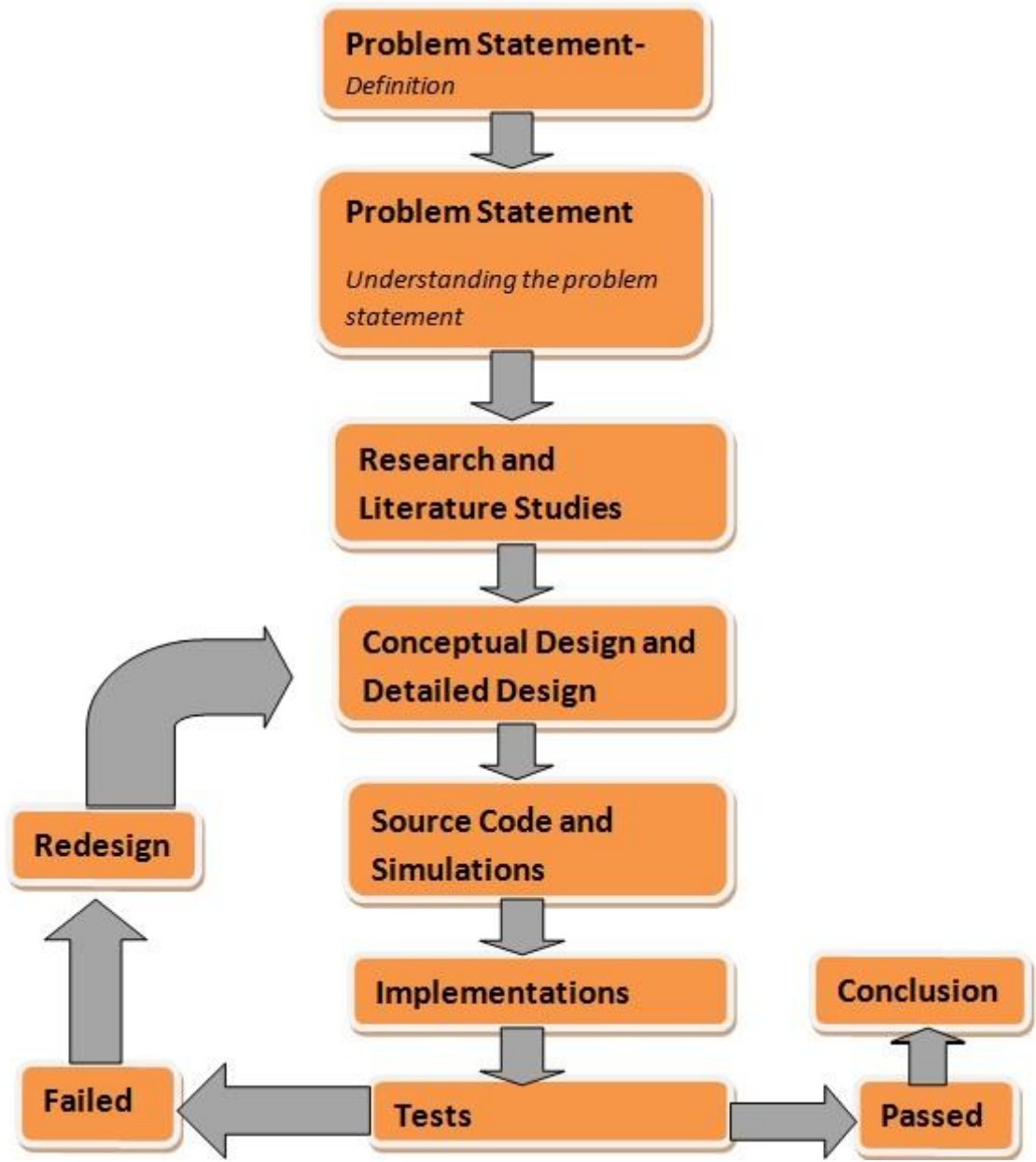

**Figure 1-1: Project Methodology**

The steps of the method to be followed in solving this engineering problem are broken down in the structure indicated in Figure 1-1.

Research will be done by looking at the work which has already been done. This information will be found mostly from journals and e-books. The library will also play a big role in writing the literature studies. The Telecommunication Research Group library at the University of Johannesburg will be visited for the past thesis on neural network which cannot be found in the library or online.

Source codes and simulations will mostly be done using MATLAB and Simulink.

## 1.6 Deliverables

Documentation with the simulations results will be handed in at the end of the academic year. The document will give details of the optimal value to be used in telecommunications error coding using neural networks. The simulation code will also be handed in together with the rest of the deliverables. The simulation code to be handed in is the code which would have been used in this project during the investigation.

## 1.7 ECSA outcomes

1. Engineering Problem Solving
   - Being able to identify and solve a not-well defined engineering problem
   - Errors are one of the main problems in engineering, so this project will help find a reliable way of solving this problem.
   - Solving a problem in engineering leads to uncovering a thousand folds of other problems, so to get to the project objective these problems need to be overcome.
2. Application of fundamental and specialist knowledge
   - The knowledge acquired from previous engineering modules must be able to be applied in this project.
   - All the mathematical and scientifically knowledge must be applied as seen fit in the project.
   - Telecommunication and information theory knowledge is very essential in this project.
   - This project will also require a lot of programming skills.

3. Engineering design and synthesis

    - Must be able to come up with a best solution for the design by following the constraints given in this thesis topic.
    - By finding the optimal value, this requirement would be met.

4. Investigation, Experiment and data analysis

    - Researches from which relevant information will be extracted and applied in this thesis.
    - Journals and articles published in the past will help to form foundation knowledge in approaching this problem.
    - Consultations with the experts in the telecommunication department will also be very helpful in this project.

5. Engineering methods, skills, tools and information technology

    - This project is an information theory project, so more information technology skill will be acquired at the end of this project.
    - Simulink and MATLAB which are simulation software will be used for the most part of the project

## 1.8 Overview of the document

Chapter 1: Introduction

- The problem statement and the project objective are discussed in full in this chapter.
- In this chapter the reason why this project is being done is discussed.
- This chapter gives more details on why this project topic is being investigated, and the disadvantages of not solving errors in telecommunications.

Chapter 2: Literature Study

- In this chapter, the works which have been done by others is discussed.
- Different methods of solving the error problems in telecommunications are discussed.
- It will mention the other methods of decoding to give a clear way to the adaption of neural network as decoders.

- These other methods will not be discussed in full length but rather they will be briefly introduced.

Chapter 3: Neural Networks

- A more detailed background of neural network will be discussed.
- More and detailed information on neural networks and their operations will be covered in this chapter.
- Different kinds of neural networks are also discussed in this chapter.
- The kind of neural network to be used in this thesis will be discussed and examples illustrating its operations will be given.
- This chapter will give an indication of how neural network are used in error correction.

Chapter 4: Detection and Correction of Inversion Errors Design

- The different designs of finding the optimal value are evaluated and the best one is chosen
- The detailed design approach will be discussed in this chapter.
- All focus will be on the kind of neural network to be used.
- The conditions and assumptions made in this project will be stated and the reasoning will be discussed.

Chapter 5: Software Implementation Design

- The block diagrams and the simulations are shown and the details of the simulations methods are discussed.
- The simulation components will be discussed and tested.
- The way to check how good the decoder is will be discussed.

Chapter 6: Results

- Results obtained from the simulations will be analysed and discussed in full details.
- The results obtained through the work done by Chapter 4 and Chapter 5 will be discussed.
- The best results will be indicated and

- The project objective will be concluded in this chapter.

Chapter 7: Conclusion

- The whole project will be discussed and any problem encountered will be discussed
- The fulfilment of the ECSA outcomes will be discussed in this chapter, and how they were fulfilled.
- Future work, if there is any will be discussed in the end of this chapter.

## 1.9 Conclusion

This chapter defines the problem statement and gives off the information of how the project design is going to be carried out. The scope of the project is included in this chapter to indicate what will be covered in this project, and then followed by the methodology which entails what steps have to be followed in this thesis.

The problem statement is the one that leads to the investigation of this topic. This section discussed the different kinds of errors in telecommunications and how these errors arise. This section lead to the next section which states how this thesis will help solve the problems discovered and stated in the problem statement.

The project objective then defined in full detailed how the problem would be solved by focusing the attention on the topic being investigated. The project scope put constraints to the project objective by stating the focus of the thesis. With the project scope defined, the method to be used in solving the problem defined in the problem statement is then discussed.

The methodology shows a block diagram indicating how the problem would be approached if the tests were to fail. When the methodology has passed, the results have to be handed in, in form of deliverables. So this chapter indicates what deliverables are to be handed in at the end of the thesis.

The expected ECSA outcomes of the project are then discussed, and how this project will help in meeting these outcomes. The overview of all chapters to be covered in this document is discussed.

The next chapter will discuss the work which have been done by others before and give an introduction to most of telecommunication terminology.

# Chapter 2: Literature Overview

## 2.1 Introduction

The objective of this chapter is to give the historical literature overview of different kinds of errors experienced in telecommunications. Synchronizations, inversion and substitution errors will be discussed, but the inversions and substitutions will just be given a brief overview of.

Block codes and Convolutional codes will be discussed as error correction schemes. The difference between Block codes and Convolutional codes will be shown in this chapter. Telecommunication has different terminologies used in error corrections; the different terminologies applicable to this thesis will be shown and explained in Section 2.4.
Lastly, the channel models that will be used for simulations will be discussed.

## 2.2 Insertions and deletions

Digital communications trail back to 1806 with telegraphy but since 1990's digital communication has rapidly grown as it started being applied in many different fields, such as computers, cellular phone and fax to name but a few. Internet packaging also use digital communications.

As the number of applications increased, telecommunications started experiencing problems when some bits were lost and some extra bits were being received increasing the size of the code. This inspired the visitation of the study of error-correction coding which Levenshtein first introduced in the 1960's [**5**] [**6**].

Uncorrected insertion/ deletion errors lead to much unnecessary data space being accumulated by wrong bits. This eventually leads to things like the internet [**7**] being slow. Insertion and deletion errors are part of a large family called synchronisation errors. They fall under a sub family branch called bit synchronisation.

Bit synchronisation is when the bit sent is the same as the bit received at the end of transmission, hence the main aim of insertion and deletion error coding is to get back synchronisation if it is lost.

Figure 2-1 is the schematic showing how information is sent and received and where errors are usually experienced in the system. Errors are mostly introduced in the channel.

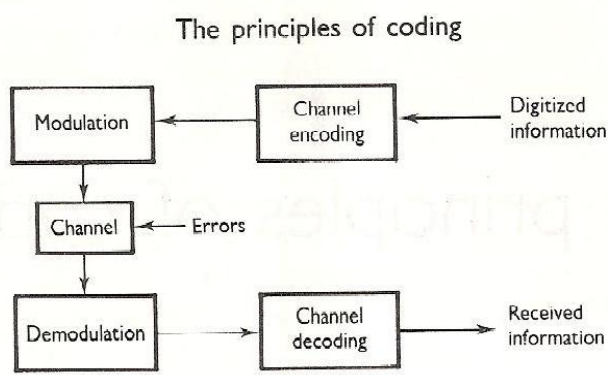

**Figure 2-1: Coding System** [8]

Just like in any other real system, in order for a system to correct errors, there must first be a system which will be able to detect these errors. The fundamental error-check system uses the redundancy. Redundancies [**9**] are made so that their size will be the same size as the sent/transmitted message. Redundancies are basically encoded in the message but they don't represent information.

Since deletion and insertion errors fall under the synchronisation errors, they are solved using synchronisation techniques. Synchronous errors either occur randomly or in a burst, where they occur close to each other. The reason why these errors are classified under synchronous errors is because of the definition of synchronisation.

Synchronisation according to [**2**] is the process where two processes occur at the same time. But the more formal definition of communication synchronisation can be obtained from [**10**] which say "*Two sequences are synchronous relative to each other if the events at the sender correspond with the events at the receiver. Synchronization is the process of establishing this situation between the sender and the receiver and maintaining it*".

Synchronisation process will be explained and illustrated with the aid of diagrams in Figure 2-2 and Figure 2-3 where the top row indicates sent data with the bottom one indicating received data.

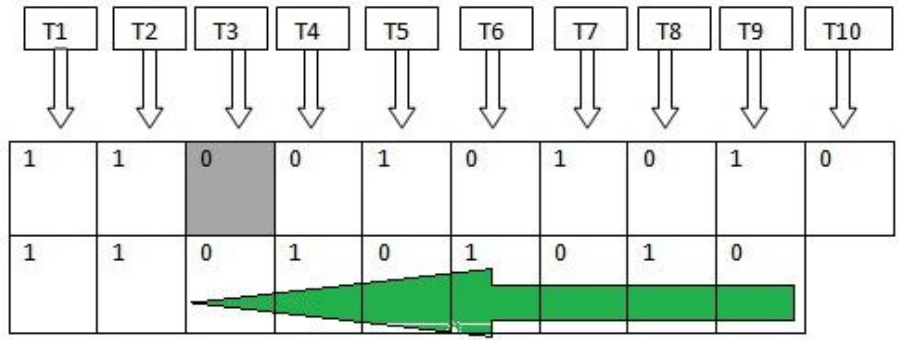

**Figure 2-2: Diagram illustrating of Loss of Synchronisation due to Bit Deletion**

At T3 of Figure 2-2 a bit will be deleted. When that bit gets deleted the bits in the blocks with the arrow move to the left replacing the bit that has been removed. It can be seen at time T10 that the received data does not occur at the same time with the sent data; hence the received data is out of synch with the sent data, hence the term synchronisation error.

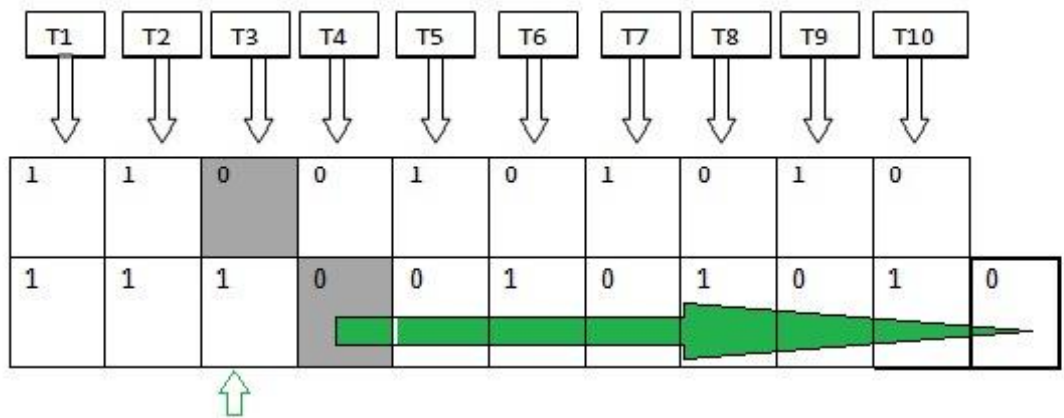

**Figure 2-3: Diagram illustrating Loss of Synchronisation due to Bit Insertion**

At T3 of Figure 2-3 a bit is inserted. The inserted bit is indicated at the bottom by a small arrow. When the bit is inserted the bits which were originally tin T3 move to T4 and with the next one moving to the next block. This is indicated by the arrow pointing in the right direction. An extra bit is introduced increasing the time to T11 which then causes the system to be out of synch.

Through these diagram illustrations it's clear that synchronisation errors have an effect on the size of the data being received with deletion reducing the size and insertion increasing the size of the data.

This whole process can be simplified and indicated in graph indicating all the insertion and deletion in terms of straight lines.

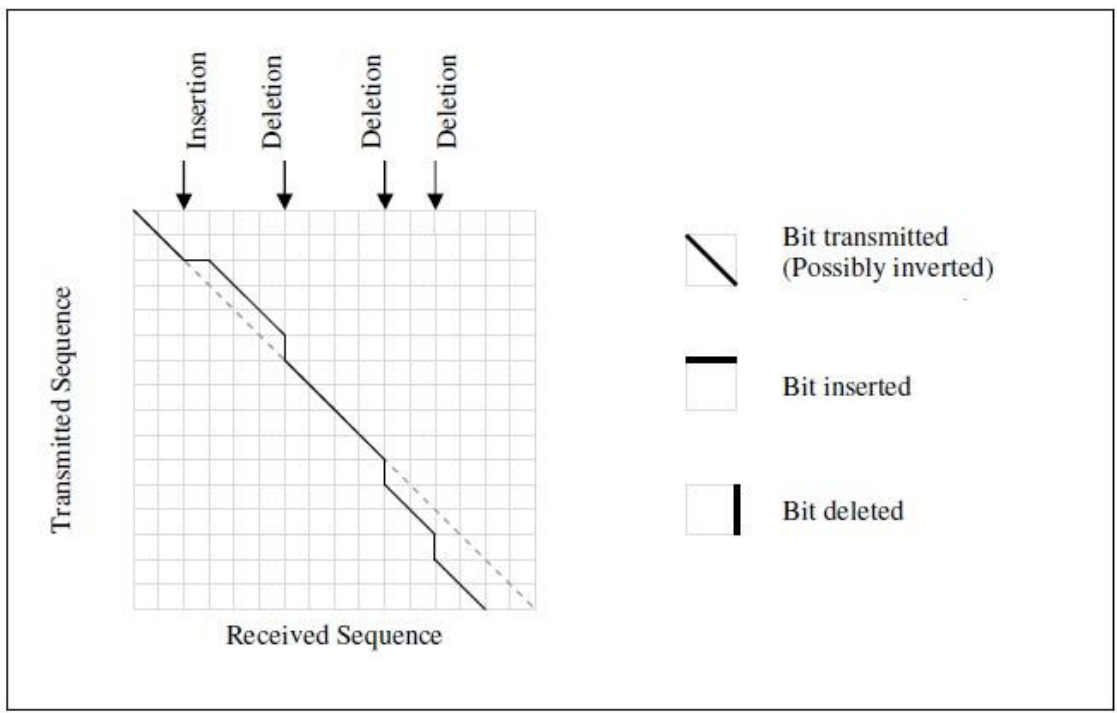

**Figure 2-4: Synchronization Graph [11]**

For the data to be synchronised, the received data (solid line) must run along the dotted line. Insertion and deletion behaviour of the graph is indicated on the graph.

## 2.3 Additive Errors

Additive errors include inversion errors and substitution errors. These errors occurred when one or more bits in the transmitted message are replaced by another bit. Unlike synchronisation errors, the number of bits when additive errors occur remains the same.

Even though the number of bits is still the same, the message is incorrect and hence a wrong message will be received by the receiver. The consequences are still the same as discussed in [**7**] for deletion errors, i.e. it slows down the internet. Most decoders used to decode synchronisation errors had been adapted to solve additive errors.

The additive errors will be explained with an aid of a diagram below. The top row indicates the transmitted message and the bottom row indicates the received message with errors.

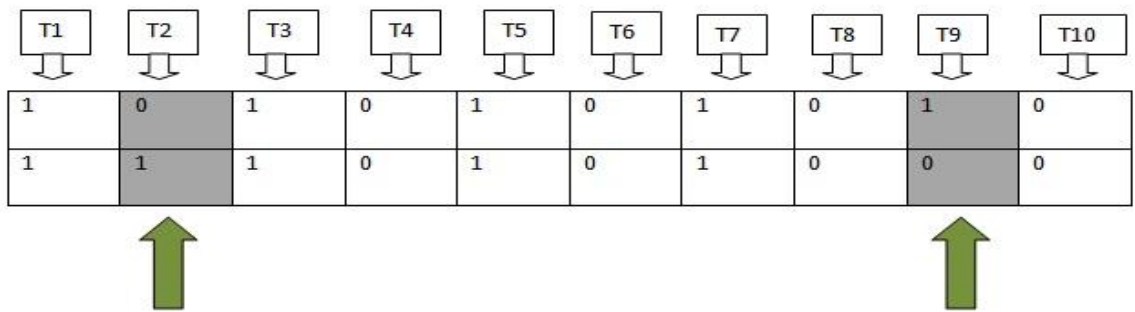

**Figure 2-5: Diagram illustrating Additive errors**

Inversion occurs at T2, where the 0 is replaced by a 1 and at T9 where 1 is replaced by a 0. This means that this message has two inversion errors.

In rare cases, additive errors can result in the received bits as a result of deletion and insertion of equal number of bits in the message, i.e. if there are two deletions and two insertions in the same message. In this special case the received message with errors will be decoded as inversion error.

This case is a very special case, as the decoder might not be able to decode the message correctly.

## 2.4 Terminology

Different people tend to use different terminologies for the same concepts in error coding, sometimes they use the same for terminology for different concepts, and that can get confusing sometimes. To minimize such confusions, this section of this chapter will be used to explain different, familiar and unfamiliar terminology and concepts as they will be applied for the rest of this thesis.

**Redundancy-** *it's the extra bits which are put in the information being sent, these extra bits are put in just in case the original ones fail. After that they are no longer useful.*

**Hamming distance**- *it's a measure of the difference between codes and is calculated by counting how many bits have to be changed to get from one code to another.*

**Parity checking**- *if even parity is used, then the extra parity bit is set so that the total number of bits in the character plus polarity that are set to one is even.*

**Hamming codes**- *often referred to as $(n,k)$ codes. Where $n$ is the total number of bits in each codeword, and $k$ is the number of data bits in a codeword.*

**Coder Rate(R) -** *is the ratio of data bits to message bits for a particular arrangement.*
- $d_{min}$ *is the minimum Hamming distance between any valid encoded messages.*
- *A low code rate indicates a high degree of redundancy*

**$(n,k)$ Block Codes**- *the code of which its decoder uses only the current frame to produce its output.*

**Convolutional Codes**- *it is the code of which the encoder remembers a number of previous frames and uses them in its algorithm.*

**Bit Error Rate (BER) -** *is the measure of the average likelihood that a bit will be in error.*

## 2.5 Block Codes

Prior to encoding and transmission the data is compiled into blocks in this error correction scheme. The data will subsequently be recompiled into blocks for decoding and error correction. Block codes have a linear algebraic structure that provides a significant reduction in the encoding and decoding complexity [**12**]. Block codes have different coding schemes such as Hamming Codes and Reed-Solomon Codes.

All this codes operate in different ways, where the marker codes use markers to detect and decode the errors. The marker must not be the same as the message being sent as to not confuse the maker with the data being sent.

Block codes encoding process breaks up the sequence of a data symbols into blocks of length $k$. Some of the block codes decoding schemes can detect the error but cannot correct the error, and some only work from insertion/deletion errors while others work for inversion errors.

The block code encoder transforms each input data into binary n-tuple with $n > k$ of which this n-tuple is referred to as the codeword or code vector [**12**].
 The term block code must be unique for the block code error correction scheme to be useful. Systematic structure is the most desirable property for linear block codes; this is when the codeword is divided into two parts viz. the data and the redundant checking parts. Block codes of minimum distance $d_{min}$ guarantees correction of all error patterns or fewer errors. [**13**].

## 2.6 Convolutional Codes

Convolutional codes were first introduced by Elias in 1955 as an alternative to Block codes. Convolutional codes differ from the block codes because the block codes are memoryless. Encoding of $(n_0, k_0, m)$ Convolutional codes can be implemented with $k_0$- input and $n_0$ - output linear sequential circuit with an input memory of $m$ words [14]. $m$ is the memory order. If the memory order is bigger the error probabilities are smaller. Convolutional codes are also liked for their less complex decoders in many applications [14].

### 2.6.1 Convolutional Encoding

The input bits are inserted into the shift register from the right hand side, all the bits then shift to the left. A register is cleared before any new bits can be inserted into the register. An example illustrating the operation Figure 2-6 is *"if a 1 is input the outputs will be 11 and the encoder will contain 1 in the right-hand shift register stage. If the next input is 0 then the encoder will output 10, the right shift register will contain 0 and the left shift resister 1."* [8]

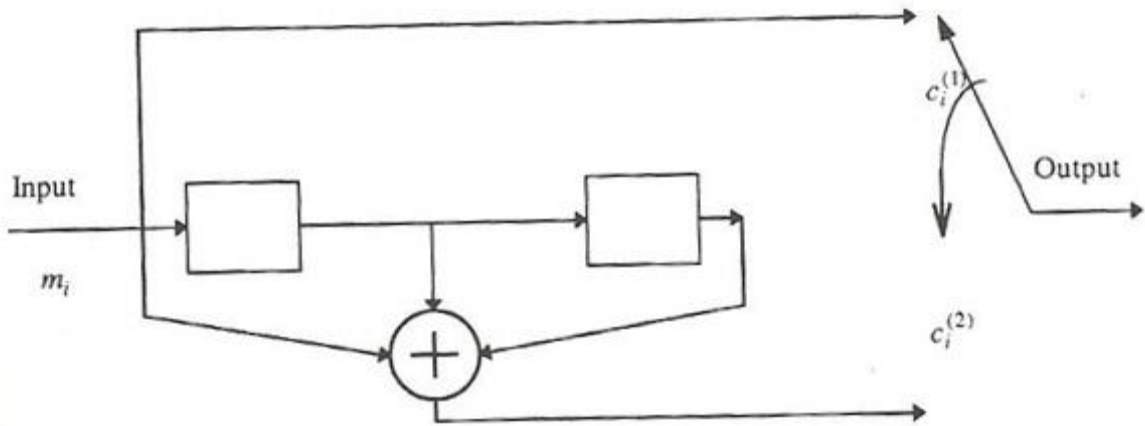

**Figure 2-6: Block Diagram of a Convolutional half rate with m=2 Encoder** [14]

The above described operation can simply be represented by Figure 2-7 which represent the state diagram of Figure 2-6.

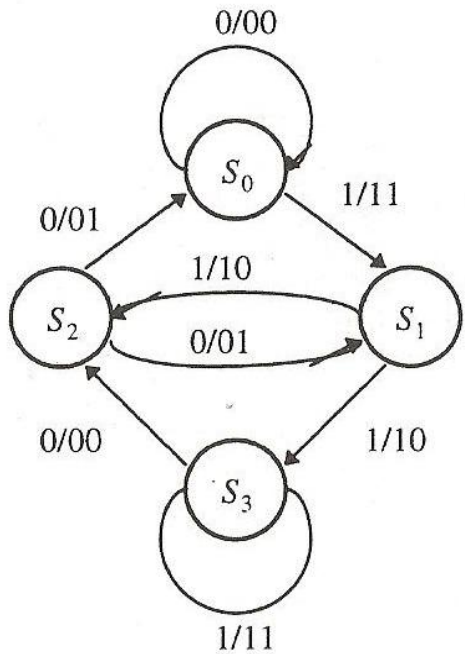

**Figure 2-7: State Diagram of a (2, 1, 2) Convolutional code** [14]

Figure 2-6 and Figure 2-7 give an idea of how Convolutional codes operate in general. The binary message sequence is presented by $\vec{m} = (\ldots, m_{-1}, m_0, m_1, \ldots)$. The codeword, $\vec{c} = \left(\ldots, c_{-1}^{(1)}, c_{-1}^{(2)}, c_0^{(1)}, c_0^{(2)}, c_1^{(1)}, c_1^{(2)}, \ldots\right)$, is the output alternatively mixed output sequence of the two binary sequences $\vec{c}^{(1)}$ and $\vec{c}^{(2)}$. The code is systematic if the output sequence of the message is clear, this occurs when the first output sequence $\vec{c}^{(1)}$ is simply equal to the input sequence $\vec{m}$.

In Figure 2-7 $S_0 = (0 \quad 0), S_1 = (1 \quad 0), S_2 = (0 \quad 1)$ and $S_3 = (1 \quad 1)$, with the labels on the branches denoting *input/output*.

### 2.6.2 Convolutional decoding

Comparing the received sequence with every possible code sequence is the best way of decoding and this process is done best by the aid of the code trellis. The state diagram information is contained in the code trellis.

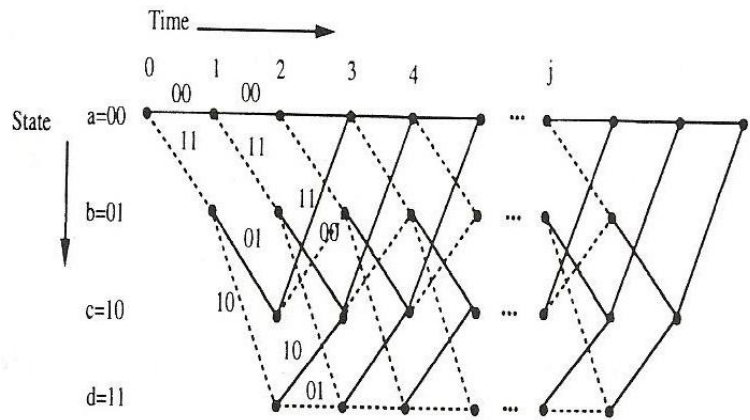

**Figure 2-8: Trellis Diagram of a (2, 1, 2) encoder** [15]

After studying the code trellis diagrams, Figure 2-8 , Viterbi came up with the method that simplifies the decoding problem without sacrificing any of the code's properties. This problem was that the trellis tends to have many paths and it becomes hard to compare the received message with every possible path.

There are four decoding schemes for Convolutional codes, namely, sequential, majority-logic, Viterbi and the look-up decoding.

### 2.6.2.1 *Viterbi Decoding*

The Viterbi algorithm was proposed by Viterbi in 1967 as an asymptotically optimum decoding technique for Convolutional codes [**16**]. It is now the most preferred Convolutional decoding algorithm and it is also used as a standard decoder when comparing other Convolutional decoding schemes.

In [**17**] and [**18**] the Viterbi algorithm is applied to naturally obtain the minimum distance decoding after the application of the distance preserved mapping technique introduced in [**17**]. Also a new concept of permutation trellis codes is introduced. Even if there are no errors the Viterbi algorithm performs a fixed number of calculations per unit time in a given received sequence.

### *2.6.2.2   Sequential Decoding*

This was the first Convolutional decoding technique introduced by Wozencraft of which after there were a series of sequential decoding techniques versions being introduced [**19**]. One was introduced by Fano, and it's called the Fano algorithm, another one by Zigangirov and Jalinek, and it's called ZJ or the stack algorithm.

Sequential decoding's encoder constraints length has no effect on the computational complexity. It adjusts the number of computational per decoding decision [**19**] according to the noise level. At heavy noise levels the sequential decoding runs the risks of failure due to the fact that it requires random variables which can be very high at high noise levels.

2.6.2.2.1   The Stack Algorithm

Figure 2-8 will be used to explain the operations of the stack algorithm. The steps that outline the operations of the stack algorithm can be read in [**19**].

2.6.2.2.2   Fano Algorithm

It has an advantage on the stack algorithm because it requires no storage but has a disadvantage because it is slow in decoding. In [**19**] the operations of the Fano algorithm are explained.

## 2.7 Channel Models

Channel models are models of transmission media which introduce errors in the transmitted data ensuring that the incorrect data should be received at the receiver for simulation purposes. It does not really introduce errors per se but its effects on the demodulator produce the error [**8**]. There is no channel model which had been decided as the standard channel model; hence everyone has usually introduced their own model.

A channel is called a memoryless channel if the output only depends on the transmitted signal and not on any previous transmissions for a given interval. The most basic discrete channel is the binary symmetric channel (BSC), diagram presented in Figure 2-9.

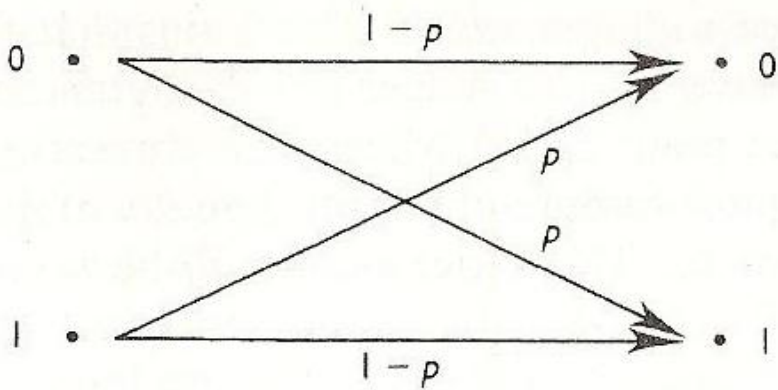

**Figure 2-9: Binary Symmetric Channel (BSC)** [8]

A bit can either be received correctly or incorrectly when transmitted through this channel [**20**]. Gilbert channel differs with BSC because it has memory and burst of errors. It also has two states, a good state and a bad state [**21**] . A good state means that no errors will occur and the bad state means that there are high probabilities of errors occurring. But the Gilbert-Elliot model introduces an error at the good state.

The statistical independent Reversal, Insertions and Deletions also known as statistical independent RID was first introduced by Swart [**22**]. Of which Knoetze introduced a simplified version of it. The simplified model is presented in Figure 2-10.

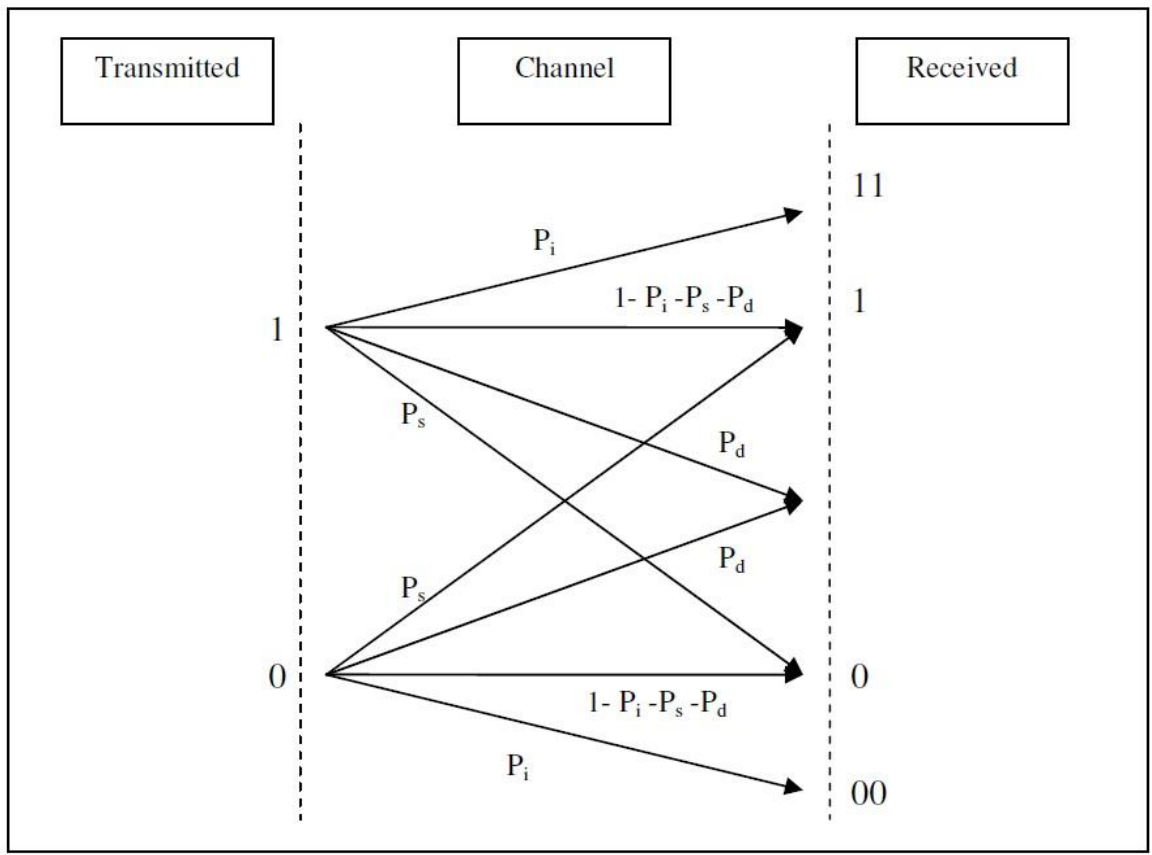

**Figure 2-10: Channel Model for Inversion, Insertion and Deletion [11]**

The probability that a bit is transmitted without an error is given by:

$$P = 1 - P_s - P_i - P_d$$

Where: *P*= Probability of transmitting a bit without any errors

$P_s$= Probability of an inversion error

$P_i$= Probability of an insertion error

$P_d$= Probability of a deletion error

In order to test specific errors, the error probabilities that one is not inserted in is set to zero [**21**].

## 2.8 Conclusion

In this chapter the works which have been done on error correction has been discussed. Both the old and modern work was introduced and discussed. The different types of channel models used for simulation purposes were also introduced.

The different kinds of errors are explained in full details. Examples to assist with more explanations were given. It is also explained how synchronisation errors and additive errors affect the received message. The most likely place for errors to be introduced is at the channel.

The channel is the medium used to transmit the message. In practical a channel cannot be perfect; it causes some sort of defects in the transmitted message. The terminologies used in error correction were also defined. These definitions are going to be used in the rest of this document as they are, since most people use different terminologies for the same concepts.

The different error correction schemes which are already in use, or maybe were once used are discussed in this chapter. These error correction schemes are divided into two main groups, viz. block code and convolutional code. There is one more scheme which had been recently introduced, and adapted in error correction, that scheme is the Neural Network decoder, which will be investigated more in the rest of this document.

The channel model to be used for simulation in this project was introduced and its components were discussed.

# Chapter 3: Neural Networks

## 3.1 Introduction

This chapter will introduce the concept of neural network. The background and mathematical theory of neural network will be discussed. The different kinds of neural network will be mentioned and how they are already being used in the information theory field to solve errors. The adaption of neural network to error decoding in telecommunications will be found in this chapter.

## 3.2 Neural Network Introduction

Neural networks have become a most interesting subject in the research field, being researched by professionals of many departments. The interest stimulating factor is the principle on which neural networks work on, they imitate the operations of the biological brain. A brain uses neuron to communicate to the rest of the body.

Biological neurons (BN), indicated in Figure 3-1, are massively interconnected and the connection is referred to as synapse. Biological neurons conduct electrical signals containing information, the information is conducted by the dendrite, processed in the cell body, sent down through the axon to the dendrite of another BN connected with the nerve terminals.

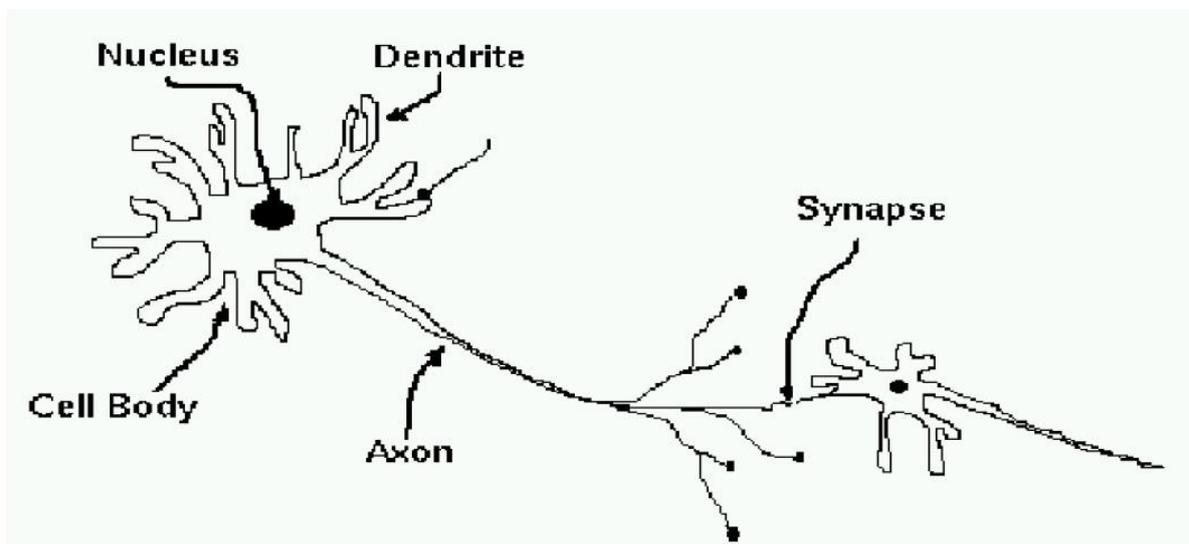

**Figure 3-1: Biological Neuron (BN) [23]**

Artificial Neurons (AN), indicated in Figure 3-2, is an artificial model of BN. AN's collect information from surrounding environment of AN's and transmits it to other AN's. Input signals are inhibited or excited through a negative and positive numerical weights associated with each connection to the AN [**23**]. The AN then collects all input signals and compute them as a function of the respective weights. The net input signal serves as input to the activation function which calculates the output signal of the AN [**23**].

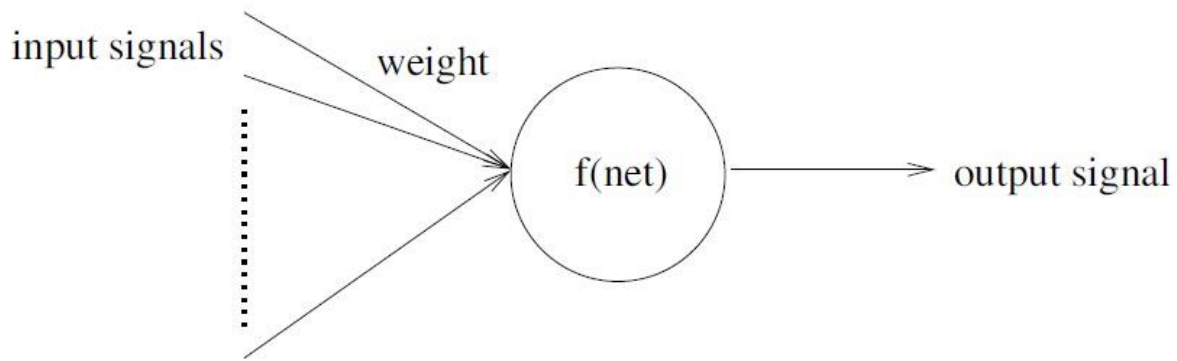

**Figure 3-2: Artificial Neurons** [23]

Artificial Neural Networks (ANN), indicated in Figure 3-3, are a group of interconnected AN's. ANN usually consists of three parts viz. input, hidden and output layers. Each AN in the input layer may be fully or partially connected to each one of the hidden layers, and each from the hidden may be connected fully or partially to the output layer's AN.

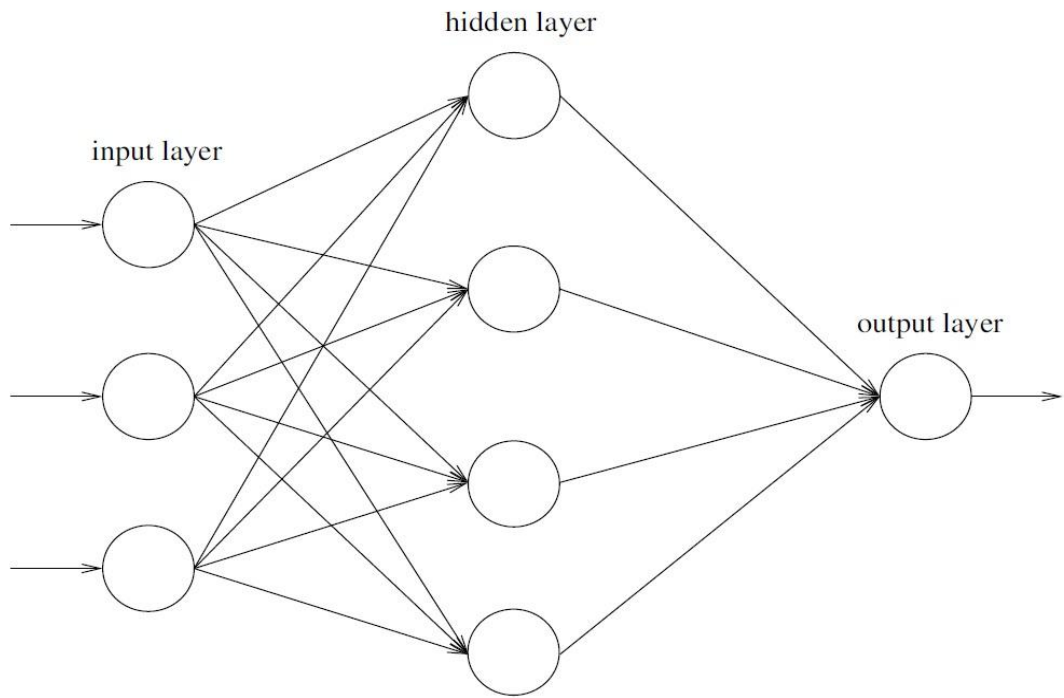

**Figure 3-3: Artificial Neural Network** [23]

In this chapter an overview of neural networks (NN) will be given, and the terminologies used in neural networks will be defined as they will be used in this thesis. The different parameters for classifying NN will be discussed and the focuses will shift more to Binary Neural Networks (BNN).

## 3.3  Neural Networks Terminologies and Definitions

**Biological Neural –** it is the neuron found in a human being, interlinked with the nervous system.

**Artificial neuron-** the mathematically defined neuron used to imitate the behaviour of a biological neuron

**Artificial neural networks-** the combination of different artificial neurons arranged and interlinked to form a network.

**Weights-** can sometimes be viewed as the summation of the inputs being fed to the neural network

**Training/ learning –** the process of teaching the neural network with certain information, e.g.: patterns.

## 3.4 Artificial Neural Networks

Artificial neural networks diagram is indicated in Figure 3-3. The diagram in Figure 3-3 indicates the ANN of a single hidden layer, an ANN can have a multi layered hidden layer as indicated in Figure 3-4.

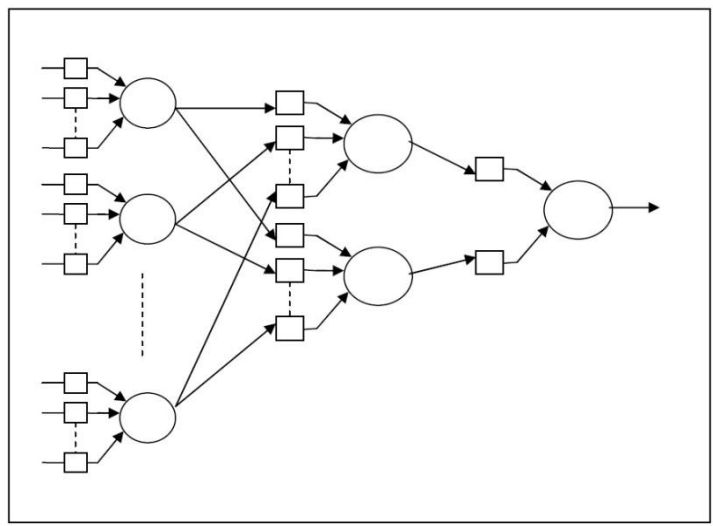

**Figure 3-4: Multi Layered Neural Network**

The diagram in Figure 3-5 will be used to indicate how the weights are obtained in NN. $I$ is the amount of input signals, $\mathbf{z} = (z_1, z_2, z_3, \ldots, z_I)$ are the input signals.

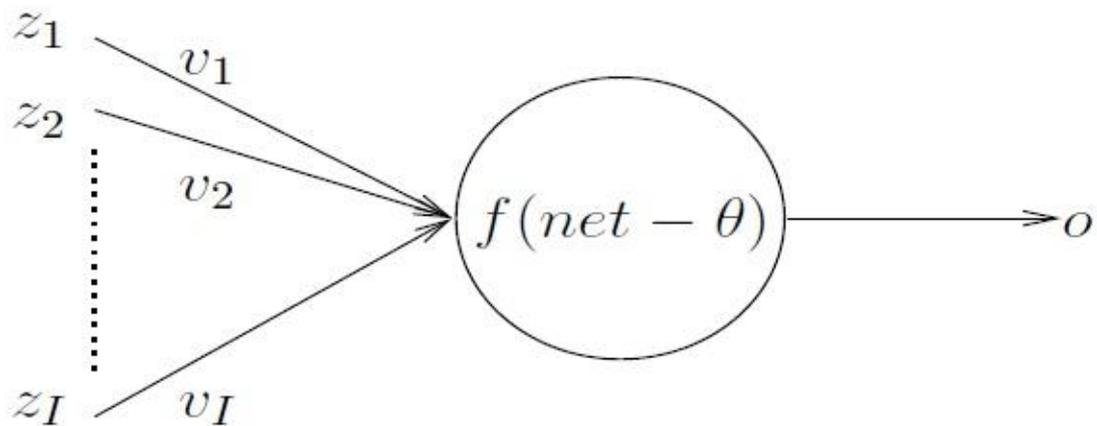

**Figure 3-5: A Simple Artificial Neuron [23]**

To compute the output, the activation function $f_{AN}$ is used to compute the net input signal. The strength of the output signal will be influenced by $\theta$, which is the threshold value. The threshold value is referred to as bias at times. The equations bellow show how the input signals are computed.

    i.    Summation Units (SU)

$$net = \sum_{i=1}^{I} z_i v_i$$

    ii.    Product Units (PU)

$$net = \prod_{i=1}^{I} z_i^{v_i}$$

If a step activation function is used, the following conditions will apply for $\theta > 0$:

$$f_{AN}(net - \theta) = \begin{cases} \gamma_1 \text{ if } net \geq \theta \\ \gamma_2 \text{ if } net < \theta \end{cases}$$

Where $\gamma_1 = 1 \text{ and } \gamma_2 = -1$ for a bipolar output.

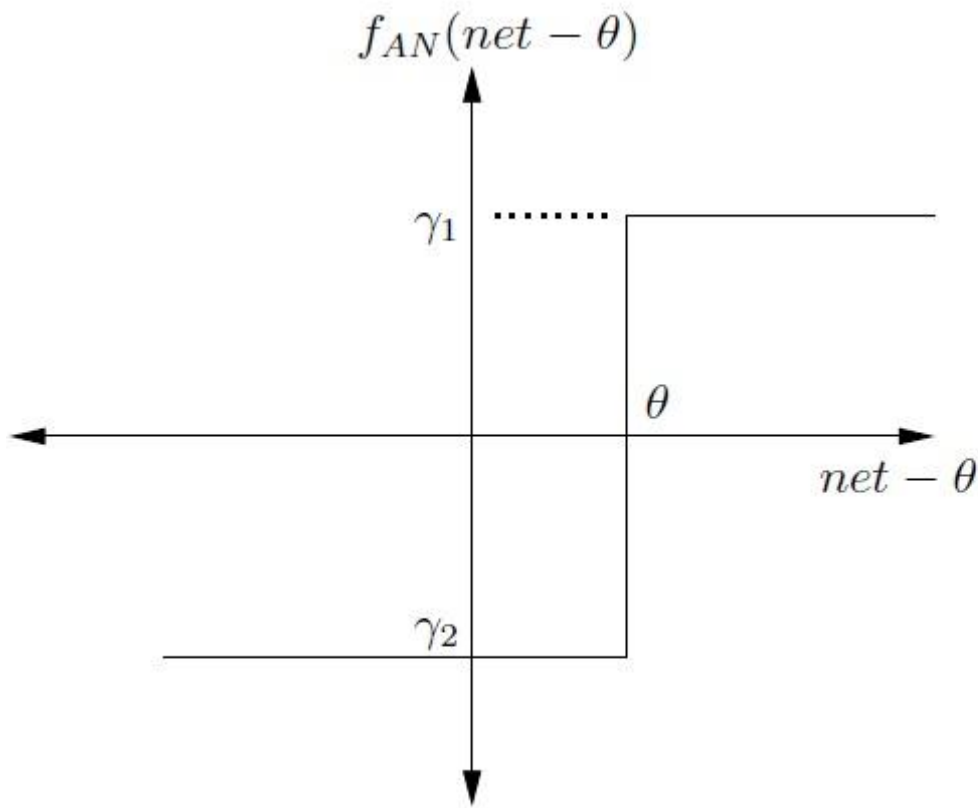

**Figure 3-6: Step Activation Function Graph** [23]

Basically an ANN design can be classified into architecture and learning. Architecture can be divided into feed-forward, feed-back and self-organising neural networks. Learning is when a neural network acquires the ability to carry out certain tasks by adjusting its internal parameters according to some learning scheme [**24**].

Learning can either be supervised or unsupervised learning depending on the architecture being used. Supervised learning trains the ANN the input and the output pattern, whereas the unsupervised learning only trains the ANN with the pattern of the input. Thus the output of the unsupervised learning can be any pattern. The unsupervised learning is due to the self-organizing neural network [**25**].

## 3.5  Binary Neural Networks

Binary Neural Networks (BNN's) are neural networks that take in the input as binary and release the output also as binary. There are sometimes referred to as Weightless Digital Neural Networks. They are called weightless neural networks because there are no weights to be summed as the input data is supplied in binary form. The inputs of the BNN are all equally important. BNN are preferred for their speed and circuit implementation simplicity. The most current use for BNN is pattern recognition and character recognition.

BNN originated from the work introduced by Bledsoe and Browning in 1959. They proposed a pattern recognition method called N-tuple method.

### 3.5.1  Tuples

Tuple, sometimes known as mapping, are the set of inputs in one neuron. In bit sequence terminology, they can be defined rather, as the bits used in the voting process [11]. Voting process is the process whereby the trained neural network will identify the pattern of the received data sequence and compare it to the transmitted data.

The set of binary functions used to recognize the pattern is known as '*identifiers*' [20]. These identifiers are the ones used to decide whether the received information is true or false.

The best way to explain the tuple will be to consider a $2^4$ bit data, and sort it in a $4 \times 4$ matrix.

|        | A | B | C | D |
|--------|---|---|---|---|
| Tuple 1 | A | B | C | D |
| Tuple 2 | E | F | G | H |
| Tuple 3 | I | J | K | L |
| Tuple 4 | M | N | O | P |

**Figure 3-7 : Tuple Diagram**

$$\therefore R = A.B.C.D + E.F.G.H + I.J.K.L + M.N.O.P$$

Tuple 1 = A B C D, Tuple 2 = E F G H, Tuple 3 = I J K L and Tuple 4 = M N O P

If an image of letter 'N' was to be presented in the above matrix, its matrix would look like follows:

| X |   |   | X |
|---|---|---|---|
| X | X |   | X |
| X |   | X | X |
| X |   |   | X |

**Figure 3-8: N- Representation**

$$\therefore R = A.\bar{B}.\bar{C}.D + E.F.\bar{G}.H + I.\bar{J}.K.L + M.\bar{N}.\bar{O}.P$$

The above expression is the Boolean algebraic expression of the image in Figure 3-8.

### 3.5.2  Training

The training process in BNN is remembering the identifiers needed for specific classes of images [**20**]. A BNN is fed a pattern in terms of the Boolean algebra as the input. Training can be referred to as teaching or learning, depending on the author. Hence in this thesis they will be used interchangeable.

Neural Networks learning can be supervised, reinforced or unsupervised. In the supervised training, a neuron is taught the pattern of the input and that of the expected output. Neurons are updated to reduce the errors between the input pattern and the output pattern.

In an unsupervised training, the neural network is taught the input pattern but not the output or expected pattern. Therefore the Neural Network will decide how the output pattern will be like depending on what pattern it sees best fit.

The next paragraph will explain the training process in Binary Neural Networks.

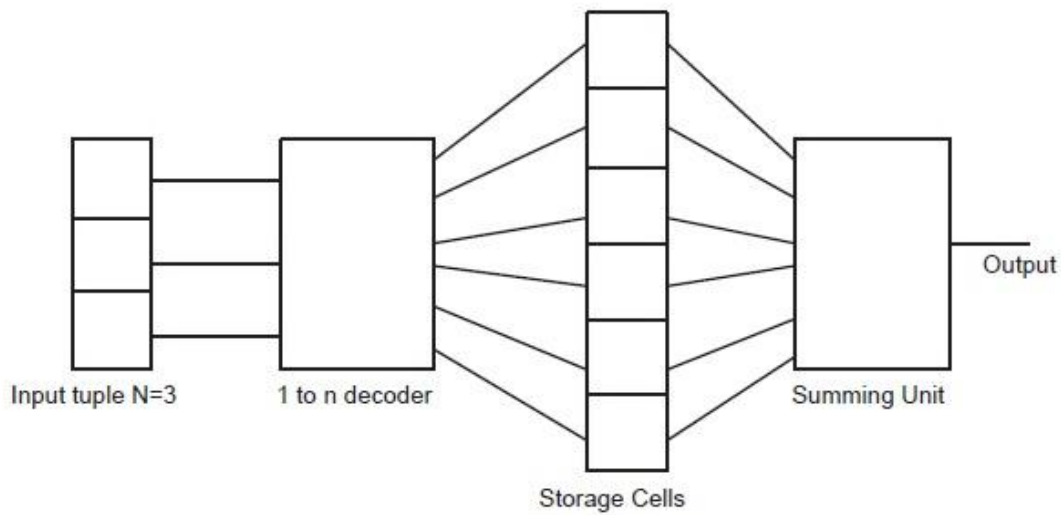

**Figure 3-9: The basic RAM node** [26]

Figure 3-9 illustrates the diagram used during the training process. The training process is done using 1 to *n* tuples and the binary input will be stored in each cell.

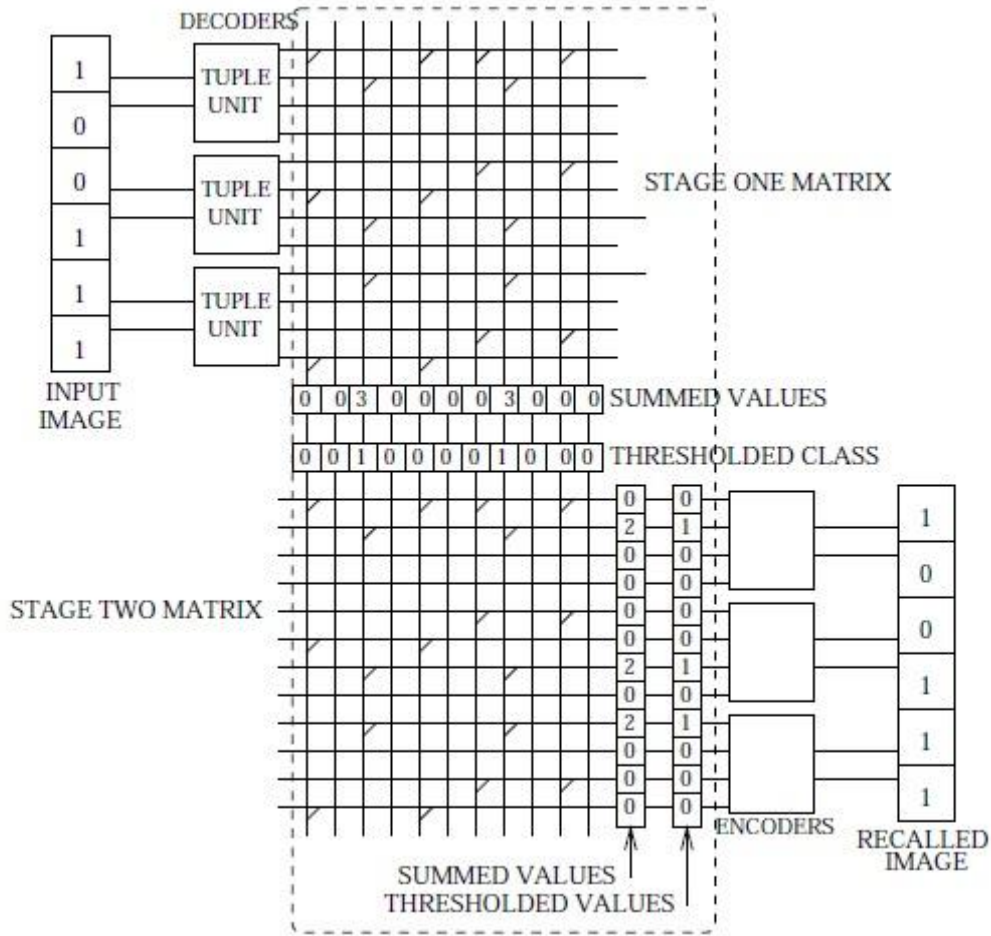

**Figure 3-10: ADAM Training Diagram** [27]

Figure 3-10 illustrates the complete architecture of the RAM Based Neural Network called the Advanced Distributed Association Memory (ADAM) [**28**]. The ADAM has advantages that it can operate in noisy data.

An input image in binary is 'fed' to the neural network in groups of two per tuple. In STAGE ONE MATRIX training process takes place, summed and then converted back to binary. The voting process takes place in STAGE TWO MATRIX and the votes are once again summed and then converted to binary.

The received pattern after the votes' summation is encoded and the output image is then extracted. The succeeding section will fully describe the ADAM memory and how it operates.

### *3.5.2.1 ADAM Memory System operations [27] [28]*

This system was developed by J Austin. Ever since then it has been used in most image analysis tasks [**29**]. It can be trained to give a known pattern when it is given an unknown pattern. This is done through training it with a known pattern. The pattern it recalls will be similar to one or a number of patterns trained into the network.

There are four stages in the ADAM system, viz.
  i.  N-Tuple pre-processor
  ii. A correlation matrix memory
  iii. An intermediate class
  iv. A second-stage correlation matrix

3.5.2.1.1 A correlation matrix memory

This part is known as the heart of the ADAM system. All possible patterns to be recognised are stored in this section. It single layer of binary 'weights' but is seen as a matrix, M. The construction of the correlation matrix will be illustrated in Figure 3-11, where A is the input message and B is the output message with errors.

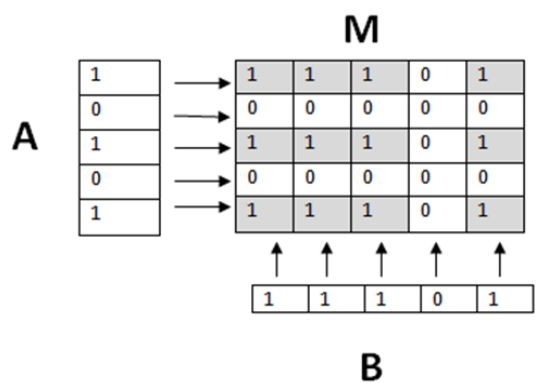

**Figure 3-11: Correlation Matrix Construction**

During training all weights are binary 0 and when each bit of logic 1 in the input is equal to that of the output, the weight is set at logic 1. This process is mathematically presented by the following equation:

$$M = \bigcap_{all\ i} A_i \otimes B_i$$

### 3.5.2.1.2 An Intermediate Class

This stage follows after the correlation matrix stage has been completed. This stage is used to help recall the known pattern form the memory using the unknown. The class are expressed in real numbers. This stage is expressed as SUMMED VALUES in Figure 3-10. These summed values will be represented as C.

Mathematically, C is obtained through the following function:

$$C = A^T M$$

For the above example

$$C = \begin{array}{|c|c|c|c|c|} \hline 3 & 3 & 3 & 0 & 3 \\ \hline \end{array}$$

The L-Max method is used to find the maximum summed value and make it the threshold value. All the summed values are checked against the maximum value and if they are equal it indicates that the corresponding class bit should be stored in here. The threshold value will be $\begin{array}{|c|c|c|c|c|} \hline 1 & 1 & 1 & 0 & 1 \\ \hline \end{array}$ for the example above. In Figure 3-10 it is expressed as THRESHOLDED CLASS.

Simple correlation matrix suffers from poor capacity. It stores less number of patterns before recall results. This is due to the following reasons

1. Excess crosstalk between stored patterns
2. The size of the memory is bounded by the input and output array sizes.

Two steps to solve this problem [**29**]
1. An orthogonalising pre-processor which ensures that the input patterns are sufficiently separated to allow the storage and recall to be more reliable.

2. Two-stage correlation memory is used to allow the size of the memory to be independent of the size of the patterns to be associated.

**This process is indicated in Figure 3-10. A worked out illustration, continuing on the example indicated above will be indicted in**
Figure 3-12. Only the stages from the second correlation matrix will be shown since the other ones had already been shown.

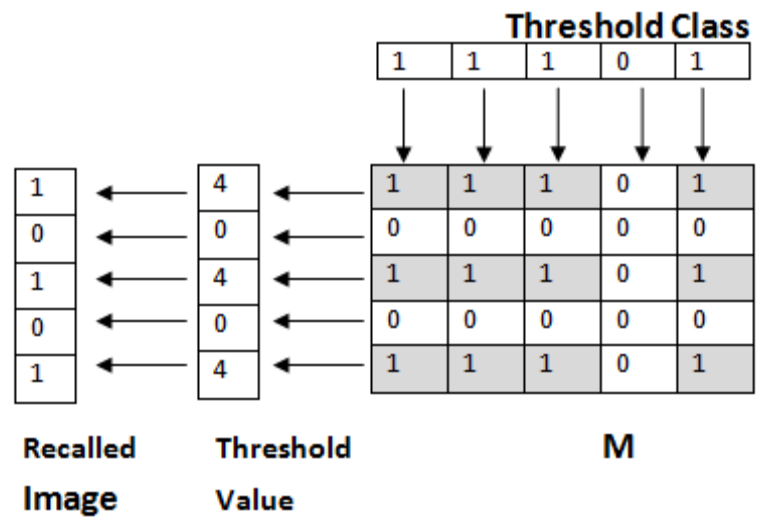

**Figure 3-12: Second Stage Correlation Matrix**

This example does not include the first step in overcoming the problem discussed above.

## 3.6 Classifiers

Classifiers consist of tuples and the voting system. The tuples have been discussed in the Section 3.5.1 and the voting system operation will be fully described in Section 3.7 and Section 3.8. Tuples are arranged in rows of matrices to describe the image in that matrix.

Tuples are then grouped together in groups, the grouping can be done in any way, up to so far, the best possible way should be explained by the end of this thesis. Each one of the groups of the tuples has its own classifier. The information found in the classifier will be used in the voting process.

The voting system arranges the votes in a histogram or a matrix. But the histogram is a more clear representation of the voting results as the image which gets more votes can be seen in a glance.

## 3.7 Operations of Binary Neural Networks

The best way to describe the BNN operations will be through explaining it in terms of images as it is mostly used to identify the image patterns.
BNN gets information as binary, encode it and transmit it. The data is then decoded and sent to the output. The received data is also in binary. Input images are trained by feeding them with images, the same images will be used later to compare the output and the input.

During the comparison, BNN votes for the closet resemblance sequence/ pattern between the inputs and outputs. The 'identifiers' are used during the voting system. BNN broad explanation of the operation can be explained or described in [**28**].

The voting system chooses the best pattern resembled by the received data. It does this by voting for that specific pattern from the trained neurons. The image pattern which gets the most votes is then the one whose image has been closely resembled by the received pattern.

## 3.8 Example of BNN Operations

An example which will fully illustrate the operations of BNN will be discussed in this section. The example considered here will be a $3 \times 3$ image. The neural network will be trained to identify 'T', 'L', 'X', and 'H' patterns. Then the neural will be input with a distorted image, and then BNN will conduct a voting process which will choose which pattern the input matches the most.

The example will be illustrated in Figure 3-13.

In this example the full operations of the N-Tuple Method as used in image processing will be briefly explained. This example is made, mostly in graphics form to better illustrate the idea previously explained in words.

The succeeding sections will explain each graphical representation in words.

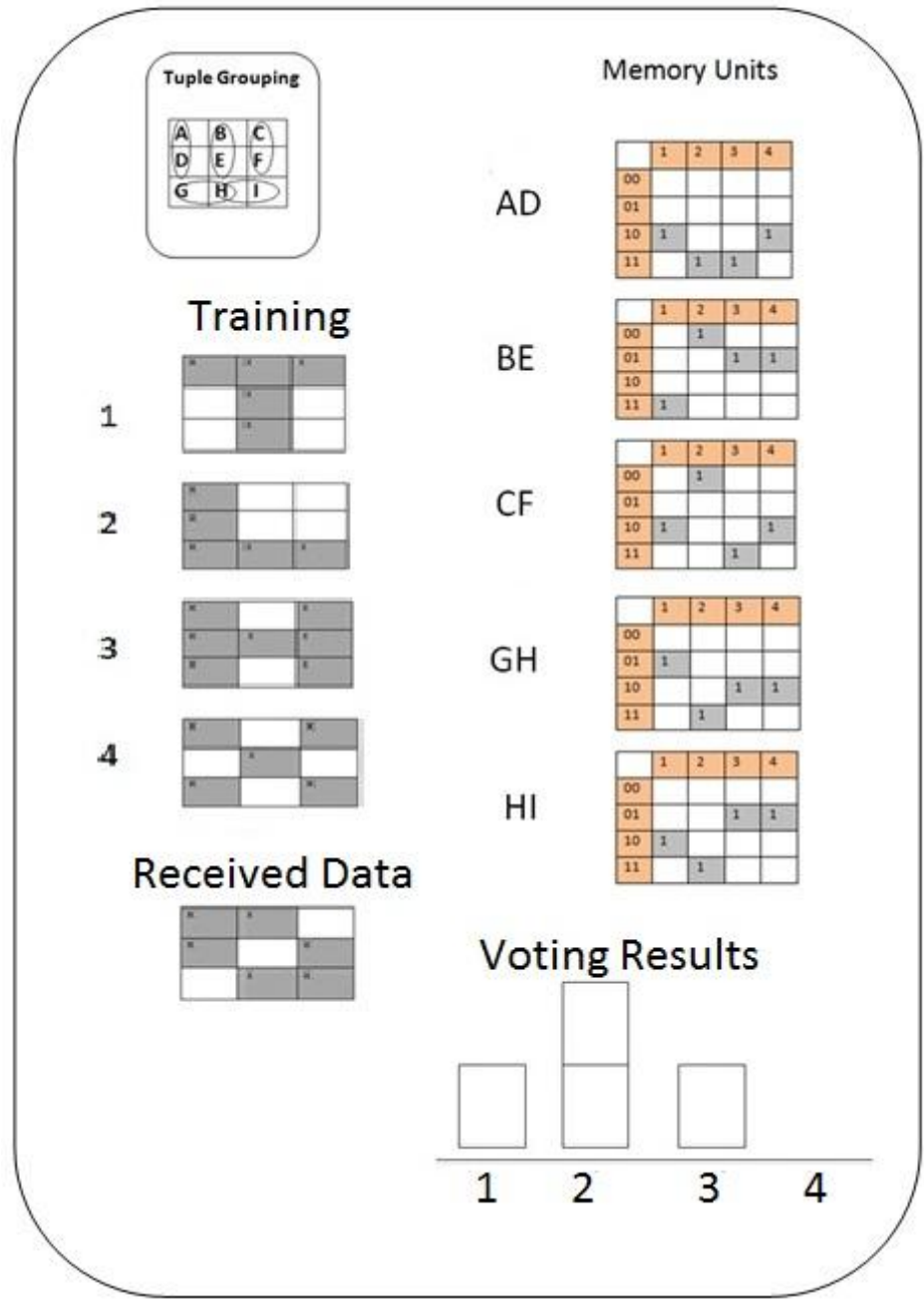

**Figure 3-13: Example Illustration** [11]

### 3.8.1 Tuple Grouping and Data Training

The tuples are grouped as indicated in Figure 3-14, i.e. tuples are grouped in groups of two.

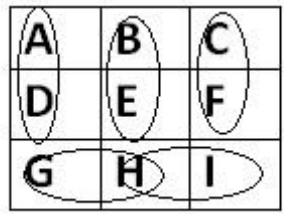

**Figure 3-14: Tuple Grouping**

Through this sorting of bits, the tuples of the images of which the neurons are being trained with will be sorted. The training bits are shown in Figure 3-15. As discussed in Section 3.7 these are the images used to train the Neural Networks.

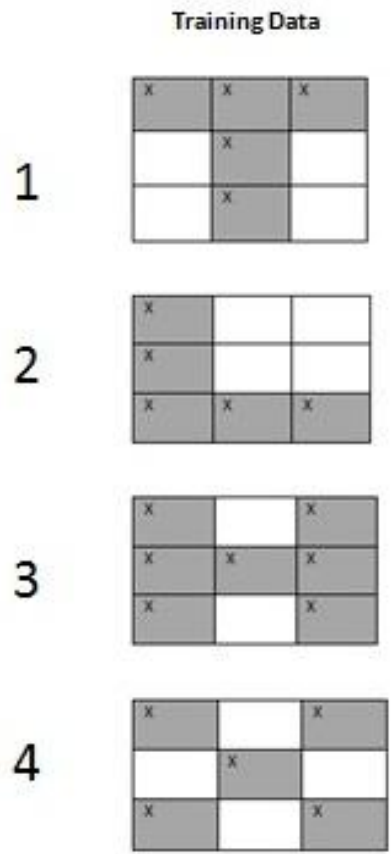

**Figure 3-15: Training Data**

After training the BNN, the received image will be input to the BNN so it may decide which image was being sent.

### 3.8.2 Received Data

The received image was assumed to have errors and distortions. This image will be illustrated in Figure 3-16. The error introduced here are arbitrary.

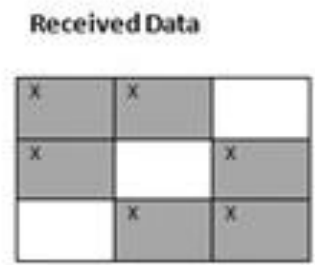

**Figure 3-16: Received Image with Errors**

After being input to the BNN, the bits of Figure 3-15 are sorted in the classifiers. Each tuple group gets allocated its own classifier. The tuple groups of the received image, Figure 3-16, compared to the classifiers and the votes are cast.

The results are as indicated in Figure 3-13 where the whole example is summarised.

### 3.8.3 Vote Results

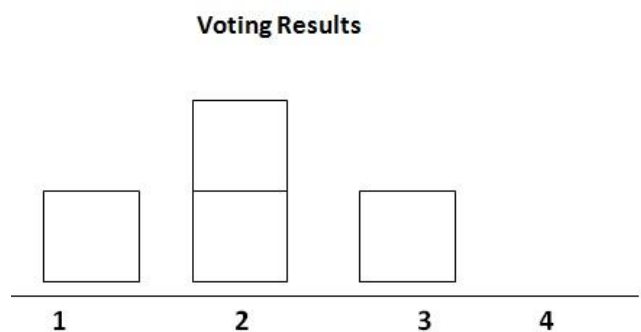

**Figure 3-17: Vote Results**

After the votes are cast, the image number with many votes gets to be selected as the image intended to be sent. In this case IMAGE 2 was intended by the image in Figure 3-16, since IMAGE 2 has the most votes, i.e. 2 votes.

## 3.9 Conclusion

By properly training a neural network, it can be used to recognise and identify different patterns. It will be able to distinguish between patterns and sequences intended to be there and the ones which were not intended to be there. The groupings of tuples are also important as it affects the efficiency of the pattern prediction results.

Binary neural networks have an advantage of using digital information over analogue information, thus making them very fast neural networks. BNN also increases its speed by its ability to perform calculations in parallel.

The example shown here indicates that the BNN can be used to decode errors, as it was able to decode the image processing error. In [**11**] and [**20**] R. Knoetze and B. Otto proved that BNN can be adapted to correcting errors. So the objective of this thesis is to narrow down the parameter used in the BNN as the decoder. This will be done following the mathematical operations stated in here.

# Chapter 4: Detection and Correction of Inversion Errors Design

## 4.1 Introduction

This chapter will discuss the design implementations towards solving inversion errors using neural networks. As it has been discussed in the previous chapters, binary neural networks are going to be used to correct errors. Information on how the message will be sent and how errors will be introduced to the message till how the message will be decoded mathematically will be discussed in this chapter.

## 4.2 Transmitting Message

The codebook to be used here contains only four codewords. These codewords are chosen randomly, but with care to make sure at high inversion error probabilities the probabilities of one of the codewords to be the same as the sent codeword are low.

The Sent message will be assumed to be encoded. A sent message will be sent using the codewords found in the codebook. The codeword to be used will be of three, four, five, six, seven or eight bits. The codeword will be arranged in a continuous bit stream of a certain bit number, n.

Sent Message= [codeword1 codeword2 codeword3 codeword4 codeword1 codeword2 ...]
The above will be done until a specific n is reached and the codewords can be arranged randomly. The sent message will be sent through a channel which will introduce inversion errors to the sent message bit stream. Depending on how many bits are in each codeword, the message passed through the channel will be split into those bits in the decoder.

*4.2.1.1   Example 1*

In this example, a four bit codeword will be used. The operations with any other bits should still be the same:

codeword1= [1 1 1 1];
codeword2= [1 0 1 0];
codeword3= [0 1 0 1];
codeword4= [0 0 0 0];
Sent Message= [codeword1 codeword3 codeword2 codeword4];
Sent Message=            [1 1 1 1 0 1 0 1 1 0 1 0 0 0 0 0];
The highlighted bits are just to indicate the different codewords.

Now the Sent message will be sent through the channel so that errors may be introduced.
Message After Channel=    [1 1 0 1 0 1 0 1 0 1 1 1 0 0 1 0];
The Message after the channel will then be split into 4 bits groups:
Group1= [1 1 0 1]; there is only one error in this "group".
Group2= [0 1 0 1]; there are zero errors in this "group".
Group3= [0 1 1 1]; this "group" has three errors.
Group4= [0 0 1 0]; this "group has one error.
"group" indicates the four bits after they have been separated from the n-bit stream.

The above example illustrates how the bits are separated after the channel has introduced errors. The separation process is done so that each group may be compared to each codeword in the codebook. This process will be explained later under Neural Networks Section. The Groups are then compared to the codeword, and if they are not the same, then it becomes clear that there is an error in the message.

A more detailed error detection technique will be discussed under the Neural Network Section, since Neural Networks are going to be used to detect and correct errors. The length of the "group" should always be equal to the length of the codeword.

## 4.3 Memory Usage

One of the most important things which have to be taken into consideration when designing decoders is memory. Therefore this section discusses the memory usage through the equation used to determine how much memory has been used per decoder.

$$Memory = \frac{2^N \times C \times m}{n}$$

$$where\ N \equiv Length\ of\ Mapper$$
$$C \equiv the\ number\ of\ codewords\ in\ a\ codebook$$
$$m \equiv number\ of\ classifiers$$
$$n \equiv length\ of\ the\ codeword$$

Memory is expressed in *bytes.*

## 4.4 Neural Networks Decoder

The same codebook used in sending the message will be used to train the neural network. Each codeword in the codebook will be compared to each bit in the "*group*". This process will be done using an N-Tuple method. The N-tuple method requires classifiers. Classifiers are made of mappers and voting system. Mappers are designed following how the tuples are arranged.

### 4.4.1 Index Set

The tuple will be arranged in such a way that all the possible combinations are considered. Thus tuple arrangements will have a total of $C_N^n$ arrangement possibilities.

Tuple arrangements will be represented as $I_i \; where \; i = 1,2,3 \ldots such \; that \; 1 \leq i \leq N$. $I$ is known as the index.

#### *4.4.1.1 Example 2*

A seven (7) bit codeword will be used. A 7-bit codeword will have a 7-bit neural network decoder.

A 3-bit tuple will be used to illustrate the tuple arrangement process.

There are $C_3^7 = 35$ arrangement possibilities.

These possibilities will be illustrated in the table below.

$I_1 = 1\ 2\ 3$  
$I_2 = 1\ 2\ 4$  
$I_3 = 1\ 2\ 5$  
$I_4 = 1\ 2\ 6$  
$I_5 = 1\ 2\ 7$  
$I_6 = 1\ 3\ 4$  
$I_7 = 1\ 3\ 5$  
$I_8 = 1\ 3\ 6$  
$I_9 = 1\ 3\ 7$  
$I_{10} = 1\ 4\ 5$  
$I_{11} = 1\ 4\ 6$  
$I_{12} = 1\ 4\ 7$  
$I_{13} = 1\ 5\ 6$  
$I_{14} = 1\ 5\ 7$  
$I_{15} = 1\ 6\ 7$  

$I_{16} = 2\ 3\ 4$  
$I_{17} = 2\ 3\ 5$  
$I_{18} = 2\ 3\ 6$  
$I_{19} = 2\ 3\ 7$  
$I_{20} = 2\ 4\ 5$  
$I_{21} = 2\ 4\ 6$  
$I_{22} = 2\ 4\ 7$  
$I_{23} = 2\ 5\ 6$  
$I_{24} = 2\ 5\ 7$  
$I_{25} = 2\ 6\ 7$  

$I_{26} = 3\ 4\ 5$  
$I_{27} = 3\ 4\ 6$  
$I_{28} = 3\ 4\ 7$  
$I_{29} = 3\ 5\ 6$  
$I_{30} = 3\ 5\ 7$  
$I_{31} = 3\ 6\ 7$  

$I_{32} = 4\ 5\ 6$  
$I_{33} = 4\ 5\ 7$  
$I_{34} = 4\ 6\ 7$  

$I_{35} = 5\ 6\ 7$  

The above example shows all the possible tuple arrangements. These arrangements are then translated into a mapper.

### 4.4.2 Mappers

The length of the mapper is defined by $N$. Each mapper has its own voting system, and together they form a classifier. When these classifiers are used together they are then known as Neural Networks Decoder (NND). The length of the mappers in each classifier must be the same, i.e. there cannot be a classifier with $N = 2$ and one with $N = 3$ mappers in one decoder.

A mapper will be used to take specific bits from the "*group*" and compare them to the same bits from the codeword. If those bits are the same, a vote will be cast, otherwise no votes. This will be done for all tuple arrangements being used, and votes will be cast and tallied every time the bits are the same. These votes will then be added together, meaning codeword1 will have specific votes, codeword2 will also have specific and so on. This process will be done the same as illustrated in 4.4.3.1: Example 4.

The codeword with most votes will be voted as the one which was originally sent. This way the error in that "*group*" has been corrected.

A mapper's size will always be the equal to $N$. The number of the classifiers will be equal to the ones in the tuple arrangements, thus, for the example above, there will be 35 classifiers used to compare "*group*" and codeword.

Mapper will be presented as $m$. The mapper is defined as:
$$m = f(codeword, I_i) = [codeword(1)\ codeword(2) \cdots codeword(N)]$$
This means that the mapper is function of tuple and codeword/ "*group*".

### 4.4.2.1 *Example 3*

codeword1= [1 0 0 1 1 0 1]

Arbitrarily choose $I_{12} = 1\ 4\ 7$ ;

$m_{12}$ = [codeword1 (1) codeword1 (4) codeword1 (7)] = [1 1 1]

The above example shows how a $m_{12}$ mapper is constructed from the $i = 12$ index set.

The mapper is then constructed for both the codeword and the Message after the channel, i.e. for the codeword and the "group".

### 4.4.3 Voting System

The mappers are then be compared, if all their elements are equal, votes will be cast and if they are not all equal, a zero vote will be cast. The votes for all equal mappers will be added for each codeword. And the codeword with most votes gets chosen as the corrected message in that specific "group" position.

#### *4.4.3.1 Example 4*

Two codewords from the codebook will be used to illustrate the mapper and voting system process.

channel, represents the message after errors had been introduced in the sent message.

codeword1= [1 0 0 1 1 0 1]
codeword2= [0 0 0 1 1 0 1]
channel= [0 0 0 1 1 1 1]

**FOR CODEWORD1:**

$$m_{12} = [codeword1(1)\ codeword1(4)\ codeword1(7)] = [1\ 1\ 1]$$
$$m_{21} = [codeword1(2)\ codeword1(4)\ codeword1(6)] = [0\ 1\ 0]$$
$$m_{33} = [codeword1(4)\ codeword1(5)\ codeword1(7)] = [1\ 1\ 1]$$

**Message after introducing errors:**

$$m_{12} = [channel(1)\ channel(4)\ channel(7)] = [0\ 0\ 1]$$
$$m_{21} = [channel(2)\ channel(4)\ channel(6)] = [0\ 1\ 1]$$
$$m_{33} = [channel(4)\ channel(5)\ channel(7)] = [1\ 1\ 1]$$

**Comparing codeword1 and the error message:**

For $m_{12}$ the bit comparison answer will be [0 0 1], it gets 0 vote

For $m_{21}$ the bit comparison answer will be [1 1 0], it gets 0 vote

For $m_{33}$ the bit comparison answer will be [1 1 1], it gets 1 vote

Total votes=1

**FOR CODEWORD2:**

$$m_{12} = [codeword2(1)\ codeword2(4)\ codeword2(7)] = [0\ 0\ 1]$$
$$m_{21} = [codeword2(2)\ codeword2(4)\ codeword2(6)] = [0\ 1\ 0]$$
$$m_{33} = [codeword2(4)\ codeword2(5)\ codeword2(7)] = [1\ 1\ 1]$$

**Message after introducing errors:**

$$m_{12} = [channel(1)\ channel(4)\ channel(7)] = [0\ 0\ 1]$$
$$m_{21} = [channel(2)\ channel(4)\ channel(6)] = [0\ 1\ 1]$$
$$m_{33} = [channel(4)\ channel(5)\ channel(7)] = [1\ 1\ 1]$$

**Comparing codeword2 and the error message:**

For $m_{12}$ the bit comparison answer will be [1 1 1], it gets 1 vote

For $m_{21}$ the bit comparison answer will be [1 1 0], it gets 0 vote

For $m_{33}$ the bit comparison answer will be [1 1 1], it gets 1 vote

Total votes= 2

From the above example, it can be seen that codeword2 gets the most votes. Therefore codeword2 is voted as the corrected message. Codeword2 can now be viewed as part of the corrected message. This will be done for all "*group*'s", and the corrected message will be reconstructed using the correct codewords.

### 4.4.4 Decoder output

The codeword with most votes is known as the decoder output. The decoder output is thus defined as:

$y = I[\max(o_1, o_2, o_3 \cdots o_C)]$, where $C$ is the number of codewords in a codebook.

### 4.4.5 Voter confidence

Voter confidence, $\delta$, is the difference between maximum vote and the highest value below the maximum votes. It is mathematically defined as $\delta = |\max(o_1, o_2 \cdots o_i \cdots o_C) - \max(o_1 o_2 \cdots o_{i-1} o_{i+1} \cdots o_C)|$ for a given classifier. The bigger the difference, the bigger the voter confidence, hence the more accurate the results will be.

#### *4.4.5.1 Example 5*

We will consider a decoder with four codewords, i.e. C=4, arranged as $\{codeword1, codeword2, codeword3, codeword4\}$. The outputs after comparing each codeword are presented as $\{3, 4, 12, 7\}$.

This means that codeword1 has 3 votes; codeword2 has 4 votes and so on. This will be illustrated in the diagram below. In this example, codeword3 will be voted as the corrected decoder output. The voter confidence, $\delta$, is then found to be 5.

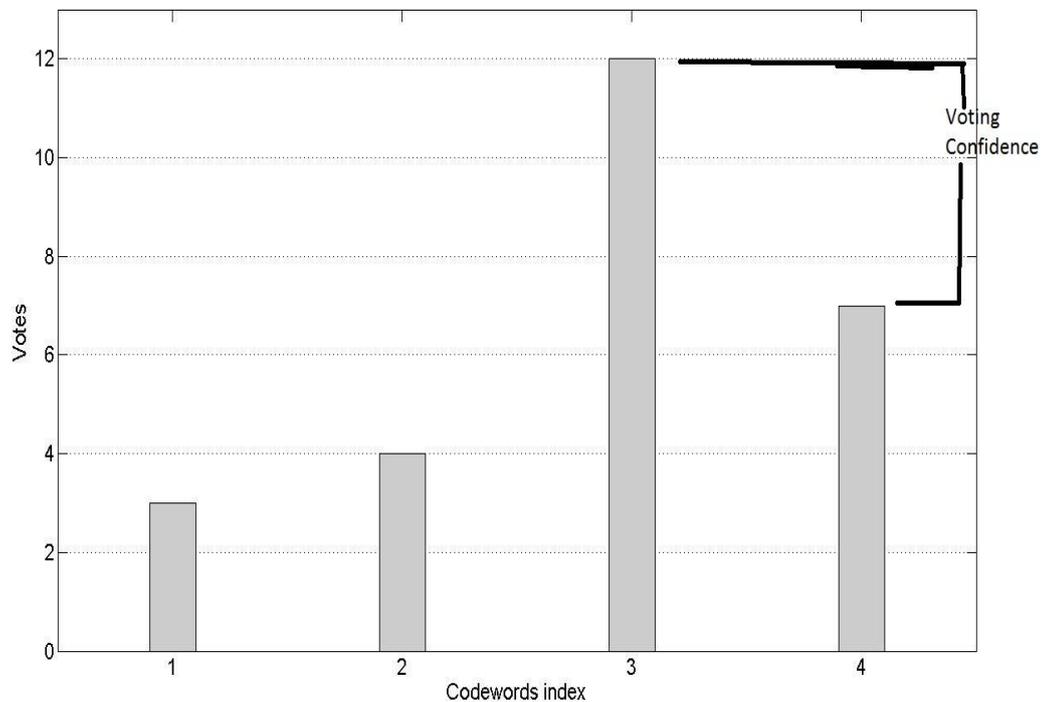

**Figure 4-1: Voter Confidence Representation**

## 4.5 Design Parameters

There are three Binary Neural Network parameters of which the design will primarily focus on, viz.

       I.    Length of the codeword, n
     II.    Length of the mapper, N
    III.    Number of Classifier, m

The design of each parameter will be discussed in the succeeding sections, respectively.

### 4.5.1  Length of the Codeword

The length of the codeword is denoted as $n$.

Since it has been mentioned earlier that four codewords will be used in the codebook, the length of the codeword must be higher than two, i.e. $n > 2$. It is important to ensure that $n \neq 2$, because the possible codeword in the codebook are $2^n$, and if $n = 2$, it means that there are already four codewords.

The possible codewords with the length of 2 are

```
0  0
0  1
1  0
1  1
```

If this is the case, that $n = 2$, it means that there cannot be errors in the message, since the NND will always pick one of them as the corrected message.

The consequences of using a small codeword length can be seen from the above condition, meaning that there will be lots of errors in the decoded message. From this observation it is clear that the codeword length of two will have very high bit error rates, hence the test will be conducted from codewords length three. Even higher lengths will be investigated, to see how the efficiency of the NND is at high codeword lengths.

The conditions under which the lengths of the codeword are investigated will be discussed below.

#### *4.5.1.1  Condition 1*

This condition is the one explained above, stating that the length of the codeword must always be greater than two.

Therefore $Condition\ 1: n > 2$

### *4.5.1.2 Condition 2*

Number of bits of the sent message must always be the same, but the message will not be the same due to the different codeword. The number of bits of the sent message was 1680 bits. 1680 bits were chosen because it is divisible by $3, 4, 5, 6, 7\ and\ 8$.

### *4.5.1.3 Condition 3*

The probabilities of inversion errors in the channel were the same for both codeword lengths to be investigated. These probabilities were 0.001, 0.005, 0.007, 0.008, 0.01, 0.02, 0.05, 0.07, 0.08, 0.1, 0.2, 0.3, 0.4, 0.5, 0.6, 0.7, 0.8, 0.9, 1.0

### *4.5.1.4 Condition 4*

For all different codeword lengths, the number of classifiers will be kept at maximum, they will not be varied.

These four conditions will be kept the same for all the tests when investigating the different values of $n$. But to complete the test and investigation of the codeword length, the following parameter must be considered at all times. This test cannot be complete without the following parameter. This parameter is the length of the mapper, N.

## 4.5.2 Length of the Mapper

Length of the mapper is denoted as N. The construction of the mapper has been explained in Section 4.4.2. The length of the mapper must be greater than one; otherwise it ends up being a simple bit to bit comparison. The length constraints of the mapper are $1 < N \leq n$. $N = 2$ will be known as $N_{min}$ and $N = n$ will be known as $N_{max}$.

Each codeword will have mappers with the length $1 < N \leq n$. Therefore a codeword with length of eight will investigated for mappers of length $N = 2, 3, 4, 5, 6, 7 \text{ and } 8$ and Condition 4 will be kept.

The testing of this parameter and the first parameter can be referred to as Test 1, since they work hand in hand. One of these parameters cannot be tested without the other.

**Table 4-1: Codeword Vs Tuple Matrix**

|       | N=2 | N=3 | N=4 | N=5 | N=6 | N=7 | N=8 |
|-------|-----|-----|-----|-----|-----|-----|-----|
| n =8  | √   | √   | √   | √   | √   | √   | √   |
| n =7  | √   | √   | √   | √   | √   | √   |     |
| n =6  | √   | √   | √   | √   | √   |     |     |
| n =5  | √   | √   | √   | √   |     |     |     |
| n =4  | √   | √   | √   |     |     |     |     |
| n =3  | √   | √   |     |     |     |     |     |

The above table indicates what parameters are to be investigated for the first test. The ticks indicate which mapper sizes are to be investigated for each particular codeword size. The best decoders from Test 1 are to be investigated under the following design parameter. The decoders where $N = n$ can be expected to be worse since it does not have much significance, because it compares the bits all at once.

### 4.5.3 Number of Classifiers

The number of classifiers is primarily determined by the number of mappers, of which is determined by the length of the mapper. The best decoders of Test 1 will be investigated for various numbers of classifiers. This will be known as Test 2. The number of classifiers will be randomly decreased to observe what happens when few classifiers are used.

The classifiers to be used will be indicated in Chapter 6. The selection of which index to get rid of will be done randomly. But this random selection of index will ensure that at least each index number is left in one of the classifiers, i.e. if reducing for $n = 7$, at least and index containing $1, 2, 3, 4, 5, 6 \text{ and } 7$ should be present in any of the classifiers.

For every $n - bit$ codeword encoder there is a $n - bit$ codeword decoder. The $n - bit$ decoder has $2 < N \leq n$ decoders in it. This means that if the codeword with $n - bits$ can be used to decode, and to make it efficient a correct $N - bit$ decoder must be chosen. As stated before, these parameters cannot be separated; they always complement each other to form one decoder. Therefore, can be said that a decoder with $n - Bit \text{ and } N - bit$ is a $n - bit \text{ } codeword \text{ } N - bit \text{ } tuple$ decoder.

### 4.5.3.1 Example 6

If an 8-bit codeword is used, there would be:
  a) $8 - Bit \text{ } codeword \text{ } 8 - Bit \text{ } tuple \text{ } decoder$
  b) $8 - Bit \text{ } codeword \text{ } 7 - Bit \text{ } tuple \text{ } decoder$
  c) $8 - Bit \text{ } codeword \text{ } 6 - Bit \text{ } tuple \text{ } decoder$
  d) $8 - Bit \text{ } codeword \text{ } 5 - Bit \text{ } tuple \text{ } decoder$
  e) $8 - Bit \text{ } codeword \text{ } 4 - Bit \text{ } tuple \text{ } decoder$
  f) $8 - Bit \text{ } codeword \text{ } 3 - Bit \text{ } tuple \text{ } decoder$
  g) $8 - Bit \text{ } codeword \text{ } 2 - Bit \text{ } tuple \text{ } decoder$

And if a 7-bit codeword is used, there would be:
  a) $7 - Bit \text{ } codeword \text{ } 7 - Bit \text{ } tuple \text{ } decoder$
  b) $7 - Bit \text{ } codeword \text{ } 6 - Bit \text{ } tuple \text{ } decoder$
  c) $7 - Bit \text{ } codeword \text{ } 5 - Bit \text{ } tuple \text{ } decoder$
  d) $7 - Bit \text{ } codeword \text{ } 4 - Bit \text{ } tuple \text{ } decoder$

e) $7-Bit\ codeword\ \ 3-Bit\ tuple\ decoder$
   f) $7-Bit\ codeword\ \ 2-Bit\ tuple\ decoder$

   And so on ...

## 4.6 Test 1

This test has two parts. The first part will be comparing the different $N$ values of each $n$ codeword and determining which $n$ codeword makes the best decoder parameter. The second test involves comparing the different $n$ values for each $N$ mapper and determining which $N$ mapper makes the best decoder parameter. This test will investigate NND of $codeword\ length = 4, 5, 6, 7\ and\ 8$.

In the conclusion of this Test, the best $n\ and\ N$ values will be deduced and then carried to Test 2. The memory usage of each pair will be considered, and the best compromise in relation to memory usage will be made, if necessary. But as this compromise is being made, a decoder that's chosen must be of high efficiency.

## 4.7 Test 2

This test uses best results of Test 1. This test investigates how the decoder's memory usage can be improved but with the efficiency still being high, or at least closer to the highest efficiency. This test should conclude on which $N\ and\ m$ values are the best for NND. The $n$ value will be investigated in Test 3. At the end of this Test, the memory usage of each decoder investigated in this test will be calculated.

## 4.8 Test 3

This test comes after looking at the memory usage formula. So this test should be able to conclude that the Memory usage at high $n$ values is low. The results obtained in this test will be able to tell what happens to the decoder's efficiency when the length of the codeword increases. The best results of Test 2 will be compared with an NND of $n = 12$.

## 4.9 Conclusion

The methods to be used in detection and correction of inversion errors were discussed in this chapter. This chapter also gave an overview of how the message will be transmitted and where errors would be introduced. The channel used to introduce error was the RID channel discussed in Chapter 2.

The message received after the channel had errors and it is received by the receiver. In the receiver, that's where the decoder is found. The decoder in this receiver is the Neural Network Decoder. The operations of this decoder were fully discussed, firstly looking at how the bits of the sent message could be compared to the ones of the original message.

Comparing the received bits with the original message was done by comparing the bits of them to the possible codeword found in the codebook. Then the Neural Network chooses the best codeword resembling those received bits. The received bits to be compared with the codeword were termed as "group".

As the codeword is found to be correct, it is reconstructed in the new message which is considered the correct message. The construction of a NND is discussed and explained. The NND is found to be made up of the Classifier. The classifier is made of the mappers and the voting system.

The mappers are designed using the index set, which it was also explained how they work. Mappers are then made for the received bits and the sent bits. These mappers are then compared to see if they are equal or not. The process that follows after this comparison is done through the voting system.

The voting system casts a vote whenever the values of the mappers are the same and no vote when they are not the same. The votes are then added, each codewords vote are added separately. Then the codeword with the most votes is considered as the correct message. If more than one codeword are correct, the NND chooses any one of those codeword as the correct one.

The concept of voter confidence was explained. Furthermore, the parameters to be investigated in this thesis were discussed. These parameters are the length of the codeword, the length of the mapper and the number of classifier per decoder. The conditions to be considered during the investigations of these parameters were also discussed.

The way in which the decoders would be called was also discussed in this chapter. This chapter was concluded by explaining the process to be followed when running the tests to test the investigated parameters. The process was discussed for Test 1, Test 2 and Test 3, stating which parameters each test will be investigating.

The objective of this project is to find the optimal parameters for Neural Network Decoders, so Test 1 investigates the length of the codeword and the mapper in an attempt to find the best parameters. Test 2 investigates the number of classifiers in the NND. Whereas Test 3 investigates the effects of LDPC in NND, but it was only proved for 12-Bits. The results obtained for the 12-Bits are assumed to be good enough to give an idea of how LDPC bits would behave in a NND.

The next chapter will discuss how the mathematical model of Chapter 4 will be implemented in software for simulation purposes.

# Chapter 5: Software Implementation Design

## 5.1 Introduction

This chapter will show how the mathematical approach of Chapter 4 will be implemented in software for simulations. The components used it the simulation will be discussed. The method to be used to test whether the decoder has decoded the message correctly and to what degree is the decoded message correct will be discussed in this chapter.

## 5.2 Simulation Procedure

This section will mention the components which are to be used in the simulation process. The simulation process will follow the block diagram below.

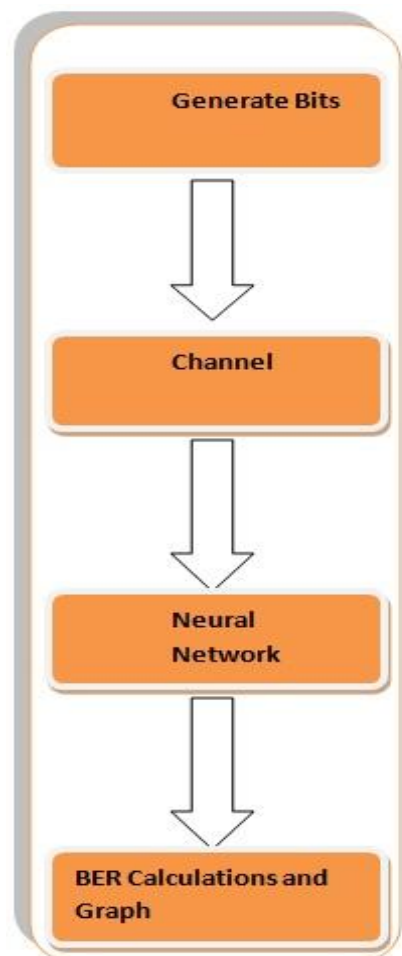

**Figure 5-1: Simulation Procedure**

Bits will be generated and sent through the channel, from the channel, the bits with errors with be decoded in the Neural Network Decoder. After the Neural Network Decoder, the bit error rate in the decoded message will be calculated and the bit error rate graph will be plotted. This block diagram does not have an encoder because, it was assumed in the previous chapters that the message will be assume to be encoded.

## 5.3 Bit Generator

The bit generator is the component that will be used to generate the bits. These bits will be generated as an encoded message. The bits to be generated will be a compilation of codewords from the codebook. The bit generator will generate the message which will be assumed encoded as indicated in Section 4.2. Random bits were not used for these tests.

## 5.4 Channel

The channel to be used in this simulation is the RID channel, which can introduce inversion errors, deletions and insertions. But only inversion errors will be used in this thesis.

The probabilities of the inversion, $P_{inv}$, errors will be varied while the probabilities of the deletions, $P_{del}$, and probabilities of insertion, $P_{ins}$, errors are kept at zero. They are kept at zero because they are not being investigated. The probabilities of receiving, $P_{rec}$, a correct bit is $P_{rec} = 1 - P_{inv}$ . The testing of this channel was done by T Theledi [**21**] and it was found to be working and correctly simulated.

## 5.5 Neural Network Decoder

This part of the simulation received bits from the channel, and detects and corrects the errors. The corrected message after this part is called decoded message. The operations of this decoder on how it will decode were discussed in Section 4.4. In these tests, the NND was used to decode the message known to the sender. It was not tested for random bits.

## 5.6 Bit Error Rate

The decoded message will be compared to the encoded message, to calculate the bit error rate. The formula to be used to calculate the bit error rate is

$$BER = \frac{number\ of\ bit\ errors}{number\ of\ generated\ bits}$$

The number of bit errors will be those of the decoded message which are not the same as those of the sent message. The bit error rate graph is then plotted for different probabilities of the channel. This graph gives a clear graphical indication of how the good or bad the decoder is for different error probabilities.

## 5.7 Testing the Simulation Components

### 5.7.1 Testing the Bit Generator

The message sent by the bit generator must look like the codeword in the codebook. To explain this statement an example below will be used.

#### *5.7.1.1 Example 1*

If using a 4-Bit code word to transmit a 20-Bit message.

The codeword in the codebook are:

$$word1 = [1 \quad 0 \quad 0 \quad 1]$$
$$word2 = [1 \quad 1 \quad 0 \quad 0]$$
$$word3 = [1 \quad 1 \quad 0 \quad 1]$$
$$word4 = [0 \quad 0 \quad 0 \quad 0]$$
$$encoded\ message = [word1 \quad word3 \quad word2 \quad word4 \quad word2]$$
$$encoded\ message = 1\ 0\ 0\ 1\ 1\ 1\ 0\ 1\ 1\ 1\ 0\ 0\ 0\ 0\ 0\ 0\ 1\ 1\ 0\ 0$$

∴ The bit generator is working.

### 5.7.2 Testing the channel

The channel used here was the same one used in [**21**], the testing of the channel can be seen there. It was concluded there that the channel simulation works.

### 5.7.3 Testing the Neural Network Decoder

To test whether the Neural Network decoder works, it will be shown by using a 7-Bit codeword as a message. An error will be introduced in this message and put through the decoder to see if it will be able to correct the message.

*5.7.3.1 Example 2*

A 7-Bit codeword with 2-Bit Tuple will be used, and error probability of 0.1 will be used in the channel.

$$word1 = [0 \quad 0 \quad 0 \quad 1 \quad 1 \quad 0 \quad 1]$$
$$word2 = [0 \quad 0 \quad 1 \quad 0 \quad 0 \quad 1 \quad 1]$$
$$word3 = [0 \quad 0 \quad 1 \quad 0 \quad 1 \quad 0 \quad 0]$$
$$word4 = [0 \quad 1 \quad 0 \quad 0 \quad 0 \quad 1 \quad 0]$$
$$encoded\ message = [word1]$$
$$error\ message = [0 \quad 1 \quad 1 \quad 1 \quad 1 \quad 0 \quad 1]$$

There are 2 inversions.

The resulting votes are:
$$vote\ for\ word1 = 10$$
$$vote\ for\ word2 = 3$$

$$vote\ for\ word3 = 6$$

$$vote\ for\ word4 = 1$$

The graphical representation of the votes is indicated in the graph below:

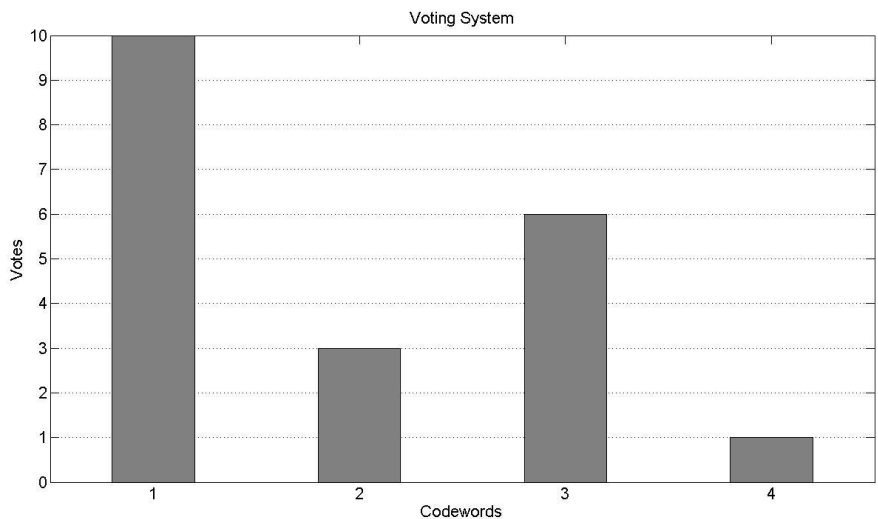

**Figure 5-2: Voting system graph for Example 2**

Therefore the correct message is then:

$$correct\ message = [0\quad 0\quad 0\quad 1\quad 1\quad 0\quad 1]$$

In the above example, the NND picked up that the message which was originally sent was from codeword1. Therefore this concludes that the Neural Network Decoder is working. The NND then chooses codeword1 as the correct message

## 5.8  Conclusion

In this chapter we saw how the NND will be implemented in the simulation. The simulation components were mentioned and discussed fully. The simulation procedure is explained, detailing how the message will be exposed to a channel with errors and how these errors in the sent message will be decoded.

This simulation procedure is done in four stages. These stages are bit generation, channel simulation, Neural Network Decoder and finally the calculation of the bit error rate and plotting it.

The bit generation stage discussed how the message was constructed. The testing of the bit generator showed that this bit generator to be used in these tests is working. The stage that follows this part is the channel.

In the channel simulation stage, the message from the bit generator is sent through and errors are introduced. The errors are introduced with different error probabilities in the channel. The error probabilities mean that depending on the probability value a certain number of errors will be introduced in the message.

The stage that follows the channel simulation is the Neural Network Decoder. The Neural Network Decoder corrects the errors in the message. The testing of the Neural Network Decoder showed that it works and how the message will be decoded.

The next chapter will discuss the results obtained after running the different tests mentioned in Chapter 4. The voting system graph will not be indicated in the following chapter, only the BER graph will be indicated.

# Chapter 6: Results and Results Analysis

## 6.1 Introduction

This chapter discusses the results obtained after running the different simulation tests. The results of Test 1, Test 2 and Test 3 will be discussed in this chapter. The best decoders will be chosen by looking at the bit error rate (BER) graphs. The best results of each test will be used in the succeeding test. The results of Test 3 will help with deducing the optimal parameters for the NDD.

## 6.2 Test 1

Classifiers of length 3, 4, 5, 6, 7 and 8 will be investigated at their maximum tuple arrangements. The bits which were sent in this test were $Number\ of\ Bits\ in\ Message = 5 \times 6 \times 7 \times 8 = 1680$. $Number\ of\ Bits\ in\ Message = 1680\ bits$ were used so that all the 3, 4, 5, 6, 7 and 8 classifiers may use the same number of bits, but each classifier had its own codebook.

### 6.2.1 Length of the Codeword

One table of results will be put in this section to help indicate how many average errors were experienced at certain probabilities. This table will also indicate the BER values of that decoder which is used as an example. These BER values are indicated in Table 6-1.

### *6.2.1.1 Eight Bits Codeword*

**Table 6-1: BER Results for 8-Bits Codeword Decoders**

| Probabilities | Total No. of Inversions | Ave. No. of Inversions | N=2 | N=3 | N=4 | N=5 | N=6 | N=7 | N=8 |
|---|---|---|---|---|---|---|---|---|---|
| 0.001 | 147 | 1.4700 | 0 | 0 | 0 | 0 | 0 | 0 | 0.0039 |
| 0.002 | 334 | 3.34 | 0 | 0 | 0 | 0 | 0 | 0 | 0.0082 |
| 0.005 | 773 | 7.73 | 0.00002381 | 0 | 0.00002381 | 0.00002381 | 0.00010119 | 0.00025595 | 0.0194 |
| 0.007 | 1198 | 11.98 | 0.000065476 | 0.000011905 | 0.0000476 19 | 0.0000654 | 0.00030952 | 0.0006667 | 0.0261 |
| 0.008 | 1347 | 13.47 | 0.00012500 | 0.00025000 | 0.00012500 | 0.0001310 | 0.00033333 | 0.00095833 | 0.0314 |
| 0.01 | 890 | 17.8 | 0.0002857 | 0.00027381 | 0.00020238 | 0.00014286 | 0.00039286 | 0.0014 | 0.0370 |
| 0.02 | 1657 | 33.14 | 0.00058333 | 0.0011 | 0.0010 | 0.00044048 | 0.0010 | 0.0042 | 0.0742 |
| 0.05 | 4283 | 84.860 | 0.0049 | 0.0055 | 0.0050 | 0.0042 | 0.0106 | 0.0281 | 0.1688 |
| 0.07 | 5929 | 118.58 | 0.0101 | 0.0111 | 0.0096 | 0.0110 | 0.0197 | 0.0520 | 0.2127 |
| 0.08 | 6768 | 135.36 | 0.0137 | 0.0127 | 0.0141 | 0.0147 | 0.0251 | 0.0601 | 0.2317 |
| 0.1 | 4099 | 163.96 | 0.0207 | 0.0195 | 0.0232 | 0.0213 | 0.0411 | 0.0901 | 0.2782 |
| 0.2 | 8430 | 337.2 | 0.0944 | 0.0934 | 0.0967 | 0.0963 | 0.1397 | 0.2469 | 0.4054 |
| 0.3 | 12657 | 506.28 | 0.2049 | 0.2019 | 0.2120 | 0.2136 | 0.2654 | 0.3617 | 0.4561 |
| 0.4 | 16795 | 671.8 | .3328 | 0.3359 | 0.3391 | 0.3481 | 0.3807 | 0.4423 | 0.4757 |
| 0.5 | 20917 | 836.68 | 0.4682 | 0.4669 | 0.4658 | 0.4740 | 0.4751 | 0.4798 | 0.4837 |
| 0.6 | 25190 | 1007.6 | 0.5805 | 0.5725 | 0.5711 | 0.5751 | 0.5340 | 0.5042 | 0.4862 |
| 0.7 | 29499 | 1180 | 0.6550 | 0.6566 | 0.6525 | 0.6515 | 0.5811 | 0.5235 | 0.4885 |
| 0.8 | 33602 | 1344.1 | 0.7121 | 0.7070 | 0.7087 | 0.7085 | 0.6402 | 0.5482 | 0.4937 |
| 0.9 | 37774 | 1511.0 | 0.7481 | 0.7520 | 0.7467 | 0.7548 | 0.7108 | 0.5982 | 0.4948 |
| 1.0 | 42000 | 1680 | 0.7851 | 0.7851 | 0.7851 | 0.7851 | 0.7851 | 0.6798 | 0.4839 |

The total number of inversion errors in Table 6-1 are the errors detected for a certain iteration while the average is the total number of inversion errors divided by the number of iterations.

The Resulting Graph:

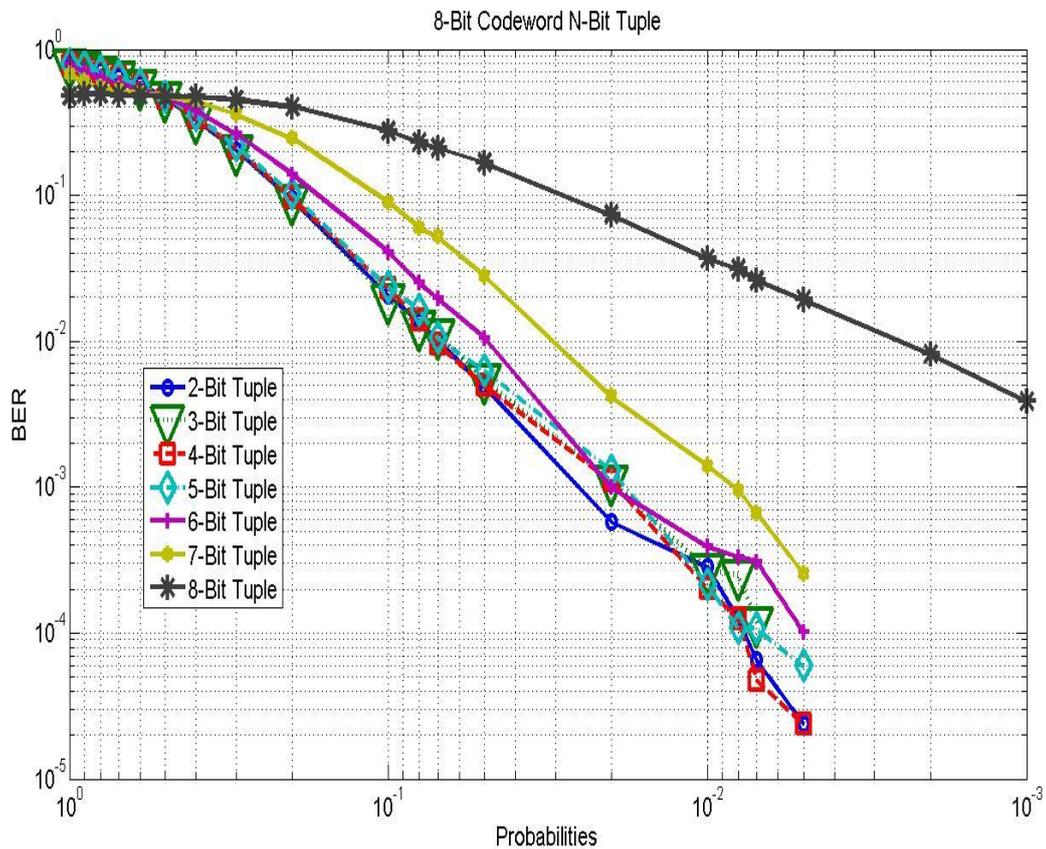

**Figure 6-1: Eight Bits Codeword Decoder**

The Eight Bit Codeword decoder has a flat slope when using tuples with higher lengths. It can be seen at the graph and results in the above table, Table 6-1, that when the tuple has 8-bits which are the same length as the codeword, the BER graph becomes very flat. The BER graph smoothens and becomes steeper as we decrease the number of bits per tuple.

2-Bits tuple, 3-Bits tuple, 4-Bits tuple and 5-Bits tuple decoder produce very good results and they are almost equal in this case. The best of these four decoders will be investigated later in some more tests to see if the current results can be improved, and if so, how.

These four decoders (2-Bits tuple, 3-Bits tuple, 4-Bits tuple and 5-Bits tuple decoders) decode the sent message better than the other three decoders. This means that the decoded message had less error compared to the sent message than the 6-Bits tuple, 7-Bits tuple and 8-Bits tuple decoder.

At the higher probabilities, the BER values of the higher bit tuple are lower compared to the ones of small probabilities. This was discovered as to being up to the choice of codebook used to train the Neural Network Decoder.

### *6.2.1.2  Seven Bits Codeword*

The general idea of the results have been shown in the previous tests in the form of tables, so from now on only the graphs will be used to indicate the obtained results. And only the best decoders will be identified since their theoretical descriptions are the same as the previously discussed.

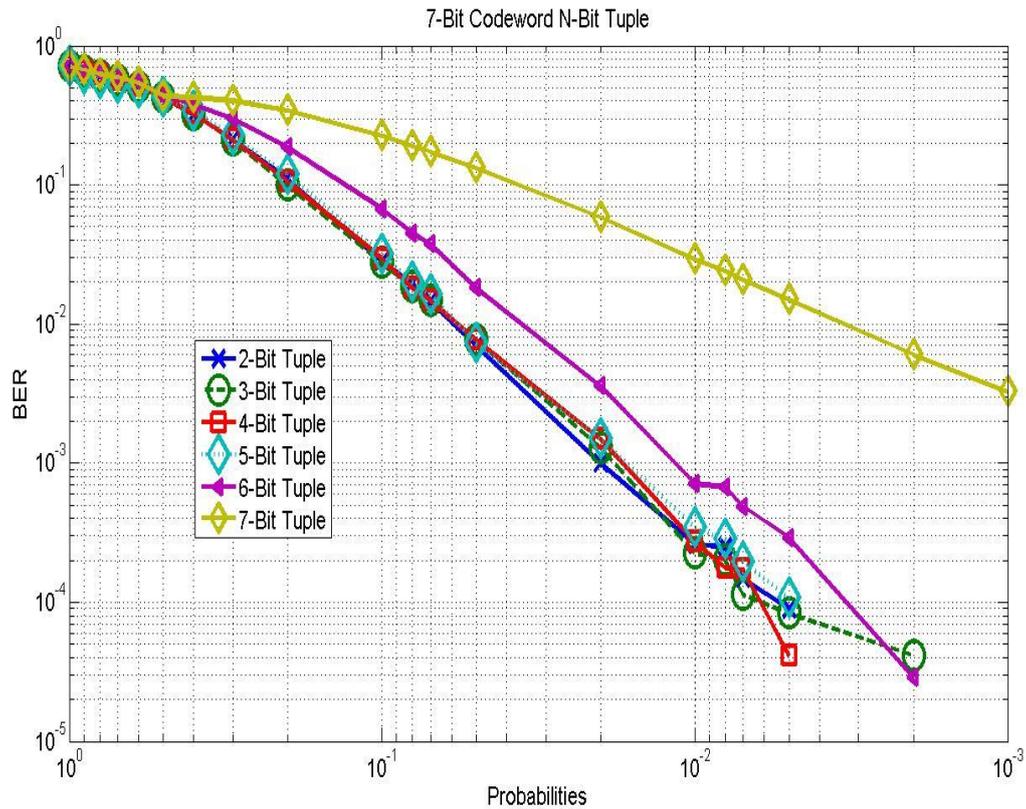

**Figure 6-2: Seven Bits Codeword**

The mapper with the same length as the length of the codeword is once again a very bad decoder; it produces too many errors in its decoded message. 2-Bits tuple, 3-Bits tuple, 4-Bits tuple and 5-Bits tuple decoders once again make a very good decoder. Their decoded massages have very low error rates as compared to the 6-Bits tuple and 7-Bits tuple decoders.

Also from the graph it can be seen that 2-Bits tuple, 3-Bits tuple and 4-Bits tuple decoders have almost same results.

### 6.2.1.3  *Six Bits Codeword*

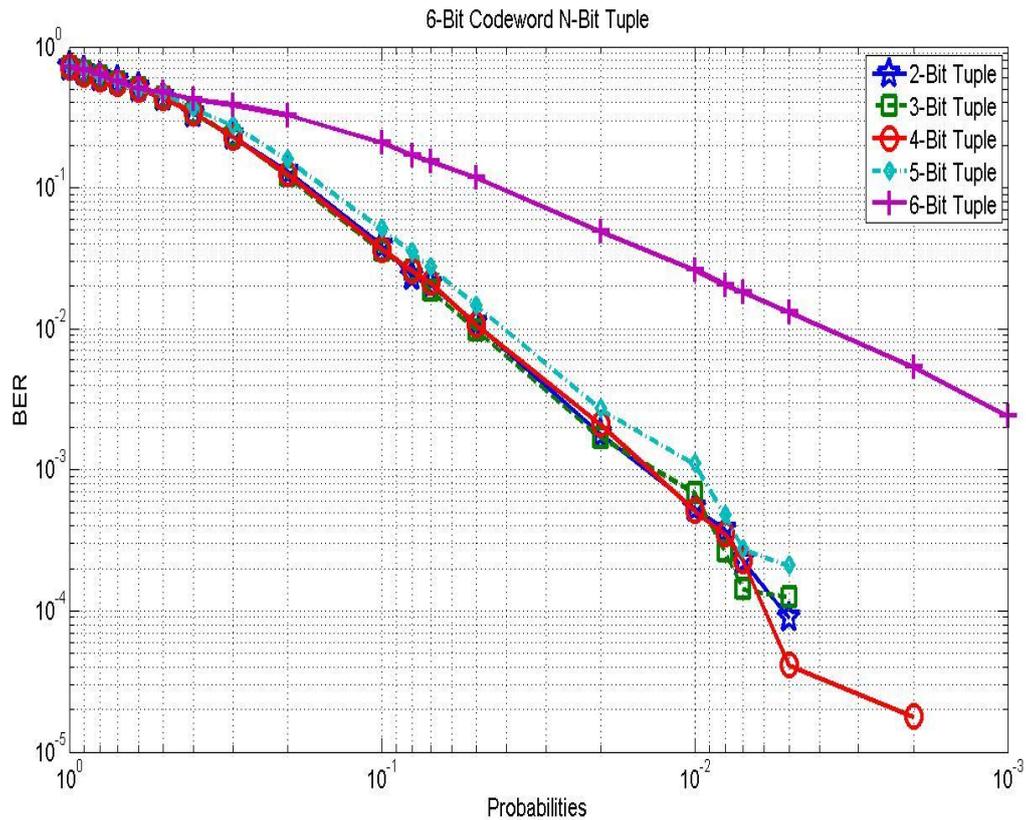

**Figure 6-3: Six Bits Codeword Decoders**

2-Bits tuple, 3-Bits tuple and 4-Bits tuples decoder make the best 6-Bits codeword decoders. Their BER graph is steeper than that of 5-Bits tuple and 6-Bits tuple decoders. Yet again the choice of codebook can be seen to have effect in this test at very high probabilities.

### *6.2.1.4 Five Bits Codeword*

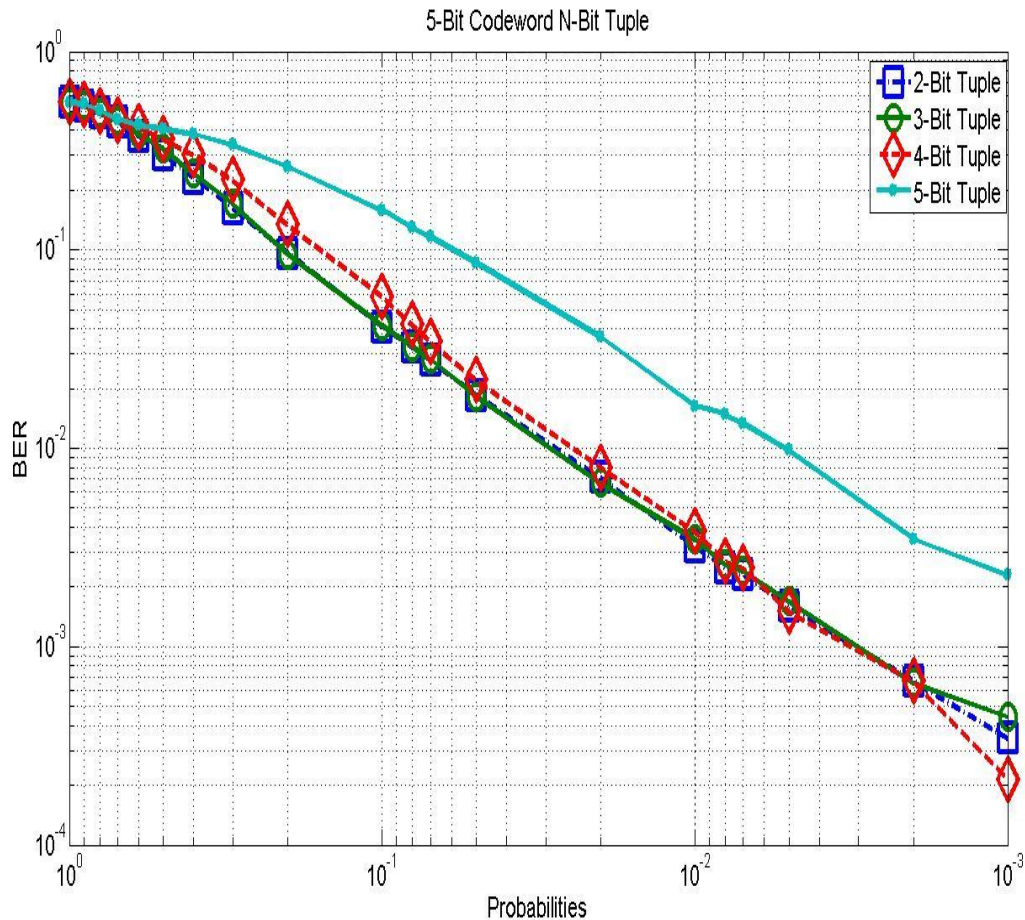

**Figure 6-4: Five Bits Codeword Decoders**

2-Bits tuple and 3-Bits tuples decoders make the best 5-Bits Codeword decoder. The 5-Bit Codeword 2-Bit Tuple and 5-Bit Codeword 3-Bit Tuple decoders have almost similar results.

A 5-Bit Codeword 4-Bit Tuple decoder gets better and similar to the 5-Bit Codeword 2-Bit Tuple and 5-Bit Codeword 2-Bit Tuple decoders at low probabilities, but very poor at high probabilities compared to the two decoders.

But even though they might be the best of the 5-bit tuple decoder, one can see that they are becoming more flat than the other BER graphs.

### *6.2.1.5   Four Bits Codeword*

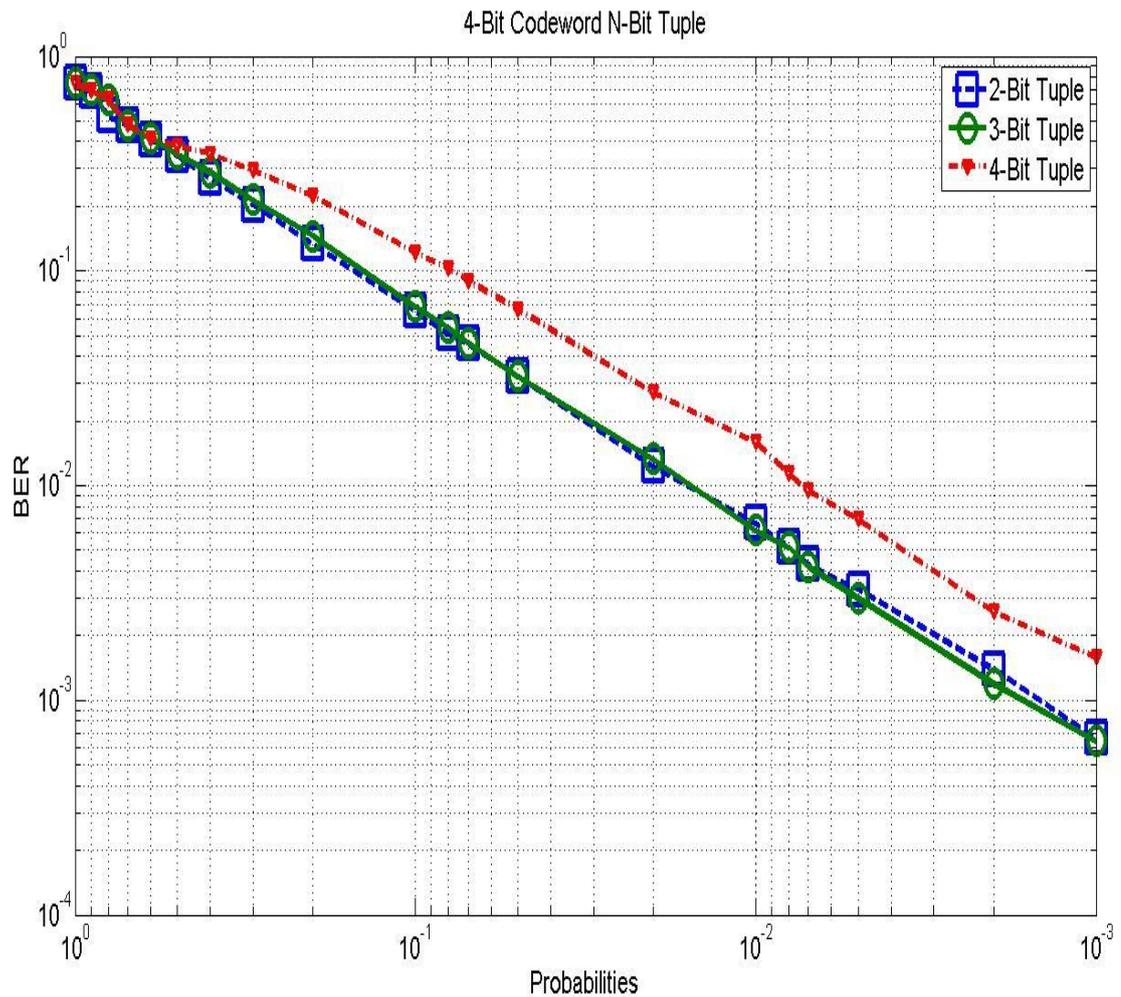

**Figure 6-5: Four Bit Codeword Decoders**

It can be seen that as we decrease the number of bits in the codeword, the results become very bad. The BER graph starts becoming very flat and this is the sign that there are too many errors in the decoded message as compared to the sent message. 2-Bit tuple and 3-Bits tuple decoders are the best decoders of the 4-Bit codeword decoders.

### *6.2.1.6 Three Bit Codeword*

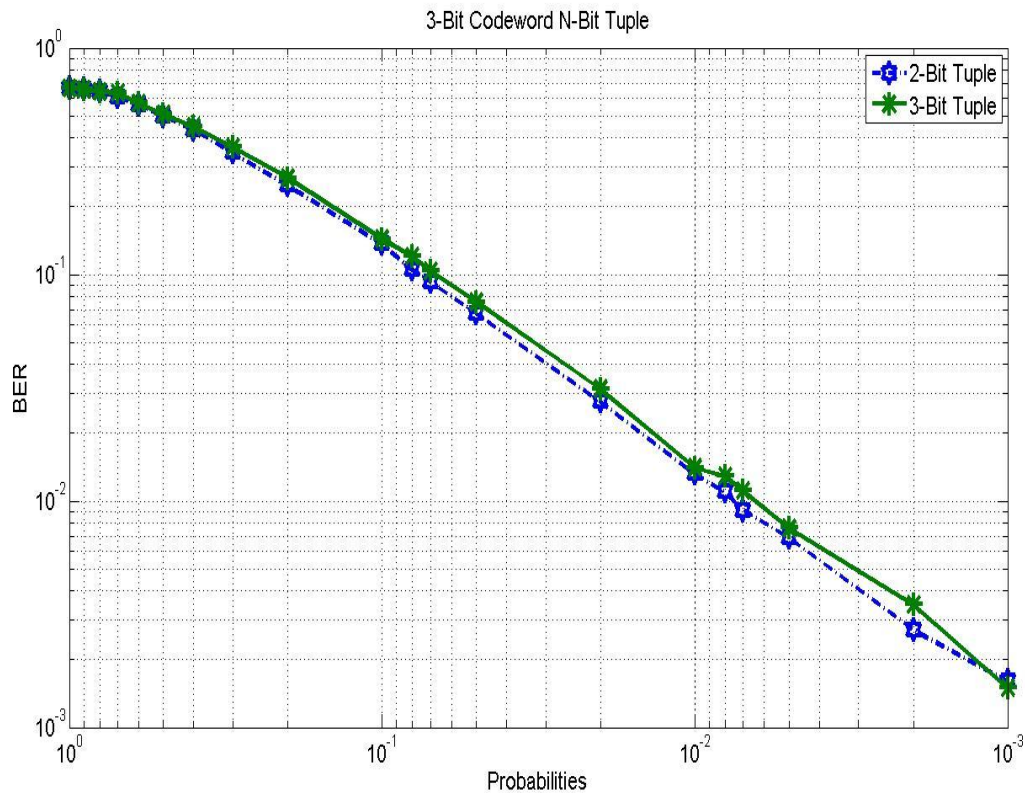

**Figure 6-6: Three Bit Codeword Decoder**

This test investigates the effect of using minimal codeword length, $n$, in transmitting message. For the codeword of this length, we can only have two decoders two compare. The best decoder for this codeword is the one made of 2-Bit tuple.

### 6.2.2  Tuple

Now that the different BER graphs have been obtained for different tuples of different codeword lengths, we can now compare the BERs by checking their tuples, i.e. check a 4-bit tuple for all codeword length in one graph. This is done so it can be seen which decoder forms the best decoder, and what parameter must that decoder have.

### *6.2.2.1 Seven Bits Tuple*

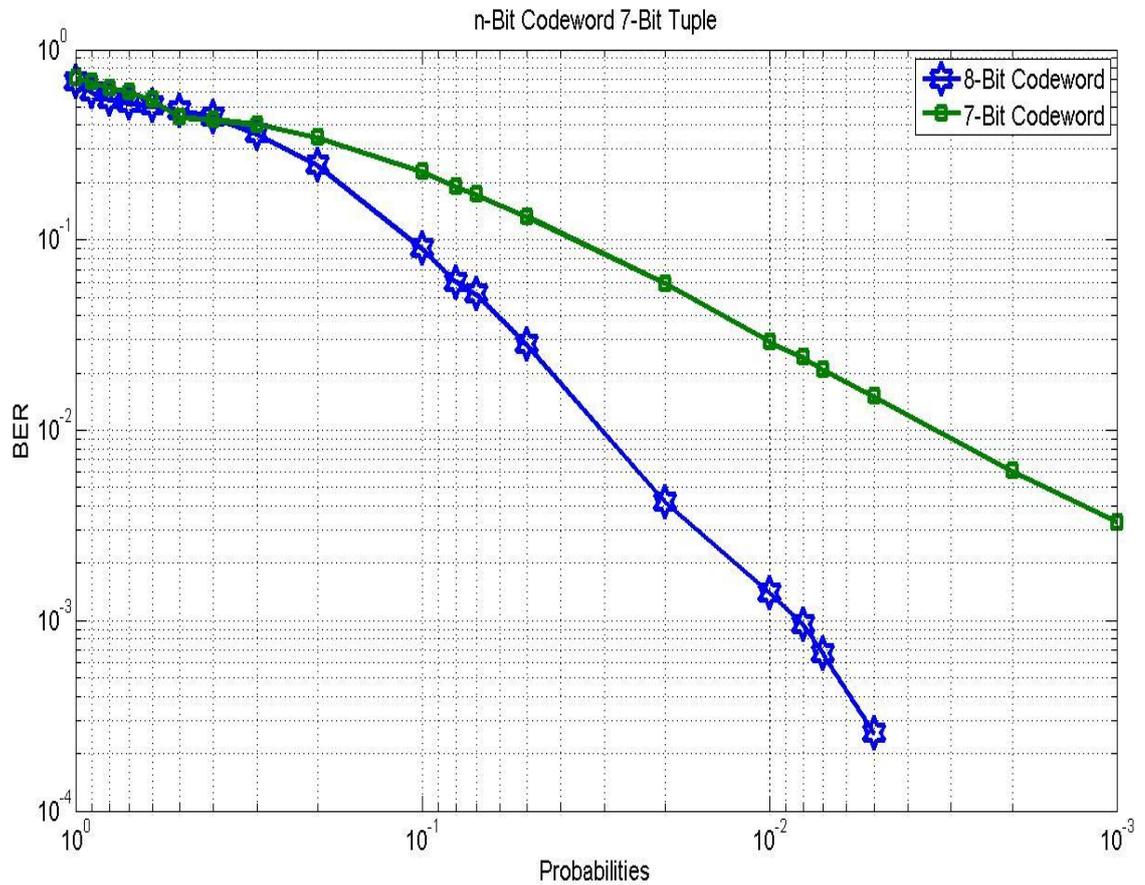

**Figure 6-7: Seven Bits Tuple Decoders**

A seven bit tuple decoder can only be found in any codeword decoder whose length is $\geq N$. In this case 7-Bit Codeword and 8-Bit Codeword are used. The BER graph of the 7-bit codeword is worse than that of the 8-Bit codeword. 8-Bit Codeword becomes a good decoder with fewer errors in the decoded message.

The efficiency of the 8-Bit Codeword 7-Bit Tuple decoder is very high compared to that of the 7-Bit Codeword 7-Bit Tuple decoder. The difference in efficiency can be deduced from graph observations. The 8-Bit Codeword 7-Bit Tuple decoder gets more effective at very low probabilities.

*6.2.2.2 Six Bits Tuple*

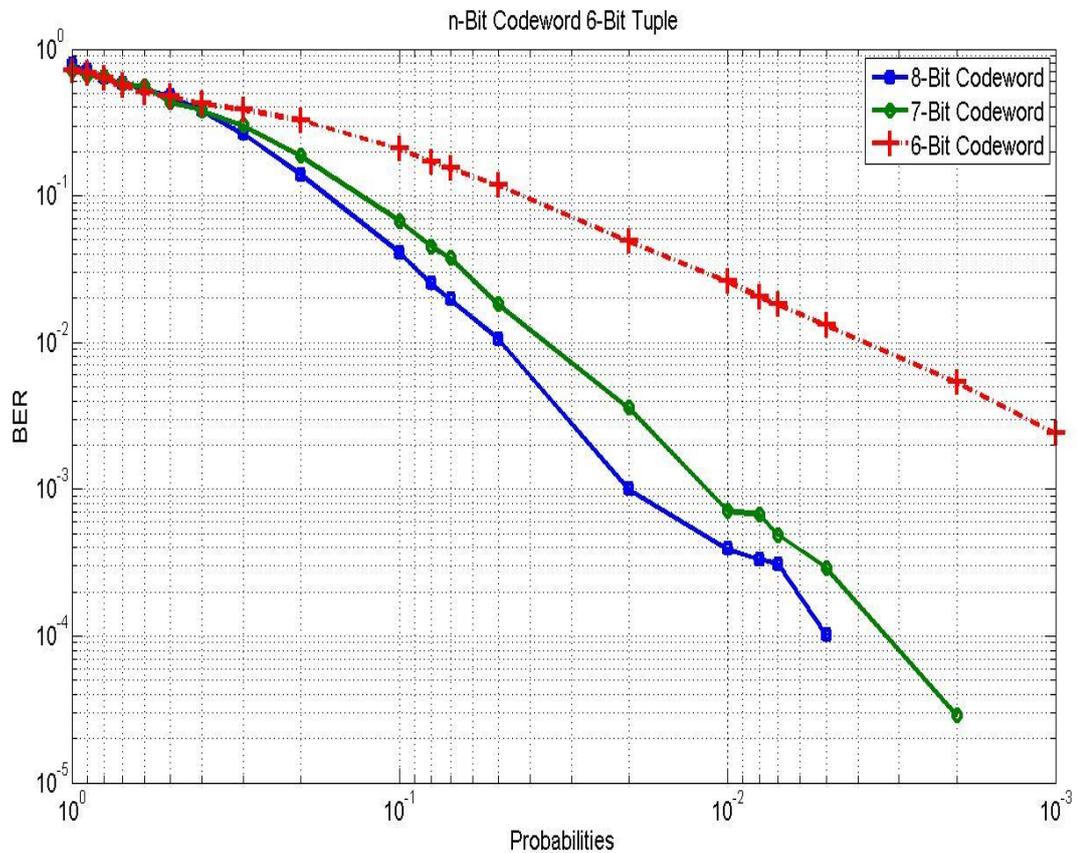

**Figure 6-8: Six Bits Tuple Decoders**

Comparing the 6-Bit Tuple decoders, shows that as in the previous one, 8-Bit Codeword 6-Bit Tuple decoder makes the best decoder compared to the 7-Bit Codeword 6-Bit Tuple and 6-Bit Codeword 6-Bit Tuple decoders.

This means that if 6-Bit Tuple decoder was to be used to transmit message, it would be an optimal decoder if an 8-Bit Codeword 6-Bit Tuple decoder was to be used.

## *6.2.2.3 Five Bits Tuple*

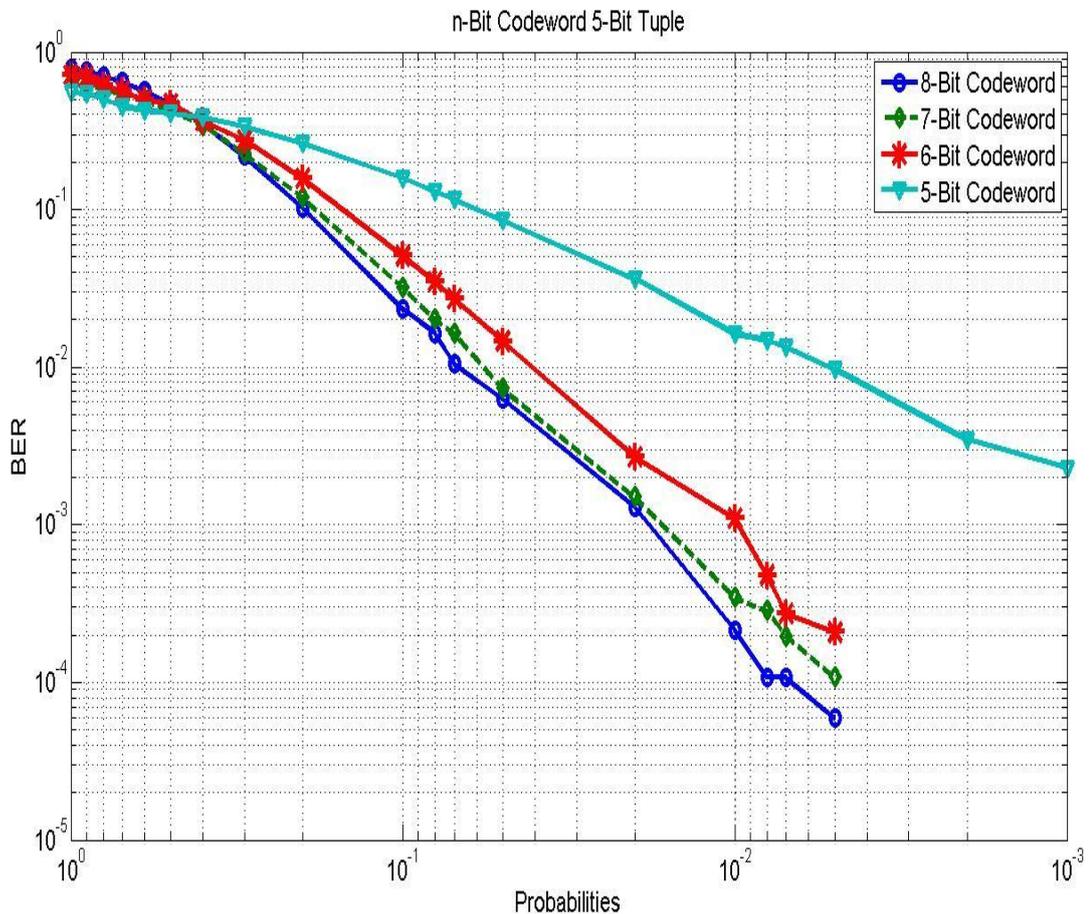

**Figure 6-9: Five Bits Tuple Decoders**

Again for the n-Bit Codeword 5-Bit Tuple decoder it can be seen that an 8-Bit Codeword 5-Bit Tuple decoder has the best BER graph. The results obtained at extremely high probabilities are strongly affected by the choice of codebook. But in this case, the 5-Bit Codeword 5-Bit decoders decodes better than the other n-Bit Codeword 5-Bit Tuple decoders.

### *6.2.2.4 Four Bits Tuple*

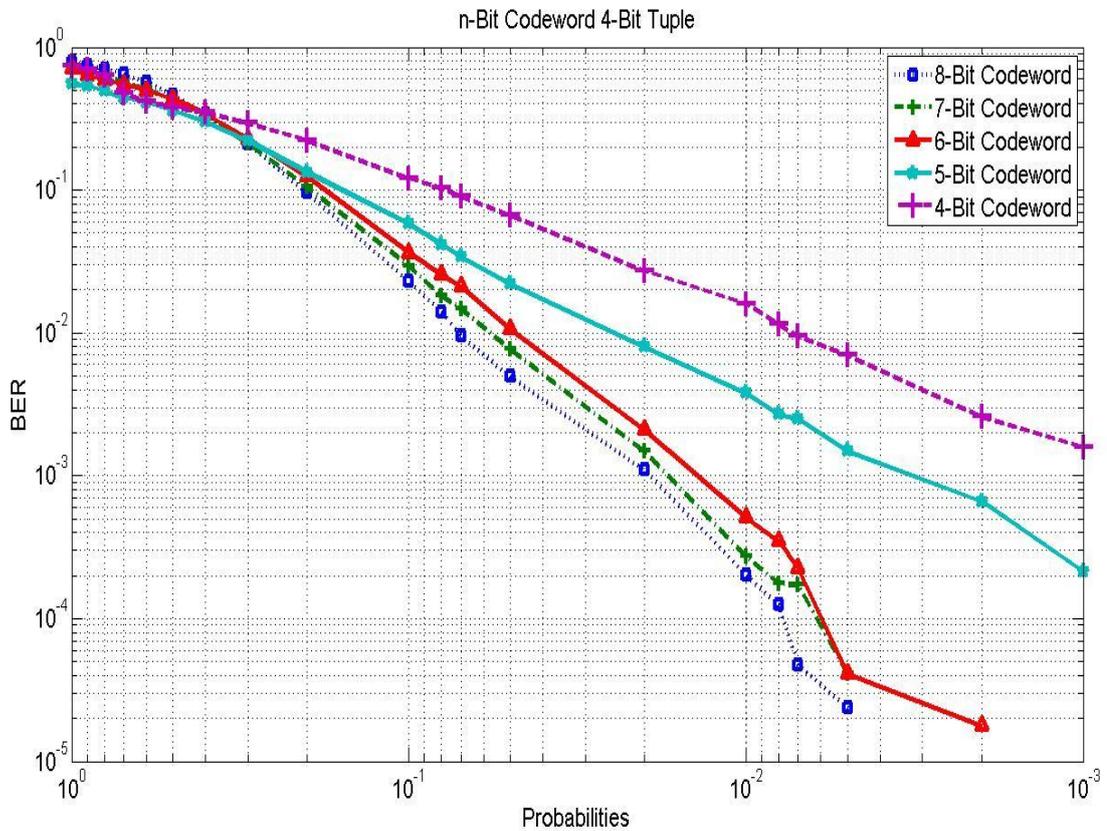

**Figure 6-10: Four Bits Tuple Decoders**

In the 4-Bits tuple of this case, the 8-Bit codeword 4-Bit Tuple decoder is once again the best decoder amongst the other decoders. The best decoders of this test are 8-Bit Codeword, 7-Bit Codeword and 6-Bit Codeword 4-Bit Tuple decoders.

### *6.2.2.5 Three Bits Tuple*

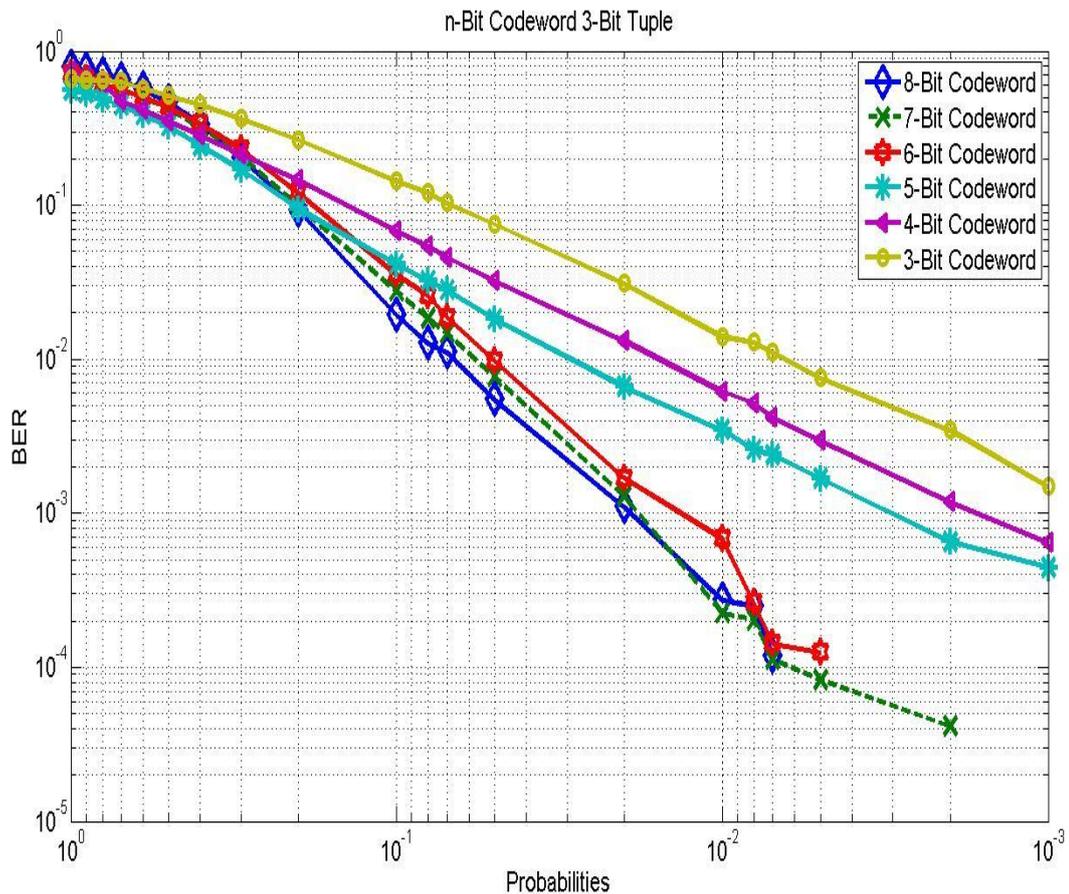

**Figure 6-11: Three Bit Tuple Decoders**

From the tuple comparisons it was noted that n-Bit Codeword 3-Bit Tuple decoder make the best decoders as compared to other decoders. These can also be seen form how close the BER graphs of each codeword length are, note that only -3-Bit Tuple decoders are shown in the above figure, Figure 6-11.

8-Bit codeword decoder is once again the preferable decoder at higher inversion probabilities. From $0.007 \leq Probability \leq 0.01$ the 7-Bit Codeword 3-Bit Tuple decoder would be the most preferable decoder. For any $Probabilities \geq 0.01$ an 8-Bit Codeword decoder would be the best decoder to use.

### *6.2.2.6 Two Bits Tuple*

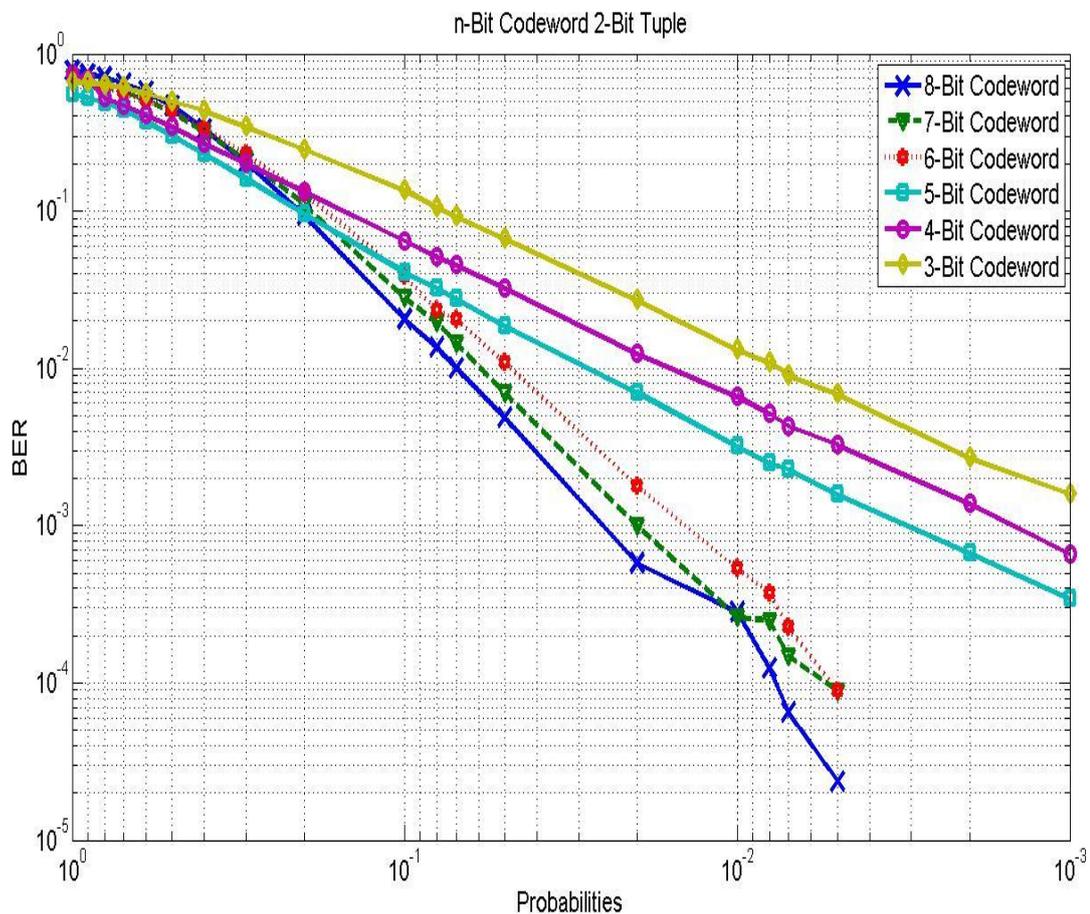

**Figure 6-12: Two Bit Tuple Decoders**

The n-Bit Codeword 2-bit Tuple decoders are not so much different from the n-Codeword 3-Bit Tuple decoders. Their efficiency is approximately the same. It will be investigated later to see which one can make the best decoder and how, and why should one use the other instead of the other one.

As such, the properties of its BER graph are the same as that of the 3-Bit tuple. Since it would be hard to have a 2-Bit codeword, a 2-bit Codeword N-Bit Tuple decoder was not investigated. Otherwise the decoded message would be wrong at high probabilities every time.

### 6.2.3  Test 1 Results Discussions

Test one was investigating the different classifier lengths at maximum tuple arrangements. From Section 6.2.1 it could be seen that, the decoders with acceptable BER graphs were mostly the ones of the tuple length of:

$$2 < N < c - 2$$

Where $N$ is the length of the tuple,

And $c$ is the length of the classifier.

The optimal decoding classifiers are those at $c \geq 7$, this leads to the deduction that at higher codeword length, Neural Networks make best decoders, but this will be verified later after Test 2 has been run in Test 3.

The best results of Test 2 will then be followed to see if the above statement can be verified to be true. The best results obtained from Test 1 will be carried further to Test 2, to see if the effect of decreasing total number of classifiers can improve the results.

So far, it has been seen that increasing the number of bits in the codeword increases the efficiency whereas increasing the number of bits in the tuple decreases the efficiency.
Therefore it can be concluded that the length of the codeword has an inverse proportional relationship with the length of tuple.

$$c \; \alpha \; \frac{1}{N} \; \alpha \; \eta$$

Where $\eta$ is the decoding efficiency.

With more iteration and more bits being sent through the channel, the results become better.

Test 1 concluded that the Binary Neural Network Decoder is best when the length of the codeword is increased and when the length of the tuple/ mapper is decreased. This actually play to the advantage since the low memory usage by these decoders is found when using high codeword length and low tuple/ mapper length. This can also be seen by looking at the memory usage formula.

The best decoders of this test are $8-Bit\ Codeword\ 2-Bit\ Tuple$ decoders and $8-Bit\ Codeword\ 3-Bit\ Tuple$ decoders. The memory usage is $56\ bytes\ and\ 224\ bytes$, respectively. During the simulations it was noted that, the less the memory usage, the faster the simulation. That is, when testing these decoders, the simulations were faster than when simulating the other decoders.

Test 2 will investigate $8-Bits\ codeword\ 2-Bits\ tuple\ and\ 8-Bits\ codeword\ 3-Bits\ tuple$ decoders will be the point of interest.

It must also be noted that using a 7-bit codeword decoder to decode a message which was sent using 8-Bit codeword encoder will lead to wrong results.

## 6.3 Test 2

In this section, the number of classifiers will be reduced to find the optimal number of classifiers for the Neural Network decoder. This test will be done for $Number\ of\ Bits\ in\ the\ Message = 1680\ bits$. This message will be decoded using the $8-Bit\ codeword\ 2-Bit\ tuple\ and\ 8-Bit\ codeword\ 3-Bit\ tuple\ decoders$, since they were chosen as best decoders from the previous test.

### 6.3.1 8-Bit Codeword Decoders

There are two tuple arrangements which will be investigated for a 8-Bit codeword decoders, viz. 2-Bit tuple and 3-Bit tuple decoders. Therefore the $8-Bits\ codeword\ 2-Bits\ tuple\ and\ 8-Bits\ codeword\ 3-Bits\ tuple$ decoders will be investigated in this test.

They will be reduced from the maximum tuples of Test 1 to different values of the number of classifier in a decoder.

*6.3.1.1 2-Bit Tuple*

This decoder has a total of 28 classifiers

- **Group A**

    This group will have 22 classifiers. The classifiers to be removed from the maximum classifier number are:

    $$I_2, I_9, I_{15}, I_{20}, I_{24}, I_{27}$$

- **Group B**

    This group has 15 classifiers. The more classifiers to be removed here will be added to those removed in group A.

    $$I_2, I_9, I_{15}, I_{20}, I_{24}, I_{27}, I_7, I_{13}, I_{16}, I_{19}, I_5, I_{18}$$

- **Group C**

    Only 10 classifiers are to be used in this part of the test. The removed classifiers are as follows:

    $$I_2, I_9, I_{15}, I_{20}, I_{24}, I_{27}, I_7, I_{13}, I_{16}, I_{19}, I_5, I_{18}, I_3, I_{10}, I_{14}, I_{23}, I_{28}$$

- **Group D**

    Only 5 classifiers will be used in this part of the test. The removed classifiers are as follows:

    $$I_2, I_9, I_{15}, I_{20}, I_{24}, I_{27}, I_7, I_{13}, I_{16}, I_{19}, I_5, I_{18}, I_3, I_{10}, I_{14}, I_{23}, I_{28}, I_{21}, I_6, I_{11}, I_{25}, I_8$$

The results obtained from all these tests are illustrated in Figure 6-13 below. The graph indicates all the groups discussed above, and show their BER graph against that of a maximum number of classifiers.

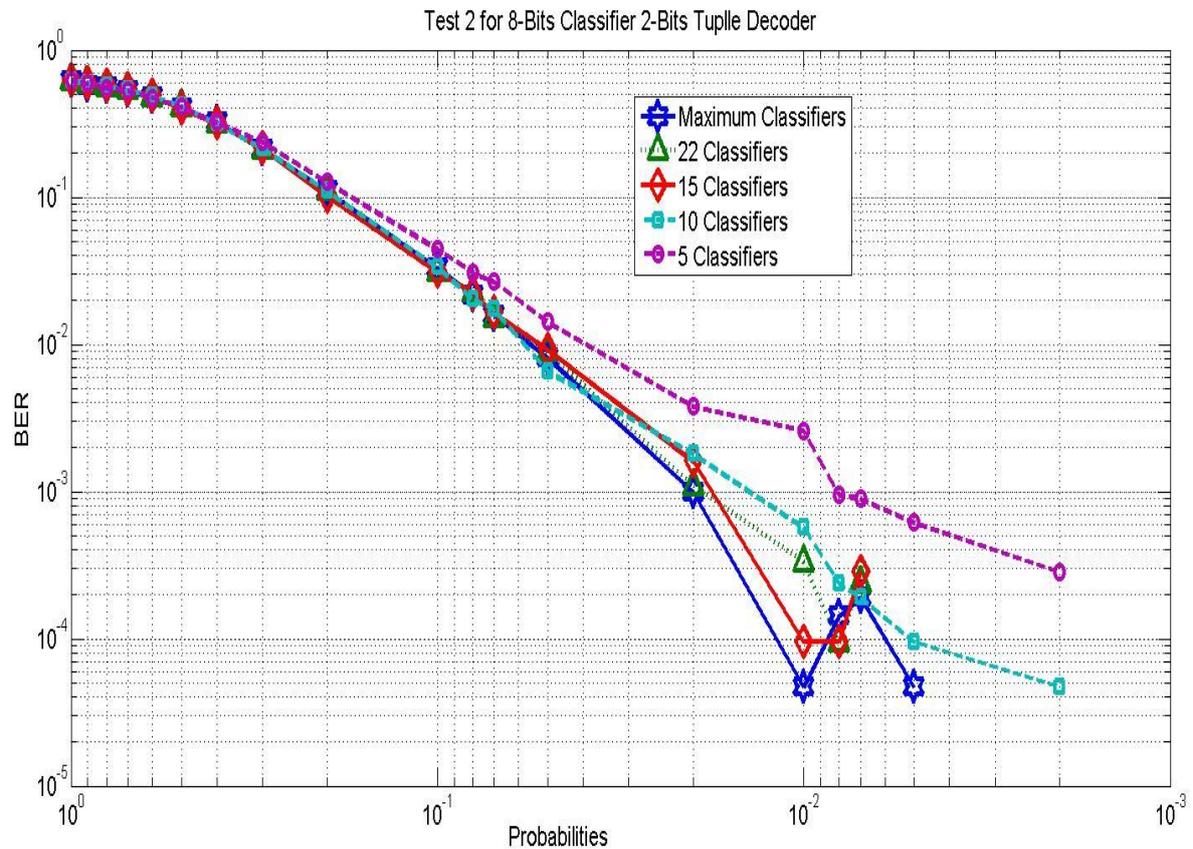

**Figure 6-13: Test 2 for 8-Bit Codeword 2-Bits Tuple Decoder**

From the graph above it can be seen that, maximum classifiers, i.e. $C_N^c$ produces a very efficient decoder at both high and low probabilities. It can also be seen that as we decrease the number of classifiers the BER graph does not improve, but worsens. This can be expected since the other groups are not considering all possible combinations of the tuple arrangements. But if memory had to be considered, Group A and Group C classifiers could be used interchangeably, since they are approximately close to the maximum classifier group.

*6.3.1.2   3-Bit Tuple*

This decoder has a total of 56 decoders.

- **Group A**

    This group will have 42 classifiers. The classifiers to be removed from the maximum classifier number are:

    $$I_8, I_{28}, I_{42}, I_{51}, I_{17}, I_{20}, I_{10}, I_{30}, I_2, I_{27}, I_{31}, I_{34}, I_{36}, I_{13}$$

- **Group B**

    This group has 29 classifiers. The more classifiers to be removed here will be added to those removed in group B.

    $$I_4, I_7, I_{15}, I_{19}, I_{25}, I_{32}, I_{38}, I_{41}, I_{47}, I_{50}, I_{54}, I_9, I_{24}$$

- **Group C**

    Only 23 classifiers are to be used in this part of the test. The removed classifiers are as follows:

    $$I_{37}, I_{56}, I_{52}, I_{40}, I_{14}, I_{33}$$

- **Group D**

    Only 14 classifiers will be used in this part of the test. The removed classifiers are as follows:

    $$I_{14}, I_{45}, I_3, I_{21}, I_{48}, I_{23}, I_{16}, I_{55}, I_6$$

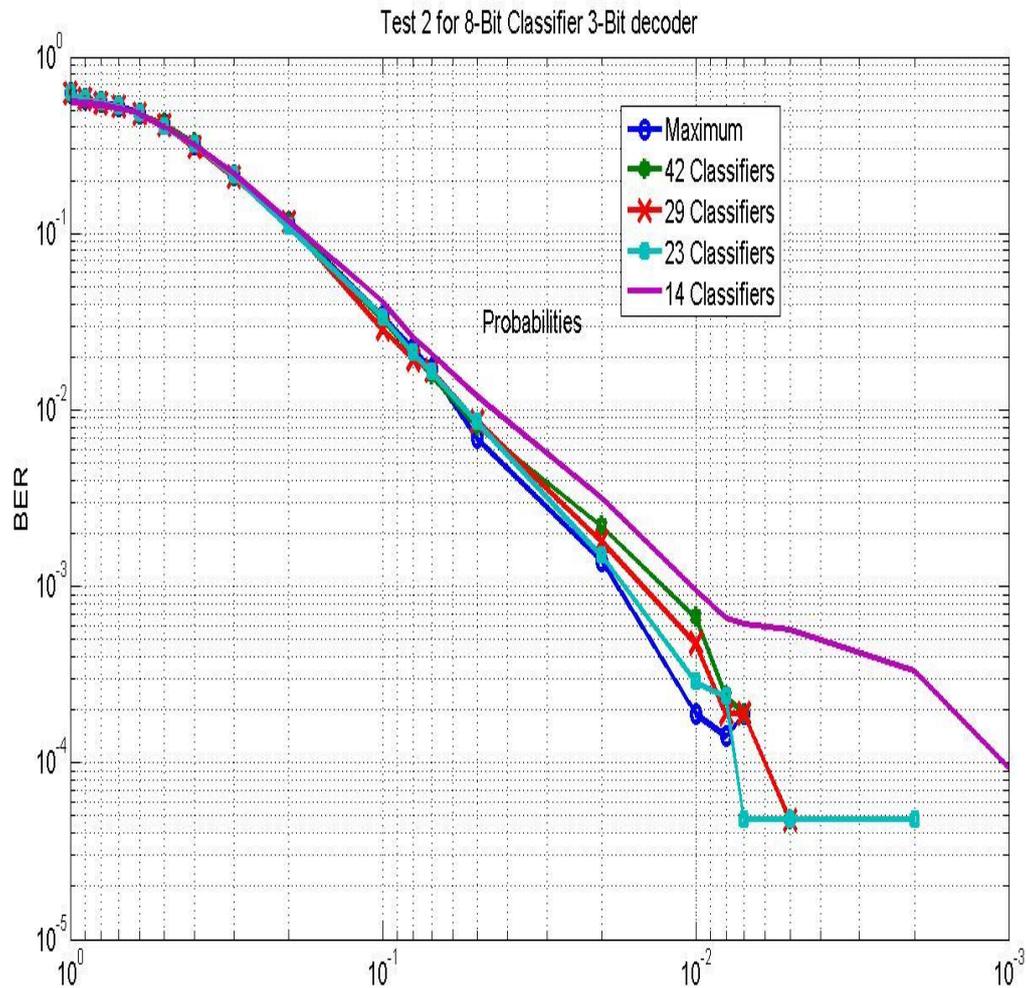

**Figure 6-14: Test 2 for 8-Bit Codeword 3-Bits Tuple Decoder**

Different decoders can be used at different probabilities in this case. At high probabilities, the 29 classifier can be used. And if memory is ever an issue the 29 classifier, i.e. Group B decoder can be the optimal decoder to use. At probabilities below 0.1, a group C decoder can be used where memory is of concern. To get a clear view on which one is better between group B and group C, the following graph can be used to check the difference.

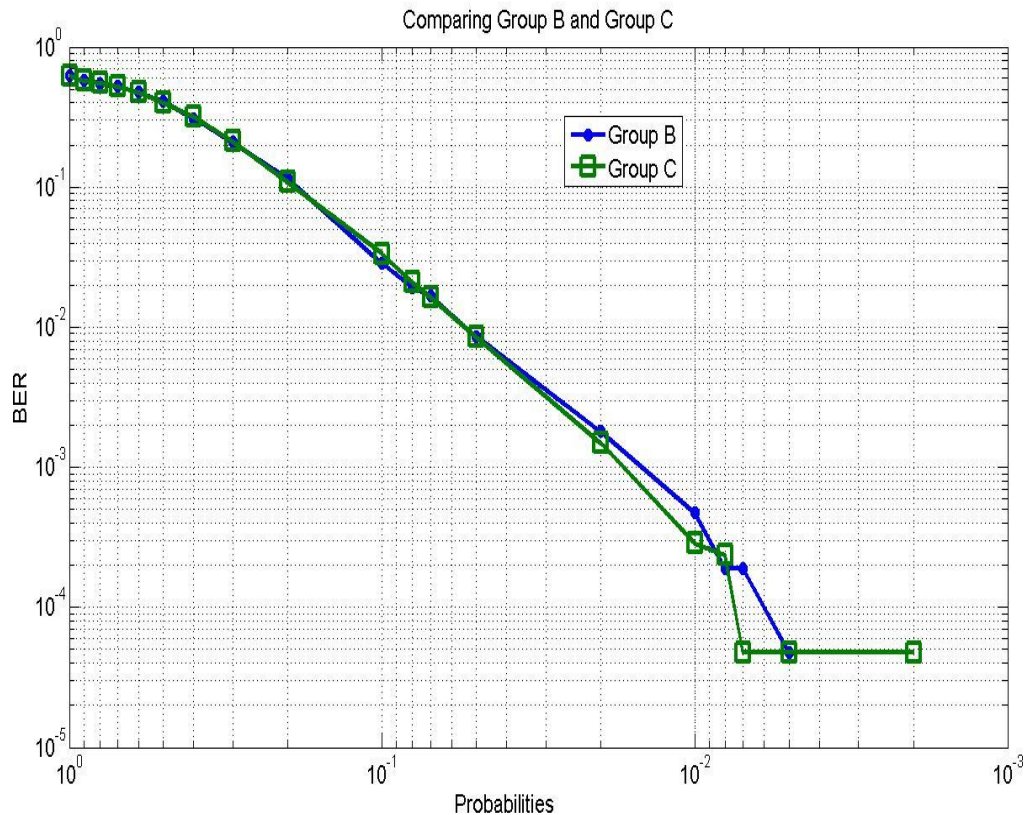

**Figure 6-15: Comparison of Group B and Group C Decoders**

As probabilities vary, they interchange, but Group C can be chosen as the best decoder because it stays below group B decoder's BER graph for most probabilities. Group B could have been statistically below group C decoder for this test, meaning that with another test it most probably will be low.

## 6.4 Test 2 Results Discussions

Test 2 investigates the effects of reducing the number of classifiers in the Neural Network Decoder. The reduction of the classifiers was done for $8 - Bits\ Codeword\ 2 - Bits\ Tuple\ decoder$ and $8 - Bits\ Codeword\ 3 - Bits\ Tuple\ decoder$. These two decoders were used because they were proved to be the best results of Test 1.

This test's results indicate that the reductions do not improve the efficiency but they don't decrease the efficiency. So the reduction of certain classifiers can be made if memory is of concern, because the reductions of the classifiers reduce the memory usage of the Neural Network Decoder.

The results obtained for the $8 - Bit\ codeword\ 2 - Bit\ tuple\ decoder$ indicate that for a decoder with maximum of 28 classifiers, the classifiers can be reduced to 15 classifiers per decoder. When the number of classifiers is below and equal to five (5), the efficiency of the decoder becomes very low.

$$\therefore 8 - Bit\ codeword\ 2 - Bit\ decoder$$
$$m \geq 15$$
$$which\ is\ approximately\ half\ of\ the\ maximum\ classifiers.$$

The results obtained for the $8 - Bits\ codeword\ 3 - Bit\ tuple\ decoder$ indicate that for this decoder of 56 maximum classifiers, the number of classifiers can be reduced to 23 classifiers per decoder. An $m = 23$ decoder has better efficiency. Any values below this will have very bad results, meaning that the decoder will not be efficient.

$$\therefore for\ 8 - Bit\ codeword\ 3 - Bit\ decoder$$
$$m \geq 23$$
$$which\ is\ approximately\ half\ of\ the\ maximum\ classifiers$$

## 6.5  Best Results between of Test 1 and Test 2

The decoders which were chosen as the best decoders in the previous tests will be compared in the following graph.
Group B of Test 1 and Group C of Test 2 will be used.

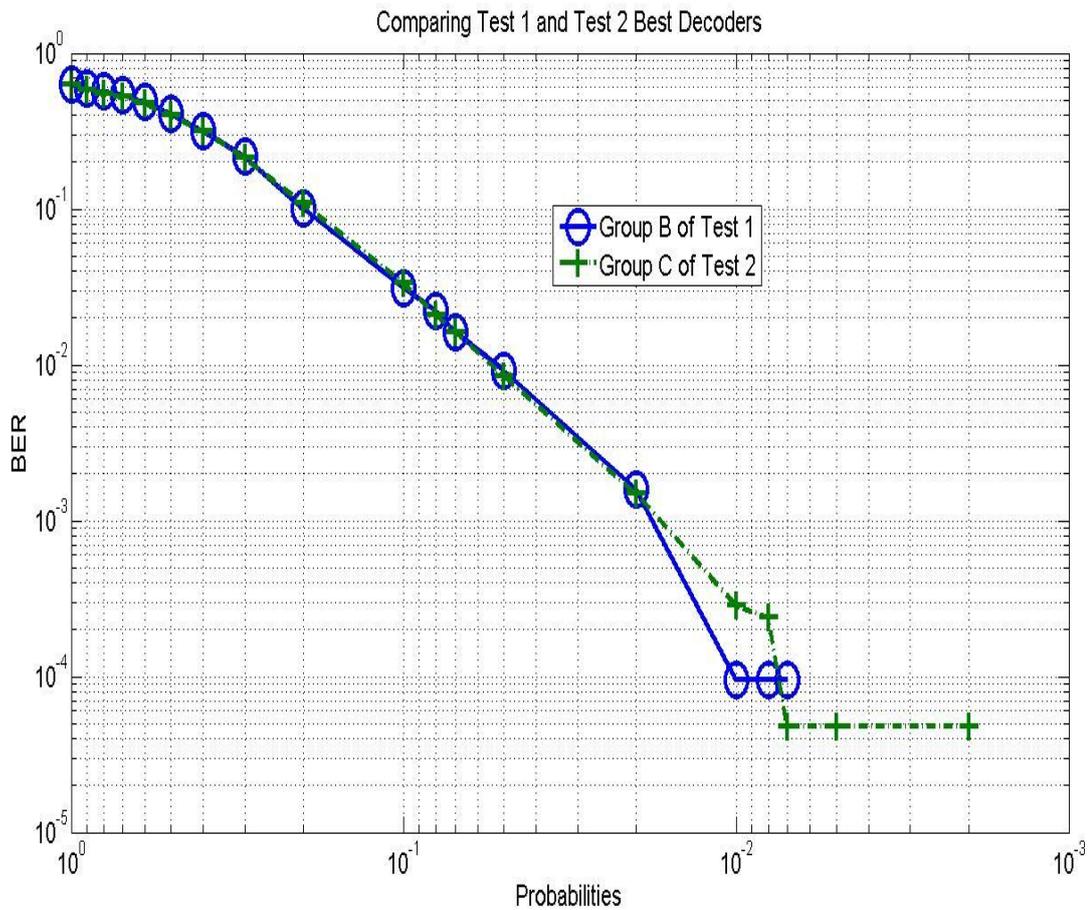

**Figure 6-16: Comparing Test 1 and Test 2 Best Decoders**

The efficiency of this graphs are approximately the same. The Test 1 group B decoder only become a little bit better at low error probabilities, from $0.01 \leq Probabilities \leq 0.02$ and Test 2 group C decoder becomes better at $0.002 \leq Probabilities \leq 0.008$.

## 6.6 Test 3

This test is to prove the hypothesis stated earlier, saying that the decoders become better as the codeword length increases. A 12-Bit codeword was tested using 3-Bit tuple decoders. The results of the 12-Bit codeword will be compared to that of 8-Bit codeword with the same mapper length.

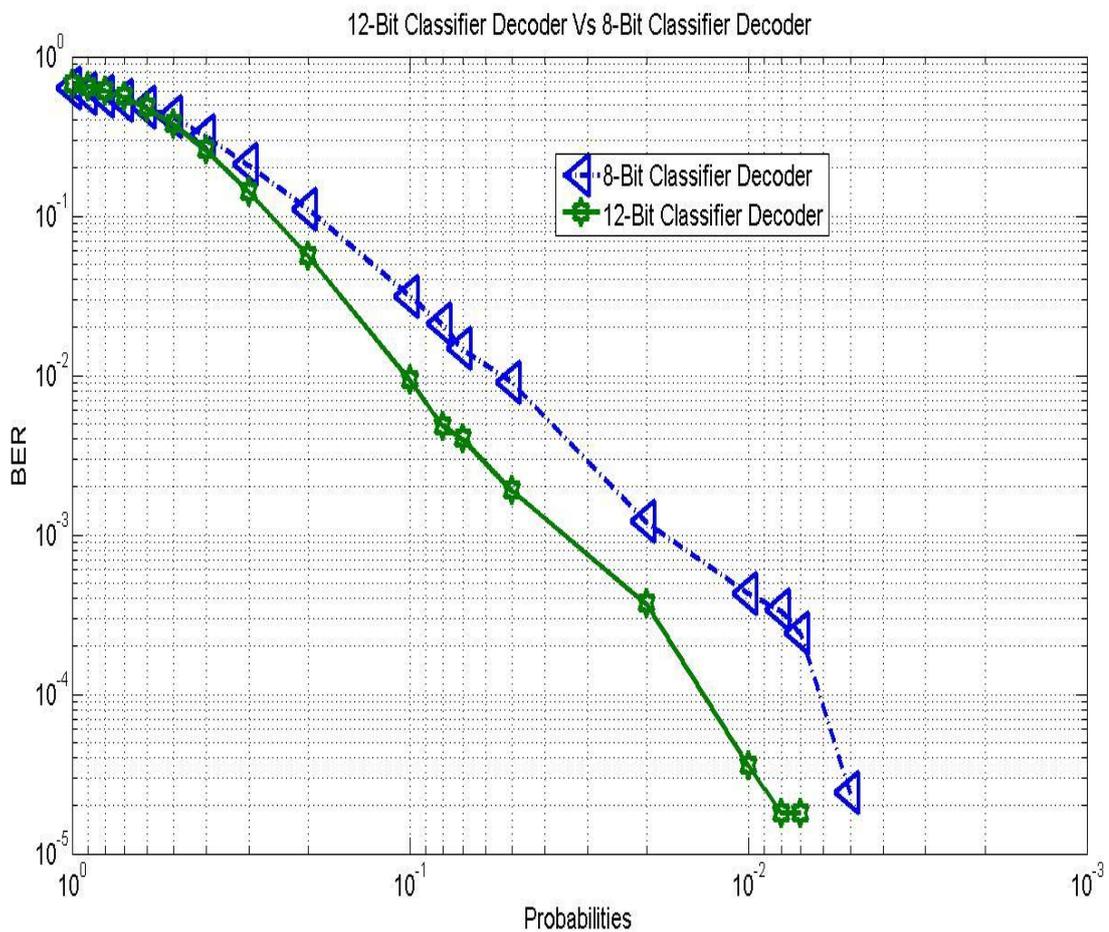

**Figure 6-17: 12 Bit Codeword Vs 8-Bit Codeword Decoder**

The above results prove that, as the size of the codeword increases, the NND becomes better. The efficiency of the 12-Bit codeword decoder is much higher than that of the 8-Bit codeword decoder.

## 6.7 Conclusion

The results obtained from the simulations were discussed in this chapter. There were three tests conducted in this chapter. All these tests were investigating different parameters of the Neural Network Decoder. Test 1 was discussing the effects of the codeword length and the mappers' length in a decoder. Test 2 was discussing the effects of the number of classifiers. Test 3 investigated the effects of LDPC bits on the Neural Network Decoder.

The parameters investigated in Test 1 are investigated as a unit. Each codeword length has certain number of mappers' length. The best length of the codeword was recognised and the best mappers' length associated with that codeword length was also noted. These best parameters were then investigated in Test 2.

Each codeword length has its associated mappers' length, and together they have a certain number of classifiers. Test 2 focused on reducing this number. The reduction of the number of classifiers assists in the reduction of memory usage and increasing the decoding speed of the decoder.

The final test, Test 3, focused on finding the effects of using a codeword with many bits. The $12 - Bits\ codeword$ was used for this purpose. This test was done with hopes that it will give an idea of how a NND will work if more bits were to be generated in the codeword.

From these tests it can be seen that the optimal Binary Neural Network Decoder parameters are:

$$length\ of\ the\ codeword:\ n \geq 8$$

$$length\ of\ the\ mapper: 2 \leq N \leq 3$$

$$m \geq \frac{C_N^n}{2}\ or\ m \cong \frac{C_N^n}{2}$$

# Chapter 7: Conclusion

## 7.1 Introduction

This chapter discusses the overview of the work done in this document and state the concluded results of the thesis. It also discusses the future work in using Binary Neural Network Decoder or any other form of Neural Network Decoders.

## 7.2 Overview

The main objective of this project was to find the optimal values of the neural networks parameters for decoding.

Chapter 1 defined the problem statement, objective and the problem scope in full details. It explained the methodology to be used when testing the different parameters when looking for the optimal ones.

The different decoding schemes which have been investigated before were stated in Chapter 2. This chapter also defined the terminologies used in this document. The introduction of the BNN decoder was also introduced. An overview of the channel to be used in the simulation was given, and the other different channel models were indicated.

More in depth overview of neural networks and their background information was given in Chapter 3. The different kinds of neural network and their mathematical design methods were discussed. An introduction to the BNN operation in error correction was mentioned and an example was given in terms of image processing.

Chapter 4 concentrated in the design of the BNN for decoding telecommunication errors. The mathematical design methods were indicated from which the design parameters which were investigated were noted. The tests which were conducted in the investigation of the effects of these parameters were mentioned. All conditions and assumptions which were considered in this project were mentioned.

Details on how each neural network parameter being investigated was going to be tested were indicated in Chapter 4. These details also included how the optimal values were going to be identified.

The software design method was indicated in Chapter 5, which explained all simulation components which were used in the simulation. Each component was individually tested to make sure that it is working. The components were proved to be working, thus the tests could then be conducted.

The results of the simulation tests were discussed in Chapter 6. These results were discussed as each test mentioned in Chapter 4 was conducted. The test results lead to the conclusion of the project as the optimal values of the project were found. The review of the BER graphs gave more details in leading to finding the optimal neural network values.

## 7.3 ECSA outcomes Achievements

### 7.3.1 Engineering Problem Solving

A creative approach to finding the solution to this project had to be used. One had to understand the problem statement of the project and think innovatively of the solution to the problem

### 7.3.2 Application of Fundamental and Specialist Knowledge

Mathematics skills were of great importance in this project as one had to think of different mathematical functions to use when implementing the proposed solution. The other skill applied was the programming skill since the mathematics skill needed to be implemented in the computer. The mathematics skill alone without the programming skill is limited to paper.

### 7.3.3 Engineering Design and Synthesis

Following the project scope, the proposed design managed to come up with the best solution for the thesis, thus meeting this requirement. The program written was able to show which parameters are the best parameters.

### 7.3.4 Investigation, Experiment and Data Analysis

There was a lot of research done, and mostly consultations, since this was a new topic in the field. Mostly the people consulted were those who have worked with Neural Networks in the control systems field.

The graphs obtained from the tests were analysed and theoretical behaviour of the decoders' BER graph was used in these analysis. From the work which was done before on decoders, a general idea of how a good decoder should be was established. This had to be used in analysing these data thus reaching the conclusion which this document provides.

### 7.3.5 Engineering Methods, Skills, Tools and Information Technology

MATLAB was used to simulate the results and the simulations were run a number of times to reduce the statistical errors made by this computer packaged software.

## 7.4 Future Work

This thesis concentrated on separate mappers' length per decoder. The next work could be to investigate the effects of using different mapper's length in one decoder.

The choice of codeword for the codebook is very important. Finding the right codeword to use in the BNN decoder might increase the decoder's efficiency.

Investigate to see if weights can be added to the votes of the combination which seems to be the most important votes in the decoder.

Investigate if synchronization errors can be solved in a similar manner, but only after synchronisation has been regained. If markers method is used to regain synchronisation, the markers positions will be skipped when comparing the codewords.

And the bits between the markers can be solved as additive errors, assuming that the markers are still correct. If the markers position have been shifted in this new synchronised message, the decoded message will be wrong.

# APPENDIX

This section will show the code used in this project. This code is the decoder's code.

```
Function
[vote1,vote2,vote3,vote4]=eight_bits_two_tuplles(word1,word2,word3,word4,rec_word_chan
nel)

   load data %% this is the data saved from the word generate function
   load data1 %%this is the data saved from the channel simulation
   load data2 %% this is the data saved from the main function

   for p=1:10 %% this represents the iterations
      o=1;
    [rec_word,numbOfErr]= channel_simulation(enc_word,Prec,Pdel,Pins,Pinv)
   for c=0:b
    c=c+1;
    e=c-1;
    rec_word_channel=[rec_word(e*d+1:c*d)] %% the function used to seperate
the bits
    N=2;
switch N==N

    case 1
        N=2;
     %% 1 and 2
    word_1=[ word1(1) word1(2)]==[rec_word_channel(1) rec_word_channel(2)];
        if (word_1==ones(1,N))
            vote1=1;
        else
            vote1=0;
        end
    word_2=[ word2(1) word2(2)]==[rec_word_channel(1) rec_word_channel(2)];
     if (word_2==ones(1,N))
            vote2=1;
        else
            vote2=0;
        end
    word_3=[ word3(1) word3(2)]==[rec_word_channel(1) rec_word_channel(2)];
     if (word_3==ones(1,N))
            vote3=1;
        else
            vote3=0;
        end
    word_4=[ word4(1) word4(2)]==[rec_word_channel(1) rec_word_channel(2)];
     if (word_4==ones(1,N))
            vote4=1;
        else
```

```
            vote4=0;
      end

    %% 1 and 3
    word_1=[ word1(1) word1(3)]==[rec_word_channel(1) rec_word_channel(3)];
        if (word_1==ones(1,N))
            vote1=1;
        else
            vote1=0;
        end
     word_2=[ word2(1) word2(3)]==[rec_word_channel(1) rec_word_channel(3)];
     if (word_2==ones(1,N))
            vote2=1;
        else
            vote2=0;
        end
     word_3=[ word3(1) word3(3)]==[rec_word_channel(1) rec_word_channel(3)];
     if (word_3==ones(1,N))
            vote3=1;
        else
            vote3=0;
        end
    word_4=[ word4(1) word4(3)]==[rec_word_channel(1) rec_word_channel(3)];
    if (word_4==ones(1,N))
            vote4=1;
        else
            vote4=0;
     end

     %% 1 and 4
     word_1=[ word1(1) word1(4)]==[rec_word_channel(1) rec_word_channel(4)];
        if (word_1==ones(1,N))
            vote1=vote1+1;
        else
            vote1=vote1+0;
        end
    word_2=[ word2(1) word2(4)]==[rec_word_channel(1) rec_word_channel(4)];
    if (word_2==ones(1,N))
            vote2=vote2+1;
        else
            vote2=vote2+0;
        end
    word_3=[ word3(1) word3(4)]==[rec_word_channel(1) rec_word_channel(4)];
    if (word_3==ones(1,N))
            vote3=vote3+1;
        else
            vote3= vote3+0;
        end
    word_4=[ word4(1) word4(4)]==[rec_word_channel(1) rec_word_channel(4)];
    if (word_4==ones(1,N))
            vote4=vote4+1;
        else
            vote4=vote4+0;
```

```
    end

%% 1 and 5
word_1=[ word1(1) word1(5)]==[rec_word_channel(1) rec_word_channel(5)];
    if (word_1==ones(1,N))
        vote1=vote1+1;
    else
        vote1=vote1+0;
    end
word_2=[ word2(1) word2(5)]==[rec_word_channel(1) rec_word_channel(5)];
if (word_2==ones(1,N))
        vote2=vote2+1;
    else
        vote2=vote2+0;
    end
word_3=[ word3(1) word3(5)]==[rec_word_channel(1) rec_word_channel(5)];
if (word_3==ones(1,N))
        vote3=vote3+1;
    else
        vote3= vote3+0;
    end
word_4=[ word4(1) word4(5)]==[rec_word_channel(1) rec_word_channel(5)];
if (word_4==ones(1,N))
        vote4=vote4+1;
    else
        vote4=vote4+0;
end

%% 1 and 6
 word_1=[ word1(1) word1(6)]==[rec_word_channel(1) rec_word_channel(6)];
    if (word_1==ones(1,N))
        vote1=vote1+1;
    else
        vote1=vote1+0;
    end
word_2=[ word2(1) word2(6)]==[rec_word_channel(1) rec_word_channel(6)];
if (word_2==ones(1,N))
        vote2=vote2+1;
    else
        vote2=vote2+0;
    end
word_3=[ word3(1) word3(6)]==[rec_word_channel(1) rec_word_channel(6)];
if (word_3==ones(1,N))
        vote3=vote3+1;
    else
        vote3= vote3+0;
    end
word_4=[ word4(1) word4(6)]==[rec_word_channel(1) rec_word_channel(6)];
if (word_4==ones(1,N))
        vote4=vote4+1;
    else
        vote4=vote4+0;
    end
```

```
%% 1 and 7
word_1=[ word1(1) word1(7)]==[rec_word_channel(1) rec_word_channel(7)];
    if (word_1==ones(1,N))
        vote1=vote1+1;
    else
        vote1=vote1+0;
    end
word_2=[ word2(1) word2(7)]==[rec_word_channel(1) rec_word_channel(7)];
if (word_2==ones(1,N))
        vote2=vote2+1;
    else
        vote2=vote2+0;
    end
word_3=[ word3(1) word3(7)]==[rec_word_channel(1) rec_word_channel(7)];
if (word_3==ones(1,N))
        vote3=vote3+1;
    else
        vote3= vote3+0;
    end
word_4=[ word4(1) word4(7)]==[rec_word_channel(1) rec_word_channel(7)];
if (word_4==ones(1,N))
        vote4=vote4+1;
    else
        vote4=vote4+0;
end
 %% 1 and 8
word_1=[ word1(1) word1(8)]==[rec_word_channel(1) rec_word_channel(8)];
    if (word_1==ones(1,N))
        vote1=vote1+1;
    else
        vote1=vote1+0;
    end
word_2=[ word2(1) word2(8)]==[rec_word_channel(1) rec_word_channel(8)];
if (word_2==ones(1,N))
        vote2=vote2+1;
    else
        vote2=vote2+0;
    end
word_3=[ word3(1) word3(8)]==[rec_word_channel(1) rec_word_channel(8)];
if (word_3==ones(1,N))
        vote3=vote3+1;
    else
        vote3= vote3+0;
    end
word_4=[ word4(1) word4(8)]==[rec_word_channel(1) rec_word_channel(8)];
if (word_4==ones(1,N))
        vote4=vote4+1;
    else
        vote4=vote4+0;
end

%% 2 and 3
word_1=[ word1(2) word1(3)]==[rec_word_channel(2) rec_word_channel(3)];
    if (word_1==ones(1,N))
```

```
            vote1=vote1+1;
        else
            vote1=vote1+0;
        end
    word_2=[ word2(2)  word2(3)]==[rec_word_channel(2) rec_word_channel(3)];
    if (word_2==ones(1,N))
            vote2=vote2+1;
        else
            vote2=vote2+0;
        end
    word_3=[ word3(2)  word3(3)]==[rec_word_channel(2) rec_word_channel(3)];
    if (word_3==ones(1,N))
            vote3=vote3+1;
        else
            vote3= vote3+0;
        end
    word_4=[ word4(2)  word4(3)]==[rec_word_channel(2) rec_word_channel(3)];
    if (word_4==ones(1,N))
            vote4=vote4+1;
        else
            vote4=vote4+0;
    end

    %% 2 and 4
    word_1=[ word1(2)  word1(4)]==[rec_word_channel(2) rec_word_channel(4)];
        if (word_1==ones(1,N))
            vote1=vote1+1;
        else
            vote1=vote1+0;
        end
    word_2=[ word2(2)  word2(4)]==[rec_word_channel(2) rec_word_channel(4)];
    if (word_2==ones(1,N))
            vote2=vote2+1;
        else
            vote2=vote2+0;
        end
    word_3=[ word3(2)  word3(4)]==[rec_word_channel(2) rec_word_channel(4)];
    if (word_3==ones(1,N))
            vote3=vote3+1;
        else
            vote3= vote3+0;
        end
    word_4=[ word4(2)  word4(4)]==[rec_word_channel(2) rec_word_channel(4)];
    if (word_4==ones(1,N))
            vote4=vote4+1;
        else
            vote4=vote4+0;
    end

    %% 2 and 5
    word_1=[ word1(2)  word1(5)]==[rec_word_channel(2) rec_word_channel(5)];
        if (word_1==ones(1,N))
            vote1=vote1+1;
        else
```

```matlab
            vote1=vote1+0;
        end
   word_2=[ word2(2)  word2(5)]==[rec_word_channel(2) rec_word_channel(5)];
   if (word_2==ones(1,N))
            vote2=vote2+1;
        else
            vote2=vote2+0;
        end
   word_3=[ word3(2)  word3(5)]==[rec_word_channel(2) rec_word_channel(5)];
   if (word_3==ones(1,N))
            vote3=vote3+1;
        else
            vote3= vote3+0;
        end
   word_4=[ word4(2)  word4(5)]==[rec_word_channel(2) rec_word_channel(5)];
   if (word_4==ones(1,N))
            vote4=vote4+1;
        else
            vote4=vote4+0;
    end

    %% 2 and 6
    word_1=[ word1(2)  word1(6)]==[rec_word_channel(2) rec_word_channel(6)];
        if (word_1==ones(1,N))
            vote1=vote1+1;
        else
            vote1=vote1+0;
        end
   word_2=[ word2(2)  word2(6)]==[rec_word_channel(2) rec_word_channel(6)];
   if (word_2==ones(1,N))
            vote2=vote2+1;
        else
            vote2=vote2+0;
        end
   word_3=[ word3(2)  word3(6)]==[rec_word_channel(2) rec_word_channel(6)];
   if (word_3==ones(1,N))
            vote3=vote3+1;
        else
            vote3= vote3+0;
        end
   word_4=[ word4(2)  word4(6)]==[rec_word_channel(2) rec_word_channel(6)];
   if (word_4==ones(1,N))
            vote4=vote4+1;
        else
            vote4=vote4+0;
    end

    %% 2 and 7
     word_1=[ word1(2)  word1(7)]==[rec_word_channel(2) rec_word_channel(7)];
        if (word_1==ones(1,N))
            vote1=vote1+1;
        else
            vote1=vote1+0;
        end
```

```
    word_2=[ word2(2) word2(7)]==[rec_word_channel(2) rec_word_channel(7)];
    if (word_2==ones(1,N))
          vote2=vote2+1;
       else
          vote2=vote2+0;
       end
word_3=[ word3(2) word3(7)]==[rec_word_channel(2) rec_word_channel(7)];
    if (word_3==ones(1,N))
          vote3=vote3+1;
       else
          vote3= vote3+0;
       end
word_4=[ word4(2) word4(7)]==[rec_word_channel(2) rec_word_channel(7)];
    if (word_4==ones(1,N))
          vote4=vote4+1;
       else
          vote4=vote4+0;
    end

    %% 2 and 8
    word_1=[ word1(2) word1(8)]==[rec_word_channel(2) rec_word_channel(8)];
       if (word_1==ones(1,N))
          vote1=vote1+1;
       else
          vote1=vote1+0;
       end
word_2=[ word2(2) word2(8)]==[rec_word_channel(2) rec_word_channel(8)];
    if (word_2==ones(1,N))
          vote2=vote2+1;
       else
          vote2=vote2+0;
       end
word_3=[ word3(2) word3(8)]==[rec_word_channel(2) rec_word_channel(8)];
    if (word_3==ones(1,N))
          vote3=vote3+1;
       else
          vote3= vote3+0;
       end
word_4=[ word4(2) word4(8)]==[rec_word_channel(2) rec_word_channel(8)];
    if (word_4==ones(1,N))
          vote4=vote4+1;
       else
          vote4=vote4+0;
    end

    %% 3 and 4
    word_1=[ word1(3) word1(4)]==[rec_word_channel(3) rec_word_channel(4)];
       if (word_1==ones(1,N))
          vote1=vote1+1;
       else
          vote1=vote1+0;
       end
    word_2=[ word2(3) word2(4)]==[rec_word_channel(3) rec_word_channel(4)];
```

```
        if (word_2==ones(1,N))
            vote2=vote2+1;
        else
            vote2=vote2+0;
        end
word_3=[ word3(3) word3(4)]==[rec_word_channel(3) rec_word_channel(4)];
if (word_3==ones(1,N))
            vote3=vote3+1;
        else
            vote3= vote3+0;
        end
word_4=[ word4(3) word4(4)]==[rec_word_channel(3) rec_word_channel(4)];
if (word_4==ones(1,N))
            vote4=vote4+1;
        else
            vote4=vote4+0;
        end

 %% 3 and 5
word_1=[ word1(3) word1(5)]==[rec_word_channel(3) rec_word_channel(5)];
    if (word_1==ones(1,N))
            vote1=vote1+1;
        else
            vote1=vote1+0;
        end
word_2=[ word2(3) word2(5)]==[rec_word_channel(3) rec_word_channel(5)];
if (word_2==ones(1,N))
            vote2=vote2+1;
        else
            vote2=vote2+0;
        end
word_3=[ word3(3) word3(5)]==[rec_word_channel(3) rec_word_channel(5)];
if (word_3==ones(1,N))
            vote3=vote3+1;
        else
            vote3= vote3+0;
        end
word_4=[ word4(3) word4(5)]==[rec_word_channel(3) rec_word_channel(5)];
if (word_4==ones(1,N))
            vote4=vote4+1;
        else
            vote4=vote4+0;
        end
 %% 3 and 6
word_1=[ word1(3) word1(6)]==[rec_word_channel(3) rec_word_channel(6)];
    if (word_1==ones(1,N))
            vote1=vote1+1;
        else
            vote1=vote1+0;
        end
word_2=[ word2(3) word2(6)]==[rec_word_channel(3) rec_word_channel(6)];
if (word_2==ones(1,N))
            vote2=vote2+1;
        else
```

```
            vote2=vote2+0;
        end
    word_3=[ word3(3) word3(6)]==[rec_word_channel(3) rec_word_channel(6)];
    if (word_3==ones(1,N))
            vote3=vote3+1;
        else
            vote3= vote3+0;
        end
    word_4=[ word4(3) word4(6)]==[rec_word_channel(3) rec_word_channel(6)];
    if (word_4==ones(1,N))
            vote4=vote4+1;
        else
            vote4=vote4+0;
    end

    %% 3 and 7
    word_1=[ word1(3) word1(7)]==[rec_word_channel(3) rec_word_channel(7)];
        if (word_1==ones(1,N))
            vote1=vote1+1;
        else
            vote1=vote1+0;
        end
    word_2=[ word2(3) word2(7)]==[rec_word_channel(3) rec_word_channel(7)];
    if (word_2==ones(1,N))
            vote2=vote2+1;
        else
            vote2=vote2+0;
        end
    word_3=[ word3(3) word3(7)]==[rec_word_channel(3) rec_word_channel(7)];
    if (word_3==ones(1,N))
            vote3=vote3+1;
        else
            vote3= vote3+0;
        end
    word_4=[ word4(3) word4(7)]==[rec_word_channel(3) rec_word_channel(7)];
    if (word_4==ones(1,N))
            vote4=vote4+1;
        else
            vote4=vote4+0;
    end

    %% 3 and 8
    word_1=[ word1(3) word1(8)]==[rec_word_channel(3) rec_word_channel(8)];
        if (word_1==ones(1,N))
            vote1=vote1+1;
        else
            vote1=vote1+0;
        end
    word_2=[ word2(3) word2(8)]==[rec_word_channel(3) rec_word_channel(8)];
    if (word_2==ones(1,N))
            vote2=vote2+1;
        else
            vote2=vote2+0;
```

```
        end
    word_3=[ word3(3) word3(8)]==[rec_word_channel(3) rec_word_channel(8)];
    if (word_3==ones(1,N))
          vote3=vote3+1;
       else
          vote3= vote3+0;
       end
    word_4=[ word4(3) word4(8)]==[rec_word_channel(3) rec_word_channel(8)];
    if (word_4==ones(1,N))
          vote4=vote4+1;
       else
          vote4=vote4+0;
     end

 %% 4 and 5
  word_1=[ word1(4) word1(5)]==[rec_word_channel(4) rec_word_channel(5)];
      if (word_1==ones(1,N))
          vote1=vote1+1;
       else
          vote1=vote1+0;
       end
    word_2=[ word2(4) word2(5)]==[rec_word_channel(4) rec_word_channel(5)];
    if (word_2==ones(1,N))
          vote2=vote2+1;
       else
          vote2=vote2+0;
       end
    word_3=[ word3(4) word3(5)]==[rec_word_channel(4) rec_word_channel(5)];
    if (word_3==ones(1,N))
          vote3=vote3+1;
       else
          vote3= vote3+0;
       end
    word_4=[ word4(4) word4(5)]==[rec_word_channel(4) rec_word_channel(5)];
    if (word_4==ones(1,N))
          vote4=vote4+1;
       else
          vote4=vote4+0;
     end

  %% 4 and 6
  word_1=[ word1(4) word1(6)]==[rec_word_channel(4) rec_word_channel(6)];
      if (word_1==ones(1,N))
          vote1=vote1+1;
       else
          vote1=vote1+0;
       end
    word_2=[ word2(4) word2(6)]==[rec_word_channel(4) rec_word_channel(6)];
    if (word_2==ones(1,N))
          vote2=vote2+1;
       else
          vote2=vote2+0;
       end
    word_3=[ word3(4) word3(6)]==[rec_word_channel(4) rec_word_channel(6)];
```

```
         if (word_3==ones(1,N))
              vote3=vote3+1;
          else
              vote3= vote3+0;
          end
     word_4=[ word4(4) word4(6)]==[rec_word_channel(4) rec_word_channel(6)];
     if (word_4==ones(1,N))
              vote4=vote4+1;
          else
              vote4=vote4+0;
     end
     
     %% 4 and 7
     word_1=[ word1(4) word1(7)]==[rec_word_channel(4) rec_word_channel(7)];
         if (word_1==ones(1,N))
              vote1=vote1+1;
          else
              vote1=vote1+0;
          end
     word_2=[ word2(4) word2(7)]==[rec_word_channel(4) rec_word_channel(7)];
     if (word_2==ones(1,N))
              vote2=vote2+1;
          else
              vote2=vote2+0;
          end
     word_3=[ word3(4) word3(7)]==[rec_word_channel(4) rec_word_channel(7)];
     if (word_3==ones(1,N))
              vote3=vote3+1;
          else
              vote3= vote3+0;
          end
     word_4=[ word4(4) word4(7)]==[rec_word_channel(4) rec_word_channel(7)];
     if (word_4==ones(1,N))
              vote4=vote4+1;
          else
              vote4=vote4+0;
     end
     
     %% 4 and 8
     word_1=[ word1(4) word1(8)]==[rec_word_channel(4) rec_word_channel(8)];
         if (word_1==ones(1,N))
              vote1=vote1+1;
          else
              vote1=vote1+0;
          end
     word_2=[ word2(4) word2(8)]==[rec_word_channel(4) rec_word_channel(8)];
     if (word_2==ones(1,N))
              vote2=vote2+1;
          else
              vote2=vote2+0;
          end
     word_3=[ word3(4) word3(8)]==[rec_word_channel(4) rec_word_channel(8)];
     if (word_3==ones(1,N))
              vote3=vote3+1;
```

```
        else
            vote3= vote3+0;
        end
    word_4=[ word4(4) word4(8)]==[rec_word_channel(4) rec_word_channel(8)];
    if (word_4==ones(1,N))
            vote4=vote4+1;
        else
            vote4=vote4+0;
    end

    %% 5 and 6
     word_1=[ word1(5) word1(6)]==[rec_word_channel(5) rec_word_channel(6)];
        if (word_1==ones(1,N))
            vote1=vote1+1;
        else
            vote1=vote1+0;
        end
    word_2=[ word2(5) word2(6)]==[rec_word_channel(5) rec_word_channel(6)];
    if (word_2==ones(1,N))
            vote2=vote2+1;
        else
            vote2=vote2+0;
        end
    word_3=[ word3(5) word3(6)]==[rec_word_channel(5) rec_word_channel(6)];
    if (word_3==ones(1,N))
            vote3=vote3+1;
        else
            vote3= vote3+0;
        end
    word_4=[ word4(5) word4(6)]==[rec_word_channel(5) rec_word_channel(6)];
    if (word_4==ones(1,N))
            vote4=vote4+1;
        else
            vote4=vote4+0;
    end

    %% 5 and 7
     word_1=[ word1(5) word1(7)]==[rec_word_channel(5) rec_word_channel(7)];
        if (word_1==ones(1,N))
            vote1=vote1+1;
        else
            vote1=vote1+0;
        end
    word_2=[ word2(5) word2(7)]==[rec_word_channel(5) rec_word_channel(7)];
    if (word_2==ones(1,N))
            vote2=vote2+1;
        else
            vote2=vote2+0;
        end
    word_3=[ word3(5) word3(7)]==[rec_word_channel(5) rec_word_channel(7)];
    if (word_3==ones(1,N))
            vote3=vote3+1;
```

```matlab
    else
        vote3= vote3+0;
    end
word_4=[ word4(5) word4(7)]==[rec_word_channel(5) rec_word_channel(7)];
if (word_4==ones(1,N))
        vote4=vote4+1;
    else
        vote4=vote4+0;
end

%% 5 and 8
word_1=[ word1(5) word1(8)]==[rec_word_channel(5) rec_word_channel(8)];
    if (word_1==ones(1,N))
        vote1=vote1+1;
    else
        vote1=vote1+0;
    end
word_2=[ word2(5) word2(8)]==[rec_word_channel(5) rec_word_channel(8)];
if (word_2==ones(1,N))
        vote2=vote2+1;
    else
        vote2=vote2+0;
    end
word_3=[ word3(5) word3(8)]==[rec_word_channel(5) rec_word_channel(8)];
if (word_3==ones(1,N))
        vote3=vote3+1;
    else
        vote3= vote3+0;
    end
word_4=[ word4(5) word4(8)]==[rec_word_channel(5) rec_word_channel(8)];
if (word_4==ones(1,N))
        vote4=vote4+1;
    else
        vote4=vote4+0;
end

%% 6 and 7
 word_1=[ word1(6) word1(7)]==[rec_word_channel(6) rec_word_channel(7)];
    if (word_1==ones(1,N))
        vote1=vote1+1;
    else
        vote1=vote1+0;
    end
word_2=[ word2(6) word2(7)]==[rec_word_channel(6) rec_word_channel(7)];
if (word_2==ones(1,N))
        vote2=vote2+1;
    else
        vote2=vote2+0;
    end
word_3=[ word3(6) word3(7)]==[rec_word_channel(6) rec_word_channel(7)];
if (word_3==ones(1,N))
        vote3=vote3+1;
    else
        vote3= vote3+0;
```

```matlab
        end
    word_4=[ word4(6) word4(7)]==[rec_word_channel(6) rec_word_channel(7)];
    if (word_4==ones(1,N))
            vote4=vote4+1;
        else
            vote4=vote4+0;
    end
    %% 6 and 8
     word_1=[ word1(6) word1(8)]==[rec_word_channel(6) rec_word_channel(8)];
        if (word_1==ones(1,N))
            vote1=vote1+1;
        else
            vote1=vote1+0;
        end
    word_2=[ word2(6) word2(8)]==[rec_word_channel(6) rec_word_channel(8)];
    if (word_2==ones(1,N))
            vote2=vote2+1;
        else
            vote2=vote2+0;
        end
    word_3=[ word3(6) word3(8)]==[rec_word_channel(6) rec_word_channel(8)];
    if (word_3==ones(1,N))
            vote3=vote3+1;
        else
            vote3= vote3+0;
        end
    word_4=[ word4(6) word4(8)]==[rec_word_channel(6) rec_word_channel(8)];
    if (word_4==ones(1,N))
            vote4=vote4+1;
        else
            vote4=vote4+0;
    end

    %% 7 and 8
    word_1=[ word1(7) word1(8)]==[rec_word_channel(7) rec_word_channel(8)];
        if (word_1==ones(1,N))
            vote1=vote1+1;
        else
            vote1=vote1+0;
        end
    word_2=[ word2(7) word2(8)]==[rec_word_channel(7) rec_word_channel(8)];
    if (word_2==ones(1,N))
            vote2=vote2+1;
        else
            vote2=vote2+0;
        end
    word_3=[ word3(7) word3(8)]==[rec_word_channel(7) rec_word_channel(8)];
    if (word_3==ones(1,N))
            vote3=vote3+1;
        else
            vote3= vote3+0;
        end
    word_4=[ word4(7) word4(8)]==[rec_word_channel(7) rec_word_channel(8)];
```

```
        if (word_4==ones(1,N))
             vote4=vote4+1;
         else
             vote4=vote4+0;
     end

end

%% Plot the Graph

    max_votes= max([vote1 vote2 vote3 vote4]);
        if vote1==max_votes
        correctMessage   =   strcat('The   corrected   message   will   be   :
',num2str(word1))
        corrected=strcat('',num2str(word1)) %% the correcet message will be
uidisplayed
        decoded_word(e*d+1:c*d)=repmat(word1,1,1);  %%  the  correct  decoded
word will now be replaced in the decoded
        end
        if vote2==max_votes
        correctMessage   =   strcat('The   corrected   message   will   be   :
',num2str(word2))
        corrected=strcat('',num2str(word2))
        decoded_word(e*d+1:c*d)=repmat(word2,1,1);
        end
        if vote3==max_votes
        correctMessage   =   strcat('The   corrected   message   will   be   :
',num2str(word3))
        corrected=strcat('',num2str(word3))
        decoded_word(e*d+1:c*d)=repmat(word3,1,1);
        end
        if vote4==max_votes
        correctMessage   =   strcat('The   corrected   message   will   be   :
',num2str(word4))
        corrected=strcat('',num2str(word4))
        decoded_word(e*d+1:c*d)=repmat(word4,1,1);
        end
        if max_votes==0
            'undicisive'
        end

   end
decoded_word
    inversion_errors=0;
error_found=0;

for j=1:length(enc_word)

word_error=enc_word(j)==decoded_word(j);
    if word_error==0
        error_found=error_found+1;
```

```
    end
        message_error=enc_word(j)==rec_word(j);
    if message_error==0
        inversion_errors=inversion_errors+1;
    end
end
BitErrorRates(p)=error_found/length(enc_word);
 inv_errors(p)=inversion_errors;
error_found;
    end
   BitErrorRate=sum(BitErrorRates)/p
     avarage_inversion=mean(inv_errors) %% it calculate the average
                                        %% inversions per sent message
   total_inversions=sum(inv_errors)
 save  data3   vote1   vote2   vote3   vote4  max_votes  decoded_word  %% this
information  will  be  used  when  plotting  the  bar  graph  and  calculating  the
voting confidence
```